\documentclass[usenatbib,lscape]{emulateapj}
\usepackage{graphicx,ifthen,url,float,morefloats,lscape}
\newcommand {\kms}{km s$^{-1}$}

\newcommand {\lya}{Ly$\alpha$ }
\def\ltsima{$\; \buildrel < \over \sim \;$}
\def\simlt{\lower.5ex\hbox{\ltsima}}
\def\gtsima{$\; \buildrel > \over \sim \;$}
\def\simgt{\lower.5ex\hbox{\gtsima}}
\newcommand {\uJy}{$\mu$Jy}
\newcommand {\um}{$\mu$m}

\def\um     {$\mu$m}
\def\ts     {\thinspace}
\def\kms    {\ifmmode{{\rm \ts km\ts s}^{-1}}\else{\ts km\ts s$^{-1}$}\fi}
\def\msol   {\ifmmode{{\rm M}_{\odot}}\else{M$_{\odot}$}\fi}
\def\lsol   {\ifmmode{{\rm L}_{\odot}}\else{L$_{\odot}$}\fi}
\def\zsol   {\ifmmode{{\rm Z}_{\odot}}\else{Z$_{\odot}$}\fi}
\def\etal   {{\rm et\ts al.}}
\def\ci     {\ifmmode{{\rm C}{\rm \small I}}\else{C\ts {\scriptsize I}}\fi}
\def\hi     {\ifmmode{{\rm H}{\rm \small I}}\else{H\ts {\scriptsize I}}\fi}
\def\hh     {\ifmmode{{\rm H}_2}\else{H$_2$}\fi}
\def\cone {\ifmmode{{\rm C}{\rm \small I}(^3\!P_1\!\to^3\!P_0)}
     \else{C\ts {\scriptsize I}{\small$(^3\!P_1\!\to\,^3\!P_0)$}}\fi}
\def\ctwo {\ifmmode{{\rm C}{\rm \small I}(^3\!P_2\!\to\,^3\!P_1)}
     \else{C\ts {\scriptsize I}{\small$(^3\!P_2\!\to\,^3\!P_1)$}}\fi}
\def\cij {\ifmmode{{\rm C}{\rm \small I}\,(^3P_i\to^3P_j)}\else{C\ts {\scriptsize I}\,{\small$(^3P_i\to^3P_j)$}}\fi}
\def\cii    {\ifmmode{{\rm C}{\rm \small II}}\else{C\ts {\scriptsize II}}\fi}
\def\tex {\ifmmode{{T}_{\rm ex}}\else{$T_{\rm ex}$}\fi}
\def\tmb {\ifmmode{{T}_{\rm mb}}\else{$T_{\rm mb}$}\fi}
\def\tkin {\ifmmode{{T}_{\rm kin}}\else{$T_{\rm kin}$}\fi}
\def\microns {\ifmmode{\mu{\rm m}}\else{$\mu$m}\fi}
\def\nhh   {\ifmmode{n({\rm H}_2)}\else{$n$(H$_2$)}\fi}

\newcommand{\sfr}{{\rm\,M$_\odot$\,yr$^{-1}$}}
\newcommand{\lsun}{{\rm\,L$_\odot$}}

\newcommand{\ha}{{\rm\,H$\alpha$}}
\newcommand{\hb}{{\rm\,H$\beta$}}
\newcommand{\hg}{{\rm\,H$\gamma$}}
\newcommand{\heii}{{\rm\,He{\sc II}}}
\newcommand{\nii}{{\rm\,[N{\sc II}]}}
\newcommand{\oii}{{\rm\,[O{\sc II}]}}
\newcommand{\oiii}{{\rm\,[O{\sc III}]}}

\newcommand{\ciii}{{\rm\,C{\sc III}]}}
\newcommand{\civ}{{\rm\,C{\sc IV}}}

\shorttitle{Redshift Survey of {\sc Spire} sources}
\shortauthors{C.~M. Casey et al.}
\begin{document}

\title{A Redshift Survey of \textit{Herschel\/} Far-infrared Selected
  Starbursts and Implications for Obscured Star Formation}
\author{C.M.~Casey\altaffilmark{1},
S.~Berta\altaffilmark{2},
M.~B{\'e}thermin\altaffilmark{3,4},
J.~Bock\altaffilmark{5,6},
C.~Bridge\altaffilmark{5},
J.~Budynkiewicz\altaffilmark{1,7},
D.~Burgarella\altaffilmark{8},
E.~Chapin\altaffilmark{9,10},
S.C.~Chapman\altaffilmark{11,12},
D.L.~Clements\altaffilmark{13},
A.~Conley\altaffilmark{14},
C.J.~Conselice\altaffilmark{15},
A.~Cooray\altaffilmark{16,5},
D.~Farrah\altaffilmark{17},
E.~Hatziminaoglou\altaffilmark{18},
R.J.~Ivison\altaffilmark{19,20},
E.~le Floc'h\altaffilmark{3},
D.~Lutz\altaffilmark{2},
G.~Magdis\altaffilmark{3,21},
B.~Magnelli\altaffilmark{2},
S.J.~Oliver\altaffilmark{22},
M.J.~Page\altaffilmark{23},
F.~Pozzi\altaffilmark{24},
D.~Rigopoulou\altaffilmark{21,25},
L.~Riguccini\altaffilmark{3,26},
I.G.~Roseboom\altaffilmark{22,20},
D.B.~Sanders\altaffilmark{1},
Douglas~Scott\altaffilmark{9},
N.~Seymour\altaffilmark{23,27},
I.~Valtchanov\altaffilmark{10},
J.D.~Vieira\altaffilmark{5},
M.~Viero\altaffilmark{5},
J.~Wardlow\altaffilmark{16}}
\altaffiltext{1}{Institute for Astronomy, University of Hawaii, 2680 Woodlawn Drive, Honolulu, HI 96822}
\altaffiltext{2}{Max-Planck-Institut f{\"u}r Extraterrestrische Physik, Giessenbachstrasse, 85748 Garching, Germany}
\altaffiltext{3}{Laboratoire AIM-Paris-Saclay, CEA/DSM/Irfu - CNRS - Universit\'e Paris Diderot, CE-Saclay, pt courrier 131, F-91191 Gif-sur-Yvette, France}
\altaffiltext{4}{Institut d'Astrophysique Spatiale (IAS), b\^atiment 121, Universit\'e Paris-Sud 11 and CNRS (UMR 8617), 91405 Orsay, France}
\altaffiltext{5}{California Institute of Technology, 1200 E. California Blvd., Pasadena, CA 91125}
\altaffiltext{6}{Jet Propulsion Laboratory, 4800 Oak Grove Drive, Pasadena, CA 91109}
\altaffiltext{7}{Department of Astronomy, University of Massachusetts, 710 North Pleasant St, Amherst, MA 01003}
\altaffiltext{8}{Laboratoire d'Astrophysique de Marseille - LAM, Universit\'e d'Aix-Marseille \& CNRS, UMR7326, 38 rue F. Joliot-Curie, 13388 Marseille Cedex 13, France}
\altaffiltext{9}{Department of Physics \& Astronomy, University of British Columbia, 6224 Agricultural Road, Vancouver, BC V6T~1Z1, Canada}\
\altaffiltext{10}{European Space Astronomy Centre, Villanueva de la Ca\~nada, 28691 Madrid, Spain}
\altaffiltext{11}{Institute of Astronomy, University of Cambridge, Madingley Road, Cambridge CB3 0HA, UK}
\altaffiltext{12}{Department of Physics and Atmospheric Science, Dalhousie University, 6310 Coburg Rd, Halifax, NS B3H~4R2, Canada}
\altaffiltext{13}{Astrophysics Group, Imperial College London, Blackett Laboratory, Prince Consort Road, London SW7 2AZ, UK}
\altaffiltext{14}{Center for Astrophysics and Space Astronomy 389-UCB, University of Colorado, Boulder, CO 80309}
\altaffiltext{15}{School of Physics and Astronomy, University of Nottingham, NG7 2RD, UK}
\altaffiltext{16}{Dept. of Physics \& Astronomy, University of California, Irvine, CA 92697}
\altaffiltext{17}{Department of Physics, Virginia Tech, Blacksburg, VA 24061}
\altaffiltext{18}{ESO, Karl-Schwarzschild-Str. 2, 85748 Garching bei M\"unchen, Germany}
\altaffiltext{19}{UK Astronomy Technology Centre, Royal Observatory, Blackford Hill, Edinburgh EH9 3HJ, UK}
\altaffiltext{20}{Institute for Astronomy, University of Edinburgh, Royal Observatory, Blackford Hill, Edinburgh EH9 3HJ, UK}
\altaffiltext{21}{Department of Astrophysics, Denys Wilkinson Building, University of Oxford, Keble Road, Oxford OX1 3RH, UK}
\altaffiltext{22}{Astronomy Centre, Dept. of Physics \& Astronomy, University of Sussex, Brighton BN1 9QH, UK}
\altaffiltext{23}{Mullard Space Science Laboratory, University College London, Holmbury St. Mary, Dorking, Surrey RH5 6NT, UK}
\altaffiltext{24}{Dipartimento di Fisica e Astronomia, Viale Berti Pichat, 6/2, 40127 Bologna, Italy}
\altaffiltext{25}{RAL Space, Rutherford Appleton Laboratory, Chilton, Didcot, Oxfordshire OX11 0QX, UK}
\altaffiltext{26}{NASA Ames, Moffett Field, CA 94035}
\altaffiltext{27}{CSIRO Astronomy \& Space Science, PO Box 76, Epping, NSW 1710, Australia}
\label{firstpage}

\begin{abstract}
We present Keck spectroscopic observations and redshifts for a sample
of 767 {\it Herschel}\footnote{{\it Herschel} is an ESA space
  observatory with science instruments provided by European-led
  Principal Investigator consortia and with important participation
  from NASA.}-{\sc Spire} selected galaxies (HSGs) at 250, 350, and
500\,\um, taken with the Keck~I Low Resolution Imaging Spectrometer
(LRIS) and the Keck II DEep Imaging Multi-Object Spectrograph
(DEIMOS).  The redshift distribution of these {\sc Spire} sources from
the {\it Herschel} Multitiered Extragalactic Survey (HerMES) peaks at
$z=0.85$, with 731 sources at $z<2$ and a tail of sources out to
$z\sim5$.  We measure more significant disagreement between
photometric and spectroscopic redshifts ($\langle\Delta z$/($1+z_{\rm
  spec}$)$\rangle$=0.29) than is seen in non-infrared selected
samples, likely due to enhanced star formation rates and dust
obscuration in infrared-selected galaxies.  The infrared data are used
to directly measure integrated infrared luminosities and dust
temperatures independent of radio or 24\,\um\ flux densities. By
probing the dust spectral energy distribution (SED) at its peak, we
estimate that the vast majority (72--83\%) of $z<2$ {\it
  Herschel}-selected galaxies would drop out of traditional
submillimeter surveys at 0.85--1\,mm.  We find that dust temperature
traces infrared luminosity, due in part to the {\sc Spire} wavelength
selection biases, and partially from physical effects.  As a result,
we measure no significant trend in {\sc Spire} color with redshift; if
dust temperature were independent of luminosity or redshift, a trend
in {\sc Spire} color would be expected.  Composite infrared SEDs are
constructed as a function of infrared luminosity, showing the increase
in dust temperature with luminosity, and subtle change in
near-infrared and mid-infrared spectral properties.
Moderate evolution in the far-infrared (FIR)/radio correlation is
measured for this partially radio-selected sample, with $q_{\rm
  IR}\propto(1+z)^{-0.30\pm0.02}$ at $z<2$.  We estimate the
luminosity function and implied star-formation rate density
contribution of HSGs at $z<1.6$ and find overall agreement with work
based on 24\,\um\ extrapolations of the LIRG, ULIRG and total infrared
contributions.  This work significantly increased the number of
spectroscopically confirmed infrared-luminous galaxies at $z\gg0$ and
demonstrates the growing importance of dusty starbursts for galaxy
evolution studies and the build-up of stellar mass throughout cosmic
time.
\end{abstract}

\keywords{
galaxies: evolution $-$ galaxies: high-redshift $-$ galaxies: infrared
$-$ galaxies: starbursts $-$ submillimetre: galaxies}

\section{Introduction}

Ultraluminous Infrared Galaxies (ULIRGs; $L_{\rm IR}>10^{12}$\,\lsun)
exhibit the most extreme star-formation rates in the Universe
\citep[see an overview in][]{lonsdale06a}.  At early epochs
($z\,{>}\,1$), ULIRG activity contributes significantly to the
build-up of stellar mass presumably through intense star-forming
bursts \citep[with $\tau\,\simlt\,100\,$Myr and
  SFR\,{\simgt}\,500\sfr, e.g. see][]{sanders96a,blain02a,smail02a}.
Since the observed properties of these starbursts are short-lived and
intense, they are thought to be triggered by the collision of gas-rich
disk galaxies \citep{engel10a} and serve as a fundamental transition
phase to luminous active galactic nuclei (AGN) or quasars
\citep{sanders88a}.  Although the merger history of high-$z$ ULIRGs
has recently come into question, with some evidence pointing to a
substantial (perhaps $>$50\%) fraction of ULIRGs building stellar mass
through minor mergers or passive gas accretion
\citep{daddi10a,elbaz11a,rodighiero11a}, there is little doubt that
ULIRGs contribute non-negligibly to the star formation history of the
Universe and the formation of massive elliptical galaxies at the
present day \citep[e.g.][]{kartaltepe10a}.  Unfortunately, much about
the infrared starburst population\footnote{In this paper, we use the
  term `starburst' to refer to high-SFR galaxies (SFR$>$100\,\sfr).
  This differs from the recent definition of `starburst' as a
  combination of SFR and stellar mass
  \citep[e.g.][]{noeske07a,rodighiero11a}.} is still unknown due to
limitations in far-infrared (FIR) observations, strong selection
biases and sample inhomogeneity.

Galaxies which have been called `Submillimetre Galaxies' (SMGs), are
selected at wavelengths around $1\,$mm, particularly in the
atmospheric window at $850\,\mu$m.  Such `classical SMGs' with $S_{\rm
  850}$\simgt5\,mJy \citep[][]{smail02a} have put powerful constraints
on galaxy evolution theories and the environments of heavy star
formation since their initial discovery a decade ago
\citep{smail97a,hughes98a,barger99a}.  However, their selection at
wavelengths 850\,\um--1.4\,mm is susceptible to strong temperature
biasing \citep{blain04a,chapman04a,casey09b,chapman10a,magdis10a}.
This leaves the possibility that a significant fraction of high-$z$
ULIRGs have yet to be discovered and characterised.  Building a
comprehensive sample of spectroscopically-confirmed submillimetre
galaxies or extreme starbursts is paramount for determining the
evolutionary histories of ULIRGs by breaking the $T_{\rm dust}/(1+z)$
degeneracy, for carrying out stellar population analysis, and
for measuring the AGN stage and contribution to luminosity.

The {\it Herschel Space Observatory\/} \citep{pilbratt10a} has
identified thousands of galaxies at 70--500\,\um, wavelengths
previously near-inaccessible from the ground, sampling galaxies'
emission at the peak of their SEDs at $z\,{\sim}\,$\,2, when the
importance of dusty starbursts in the global context of the Universe's
star formation is most evident \citep{chapman05a}.  The {\sc Spire}
instrument \citep{griffin10a} will map $\sim$350\,deg$^{2}$ in sky
near the confusion limit at 250, 350 and 500\,\um\ as part of the
Herschel Multi-tiered Extragalactic Survey (HerMES; Oliver \etal\ in
press), covering areas much larger than SCUBA, MAMBO, AzTEC or LABOCA.
With much larger areas, the rarest sources can be uncovered and the
dynamic range of sources thereby expands, from nearby luminous
infrared galaxies (LIRGs; $>10^{11}$\lsun) to distant hyper-luminous
infrared galaxies (HyLIRGs; $>$10$^{13}$\lsun) and lensed sources.
Working towards completeness in high redshift starburst samples, and
removing the impact of selection biases introduced by prior starburst
selection techniques, is a key long term goal.

Understanding ULIRG completeness and, in turn, constraining key
astrophysical quantities of the luminous starburst population, is only
possible with a spectroscopic census of a diverse population of
FIR-luminous galaxies.  Redshift identification is a crucial piece of
information for a high-redshift dusty galaxy, since it allows the
measurement of its luminosity and star-formation rate, and is a
prelude to subsequent interferometry (often dependent on a known
redshift) in order to constrain the vast reservoirs of molecular gas
which fuel extreme starbursts.  Such subsequent studies cannot be
completed using photometric data alone, and to date, Keck LRIS and
DEIMOS multi-slit spectroscopy is the most efficient method for
uncovering large samples of galaxy redshifts, for both normal
star-forming $z>$\,1 galaxies
\citep{cowie98a,cowie99a,cowie01a,steidel96a,steidel99a} and heavily
dust-obscured ULIRGs
\citep{barger98a,barger99a,barger00a,cowie02a,chapman05a}.

This paper presents the first results from a large spectroscopic
redshift survey of 1594 {\it Herschel}-{\sc Spire} selected galaxies
(HSGs).  We measure redshifts for 767 of 1594 targeted HSGs, describe
their bulk infrared properties, address their relationship to the now
well-studied SMGs, and assess their contribution to cosmic star
formation.  The results of this spectroscopic survey have been split
into two papers of which this is the first, presenting the details of
source selection, completeness, spectroscopic confirmations and
associated results for $z<2$ sources.  An accompanying paper presents
the $2<z<5$ sub-sample in more detail.  Throughout we use a
flat $\Lambda$CDM cosmology \citep{hinshaw09a} with $H_{\rm
  0}$=71\kms\,Mpc$^{-1}$ and $\Omega_{\rm M}$=0.27.

\section{\textit{Herschel}-Selected Galaxy Sample}\label{sec:selection}

The sources observed in this paper were detected by the {\it Herschel
  Space Observatory\/} {\sc Spire} instrument as part of the {\it
  Herschel\/} Multi-tiered Extragalactic Survey (HerMES; Oliver
\etal\ in prep.).  {\sc Spire}, the Spectral and Photometric Imaging
Receiver \citep{griffin10a}, is designed for wide-field mapping at
250, 350, and 500\,\um.  The beamsizes at these respective wavelengths
are 18\arcsec, 25\arcsec, and 36\arcsec with measured mean
point-source confusion noise uncertainties of $\sigma_{\rm
  250}=$\,3.8\,mJy, $\sigma_{\rm 350}\,=\,4.6\,$mJy, and $\sigma_{\rm
  500}=$\,5.2\,mJy which dominates over instrumental noise
\citep[these values are for 3$\sigma_{\rm conf}$ cuts used for
  deboosted photometric measurements, see the {\sc Spire} Observers'
  Manual and][]{griffin10a,nguyen10a}.  We make use of {\sc Spire}
maps as described by \citet{levenson10a}.

In this paper, deep ancillary data, particularly radio and 24\,\um,
are essential for optical spectroscopic surveying.  We observe sources
in the Lockman Hole North (LHN) whose {\it Spitzer\/} imaging comes
from the {\it Spitzer\/} Wide-Area Infrared Extragalatic (SWIRE)
survey \citep{lonsdale03a} and GO MIPS programs (PI Owen) and very
deep 1.4\,GHz mapping from the Very Large Array (VLA) \citep{owen08a}.
LHN has additional coverage with {\it Herschel}-{\sc Pacs} from HerMES
(PI G. Magdis).  In the Great Observatories Origins Deep Survey North
(GOODS-N) field, deep 1.4\,GHz radio mapping comes from the VLA
\citep{morrison10a} and {\it Spitzer\/} coverage of the GOODS-N center
is from FIDEL (Dickinson \etal\ in prep.) and some of the extended
area is from {\it Spitzer} program ID83 (PI Rieke; Shupe, private
communication).  We also observe sources in the ELAIS-N1 (EN1) and
extended UKIDSS Ultra Deep Field (UDS)/XMM fields, both extragalactic
areas in the SWIRE survey.  The UDS has additional coverage from the
{\it Spitzer\/} Legacy Program (SpUDS; PI Dunlop).  Radio coverage of
the EN1 field is substantially more sparse than in GOODS-N or LHN,
with the only mapping taken with the GMRT at 610\,MHz \citep{garn08a}
and 325\,MHz \citep{sirothia09a}, whose depths effectively translate
to 100\,\uJy\ RMS at 1.4\,GHz assuming a synchrotron slope of
$\alpha$=0.75, where $S\propto\nu^{-\alpha}$ \citep[e.g.][]{dale07a}.
Our UDS observations sit on the edge of a new deep VLA radio map with
$\sim$7\,\uJy\ RMS (Arumugam \etal\ in prep.).  In the Cosmic
Evolution Survey field \citep[COSMOS;][]{scoville07a}, radio coverage
in the central 1\,deg$^2$ has a depth of $\sim$10.5\,\uJy\ RMS
\citep{schinnerer07a} and {\it Spitzer} coverage is described in
\citet{sanders07a}, \citet{le-floch09a} and \citet{frayer09a}. COSMOS
is also covered by {\sc Pacs} as part of the PEP program (PI D. Lutz).
We additionally observed sources in the Extended Chandra Deep Field
South or ECDF-S region which has deep radio coverage
\citep{miller08a,biggs11a}, as well as {\it Spitzer}-MIPS
24\,\um\ from FIDEL and IRAC \citep{damen11a}.

We use the photometric redshift catalogs from SWIRE (EN1) described by
\citet{rowan-robinson08a}, the deep LHN catalog in
\citet{strazzullo10a}, and the extensive COSMOS \citep{ilbert10a} and
ECDFS \citep{cardamone10a} catalogs.  In GOODS-N, since we survey
radio galaxies outside of the central deep region, the photometric
catalog is limited, so we exclude it from photometric redshift
analysis.

\subsection{Source Extraction and Photometry}

{\sc Spire} point source photometry is performed by flux extraction at
positions of known 24\,\um\ sources or radio 1.4\,GHz sources.  This
cross-identification prior source extraction (XID) method is described
in detail in \citet{roseboom10a} with some follow-up discussion in
\citet{roseboom12a}.  The disadvantage of the XID technique is that it
excludes any sources which are not 24\,\um\ or radio identified.  This
is particularly problematic for potentially high-redshift sources
which drop out of both 24\,\um\ and radio surveys, and might have
ambiguous near-IR counterparts; since 24\um\ and radio source dropouts
are excluded from this sample, and their influence is undoubtably more
significant at high-redshifts than at $z<2$
\citep[e.g.][]{magdis11a,bethermin12a}, the high-$z$ sample is treated
in a separate paper \citep[][henceforth C12]{casey12c}.

The advantage of cross-identification with 24\,\um/IRAC and radio
sources is that it can correct for confusion boosting in the extracted
{\sc Spire} flux densities by estimating the flux contributions from
nearby sources within one beamsize.  It also reduces the confusion
noise by a factor of $\sim$2 by pushing slightly below the nominal
confusion limit using the Least Absolute Shrinkage and Selection
Operator (LASSO) method to assign {\sc Spire} flux densities to an
overdense prior source list \citep[see ][ for method
  details]{roseboom10a}.  The LASSO algorithm combines strengths of
model prediction and filter prediction source identification
(i.e. balancing source priors to the {\sc Spire} map flux distribution
with choosing the brightest source correspondence between
24\,\um/radio and {\sc Spire}). The algorithm upweights `rare'
sources, therefore radio sources are prefered {\sc Spire} counterparts
over 24\,\um\ sources.  This makes sense given the expectation that
radio sources are FIR-luminous \citep{helou85a,condon92a}.  The
procedure assumes that the ancillary data are of adequate depth to
identify the vast majority of FIR emitting sources, so it is only
practical in deep legacy survey fields.

\subsection{Completeness of Source Catalog}\label{sec:xid}

The most crucial aspect of a redshift survey is having a clear
understanding of survey completeness and biases.  This subsection
addresses our survey's completeness in identifing strong {\sc Spire} sources
using the XID flux extraction method.

Since this is a test of the robustness of the XID technique, the
results will vary by field (LHN, GOODS-N, EN1, COSMOS, CDFS and UDS),
based on survey depths, and is a function of {\sc Spire} flux density.
\citet{roseboom10a} demonstrate that the robustness of the XID catalog
depends greatly on the depth of ancillary data available in the field
to act as source priors.  Beyond a fairly standard depth at
24\,\um\ of $S_{\rm 24}\sim$\,150\,\uJy\ (or a sky density of
$\simgt$3000\,deg$^{-2}$), XID source flux extraction will be
$>$95\%\ complete at the 3$\sigma$ limit of {\sc Spire} (where
$\sigma$ includes instrumental and confusion noise, and 3$\sigma$
roughly corresponds to a \simgt15\,mJy cut-off).  At higher source
densities XID is more robust, even when the source density of priors
exceeds the number of {\sc Spire}-bright sources.

\begin{figure}
  \centering
  \includegraphics[width=0.9\columnwidth]{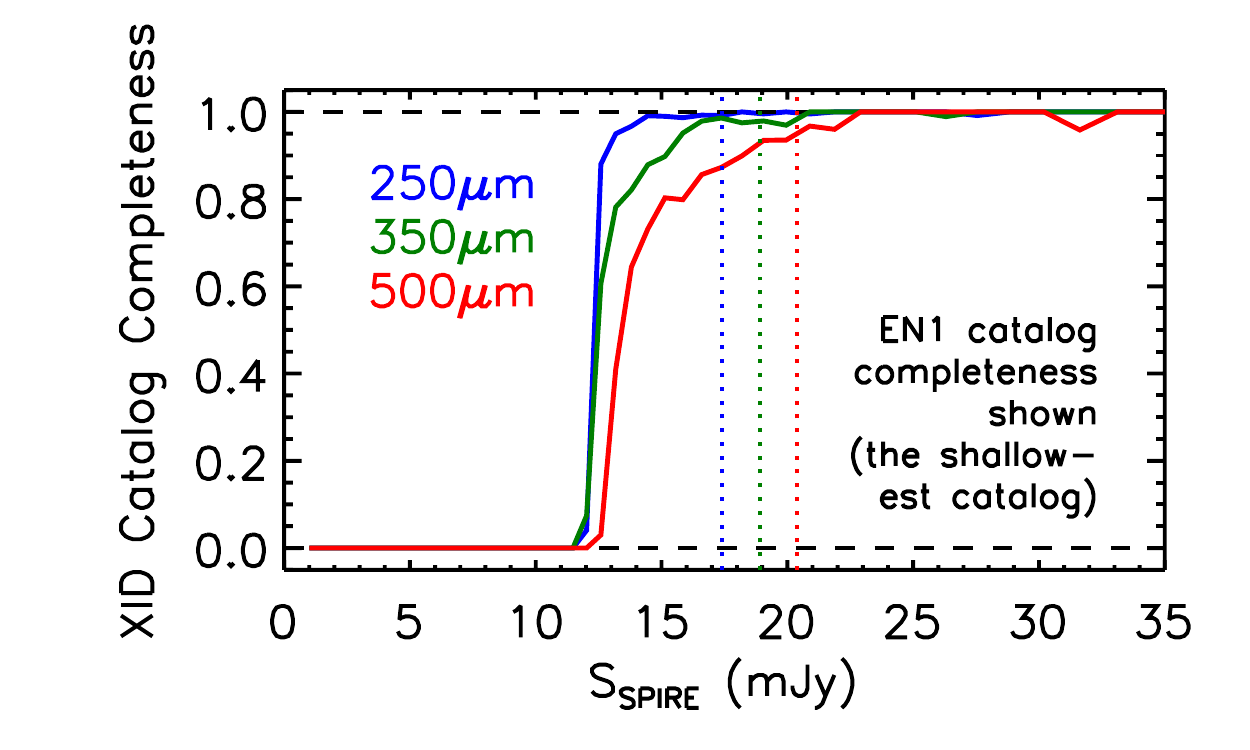}
  \caption{ 
    XID catalog completeness curves in the EN1 field as a function of flux
    density at {\sc Spire} wavelengths 250\,\um\ (blue), 350\,\um\ (green) and
    500\,\um\ (red).  The dotted vertical lines mark the 3$\sigma$
    noise limits at the respective wavelengths.  This completeness measure
    is a reflection of the XID catalog's ability to assign and identify
    24\,\um\ or radio counterparts for all {\sc Spire} sources at a given
    flux density.
  }
  \label{fig:xidcomp}
\end{figure}

The completeness of the XID process is tested by first producing
residual maps at the {\sc Spire} wavelengths using the best fit XID
solution, and then re-injecting sources into these maps using a number
count and clustering model consistent with the real data.  The XID
process is repeated on these simulated images and the results are
assessed to determine the number of sources returned at $>3\sigma$ as
a function of injected flux density.  This process is repeated several
times for different realisations of the {\sc Spire} maps to build up
suitable statistics across a wide range of {\sc Spire} flux densities.
The results of this process are shown in Figure~\ref{fig:xidcomp} for
the field with the shallowest ancillary data, EN1; deeper fields have
completeness curves approaching boxcar functions.  This completeness
curve is generated using the same analysis from \citet{roseboom10a}
with improvements described in \citet{roseboom12a}.  An $S_{\rm
  SPIRE}>$\,15\,mJy limit should be $>$80\%\ complete at
500\,\um\ ($>$95\%\ for 250\,\um\ and 350\,\um), and a 20\,mJy limit
should be $>$95\%\ complete in all fields.  Additional sources are
identified down to flux densities comparable to the confusion limits
(5--6\,mJy).  This is below the nominal 3$\sigma$ cut-off and is
achieved by using positional priors of many galaxies thought only to
emit at {\sc Spire} wavelengths at levels $\sim0.1$--1\,mJy; the flux
density of {\sc Spire}-bright sources is then `deboosted' using the
density of source priors which are thought to be {\sc Spire}-faint.

Note that the ability to match counterparts down to 15\,mJy does not
necessarily mean that counterpart matching is always correct \citep[an
  issue which is so far largely unconstrained, but is starting to be
  addressed through systematic interferometric work, e.g.][]{wang11a}.
This question, particularly as it applies to the application of the
XID extraction technique, is discussed in \citet{roseboom10a}.

\subsection{Completeness in Radio, 24\,\um\ Samples}\label{sec:radio24comp}

\begin{figure*}
  \centering
  \includegraphics[width=1.65\columnwidth]{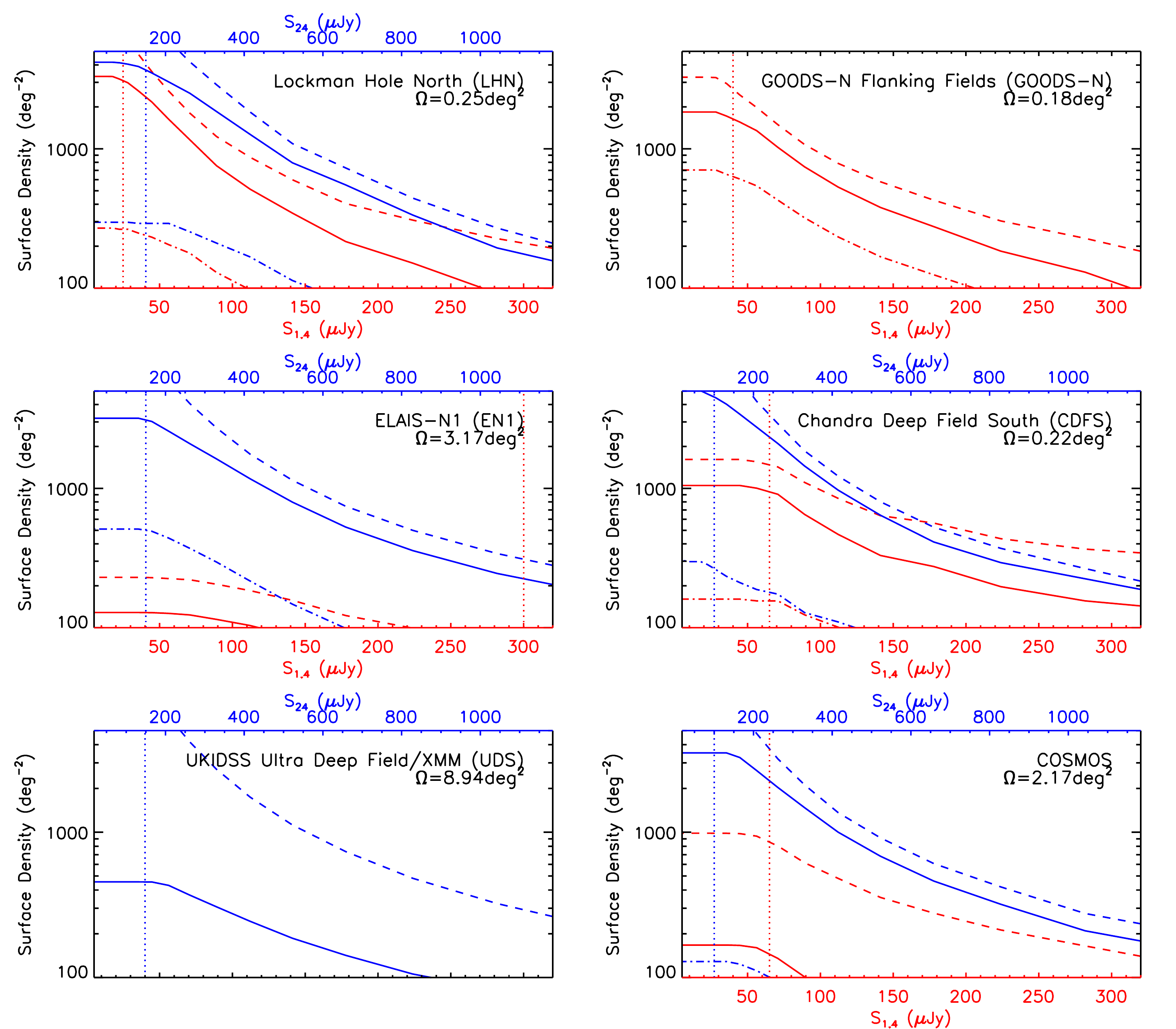}
  \caption{Cumulative surface density of 24\,\um\ and radio sources with
    flux density.  In other words, as a function of radio \emph{or}
    24\,\um\ flux density, we plot the surface density of sources with
    flux densities above that given flux density.  Dashed lines
    represent the parent population of all 24\,\um\ sources ($blue$) and
    radio sources ($red$) in the whole field.  Each panel is labeled at
    top with a 24\um\ flux density scale and at bottom with a 1.4\,GHz
    flux density scale (the scales are omitted if the data are
    insufficient/do not exist).  Solid lines represent the source
    density for sources which are $>$3$\sigma$ significant in at least
    one of the three {\sc Spire} bands.  Dot-dashed lines represent
    sources $>$3$\sigma$ significant in all three {\sc Spire} bands.
    Vertical dotted lines mark where 24\,\um\ or radio catalogs become
    incomplete in flux density.  }
  \label{fig:surfacedensity}
\end{figure*}

\begin{figure}
  \centering
  \includegraphics[width=0.99\columnwidth]{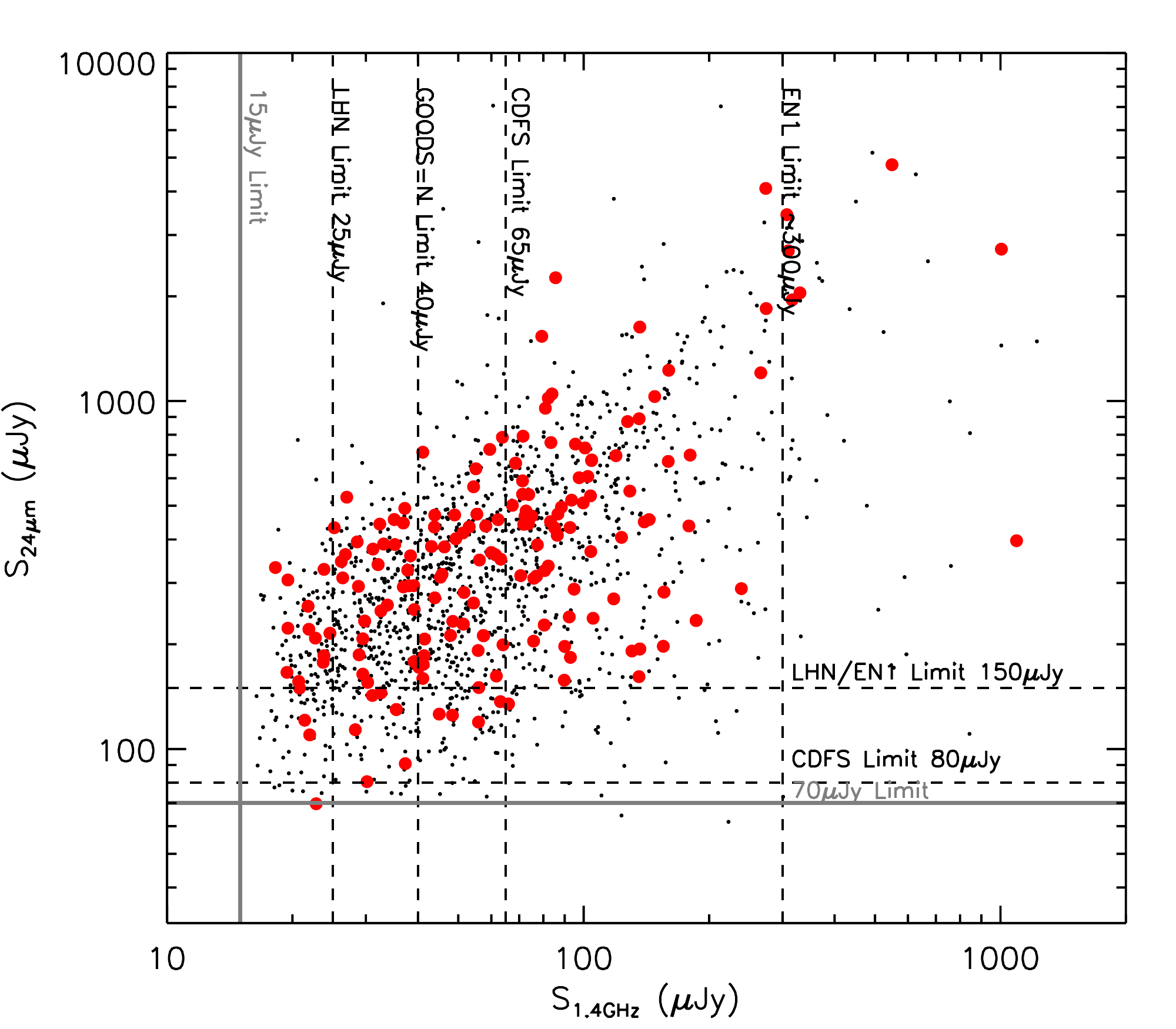}\\
  \caption{ 
    The distribution of LHN targets in $1.4\,$GHz radio flux density
    ($S_{\rm 1.4}$) versus 24\,\um\ flux density ($S_{\rm 24}$).  The
    various flux limits in the three different fields are marked.  Black
    points represent field galaxies while red points denote {\sc
      Spire}-selected galaxies.  Note that {\sc Spire} detection is
    largely uncorrelated with 24\,\um\ or radio flux density.  However,
    faint 24\,\um\ sources tend also to be faint in the radio (roughly
    corresponding to $S_{\rm 24}\sim3.7\,\times\,S_{\rm 1.4}$ as measured
    here), leading us to conclude that the majority of faint 24\,\um\ EN1
    sources are drawn from the same population as the GOODS-N radio
    sources \citep[also see the work of ][]{magdis11a}.  }
  \label{fig:sradios24}
\end{figure}

The source density of $>$3$\sigma$ {\sc Spire}-significant XID sources
is constant within a factor of $\sim$2 (1500--3000\,deg$^{-2}$)
despite large variations in ancillary data across the different
fields.  For example, the Lockman Hole North $-$ a field with
ultra-deep radio coverage and deep {\it Spitzer\/} mid-infrared
coverage $-$ has a source density of $\sim$3000\,deg$^{-2}$ for
sources above $S_{\rm 1.4}\sim$\,25\,\uJy\ or $S_{24}\sim$\,150\,\uJy,
which is the same density measured in ELAIS-N1 $-$ a field with only
very shallow radio coverage $-$ and in the GOODS-N flanking fields $-$
an area not completely covered by {\it Spitzer}.
Figure~\ref{fig:surfacedensity} shows the cumulative surface densities
of sources selected at 24\,\um\ and radio, then selected to be
$>$3$\sigma$ in at least one of the {\sc Spire} bands.  The surface
density of 24\,\um\ {\sc Spire}-significant sources is about a factor
of 2 times the surface density of radio {\sc Spire}-significant
sources, assuming a rough correlation between 24\,\um\ and $1.4\,$GHz
radio flux density of $S_{\rm 24}\approx3.7\,S_{\rm 1.4}$ (which we
measure from Figure~\ref{fig:sradios24}).
Figure~\ref{fig:surfacedensity} also highlights that the surface
density of any one field is dependent on the depth of the prior
catalog, although across all fields this is consistent within a factor
of about 2.  With comparable surface densities
$\sim$\,1500--3000\,deg$^{-2}$, one might then ask whether this means
that 24\,\um\ identified sources in EN1/COSMOS/CDFS/UDS and the
radio-identified ($>$40\uJy) sources in LHN/GOODS-N/COSMOS/CDFS are
drawn from the same population of IR galaxies.

\begin{figure}
  \centering
  \includegraphics[width=0.99\columnwidth]{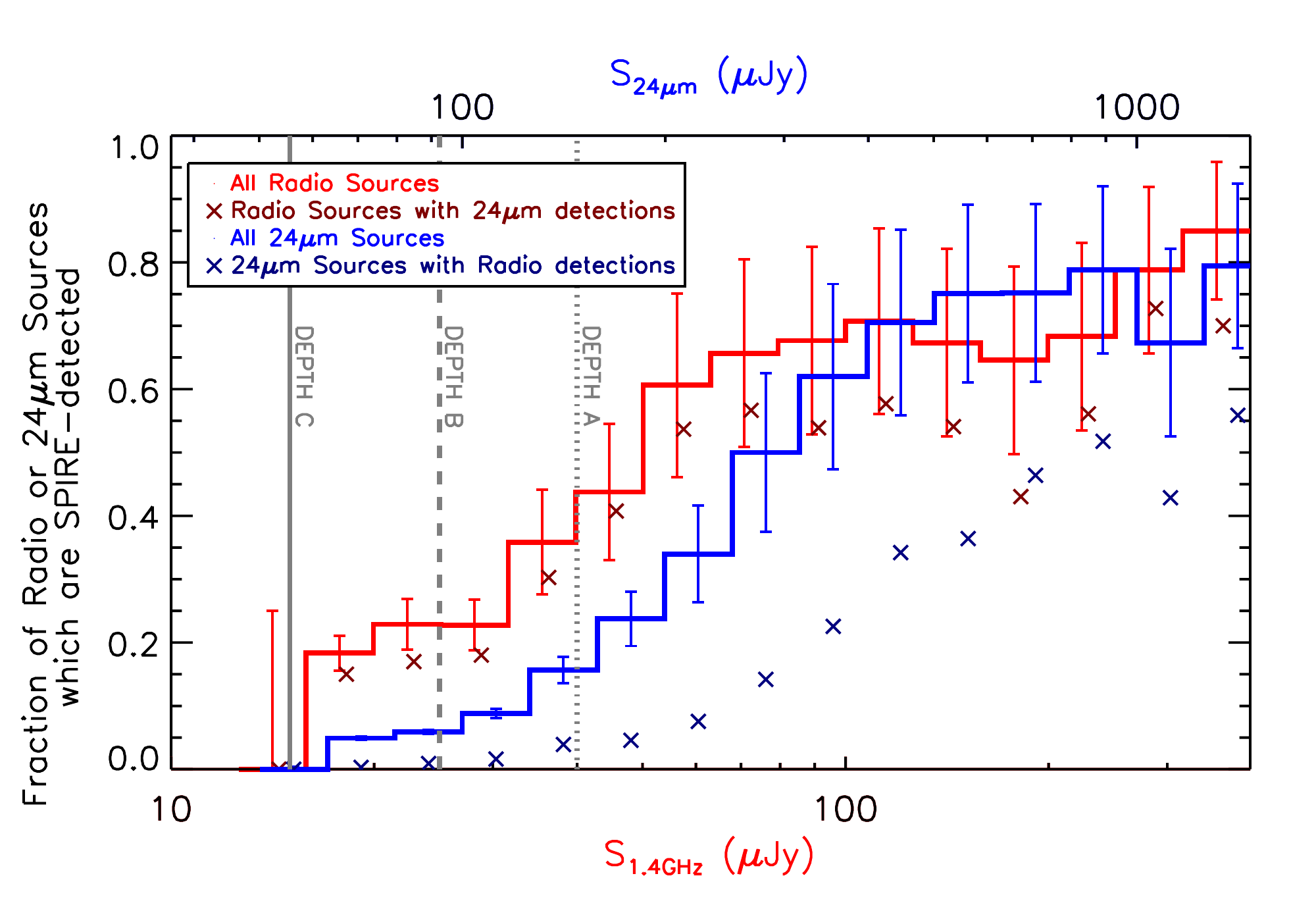}\\
  \caption{ The fraction of radio sources (red) or 24\,\um\ sources
    (blue) which are {\sc Spire}-detected as a function of radio or
    24\,\um\ flux density.  The `x's represent the fraction of
    24\um\ {\sc Spire} sources which are also radio-detected (blue) and
    radio {\sc Spire} sources which are also 24\um-detected. Here we use
    $S_{\rm 24}\sim3.7\,\times\,S_{\rm 1.4}$, which is the rough
    relation we observe for field galaxies in
    Figure~\ref{fig:sradios24}; this scaling allows us to judge the
    relative depths of the catalogs.  `Depth A' corresponds roughly to
    $S_{\rm 1.4}\sim$40\,\uJy\ and $S_{\rm 24}\sim$150\,\uJy , `Depth B'
    corresponds roughly to $S_{\rm 1.4}\sim$25\,\uJy\ and $S_{\rm
      24}\sim$100\,\uJy, and `Depth C' corresponds roughly to $S_{\rm
      1.4}\sim$15\,\uJy\ and $S_{\rm 24}\sim$70\,\uJy .  `Depth A'
    corresponds to the EN1 and UDS 24\,\um\ limit, `Depth B' corresponds
    to LHN 24\um\ and COSMOS/GOODS-N/CDFS radio limits, while `Depth C'
    corresponds to the LHN radio limit and GOODS-N(center)/CDFS
    24\,\um\ limit.  }
  \label{fig:sradios24-2}
\end{figure}

Figure~\ref{fig:sradios24} investigates the relation between
24\,\um\ and 1.4\,GHz flux density for the LHN subsample which
contains the deepest ancillary data.  Our {\sc Spire} targets are
shown against normal field galaxies, both carving out a similar
parameter space, indicating no clear {\sc Spire}-bright/faint bias.
Using our LHN sample as a guide, we find that 70$\pm$20\%\ of $S_{\rm
  1.4}\ge40$\uJy\ sources have $S_{\rm 24}\ge150$\uJy\ (see
Fig~\ref{fig:sradios24-2}), suggesting that most sub-mJy radio sources
are sub-mJy 24\um\ sources.  This highlights that: (a) the incidence
of {\sc Spire}-detection is higher in faint radio sources than in
24\um\ sources; and (b) that nearly all radio sources ($\sim$70\%) are
24\,\um\ detected and a large fraction of 24\,\um\ sources are radio
detected (from Fig~\ref{fig:sradios24-2}).  Since our sample is
rest-frame FIR selected and is likely to obey the FIR/radio
correlation \citep{helou85a}, our sources are more likely to be
radio-detected than random field galaxies.  Also, given the beamsize
of {\it Herschel} to be 18--36\arcsec, the probability of random
coincidence with a radio galaxy is $>$8 times lower than with a
24\um\ galaxy, thus the probability of correct counterpart
identification is $>$8 times higher in radio galaxies than it is in
24\um\ galaxies (although that probability itself cannot be
constrained without interferometric infrared observations).

The typical range of surface densities of {\sc Spire} galaxies,
$\sim$1500--3000\,deg$^{-2}$, translates to an expected number of {\sc
  Spire}-detected sources per slit-mask of N$_{\rm
  DEIMOS}=33^{+15}_{-8}$\,sources per mask and N$_{\rm
  LRIS}=\,17^{+8}_{-4}$\,sources per mask across all fields (derived
from mean and standard deviations of source densities between the six
survey fields).  This agrees with our actual spectroscopic sampling
per slit-mask; we average 33$\pm$7 significant {\sc Spire}-sources per
DEIMOS slit-mask and 20$\pm$6 sources per LRIS slit-mask.

It is important to note that this analysis only tests the completeness
and source density for galaxies already detected at mid-infrared or
radio wavelengths.  Some sources will be excluded from the XID catalog
since they will not be detected in the ancillary data.  Important
examples are very high redshift sources which are radio-faint and
24\,\um-faint and cannot be identified due to a lack of a
multi-wavelength counterpart(s).  These high-redshift sources will
have a profound effect on the derived IR luminosity function at
$z\,{>}\,3$; therefore, this is a significant limitation of this HSG
sample.  However, at lower redshift we suspect that, at the least, the
XID catalog is 80\%\ complete out to $z\,{=}\,2$ \citep[again, see
  more details on estimated completeness as a function of redshift
  in][]{roseboom10a}.  The completeness of the XID catalog with
respect to the low redshift sources, $z$\simlt2, has also been shown
to be $\approx$95\%\ \citep{magdis11a}.

\subsection{Spectroscopic Target Characteristics}

\begin{figure}
  \centering
  \includegraphics[width=0.99\columnwidth]{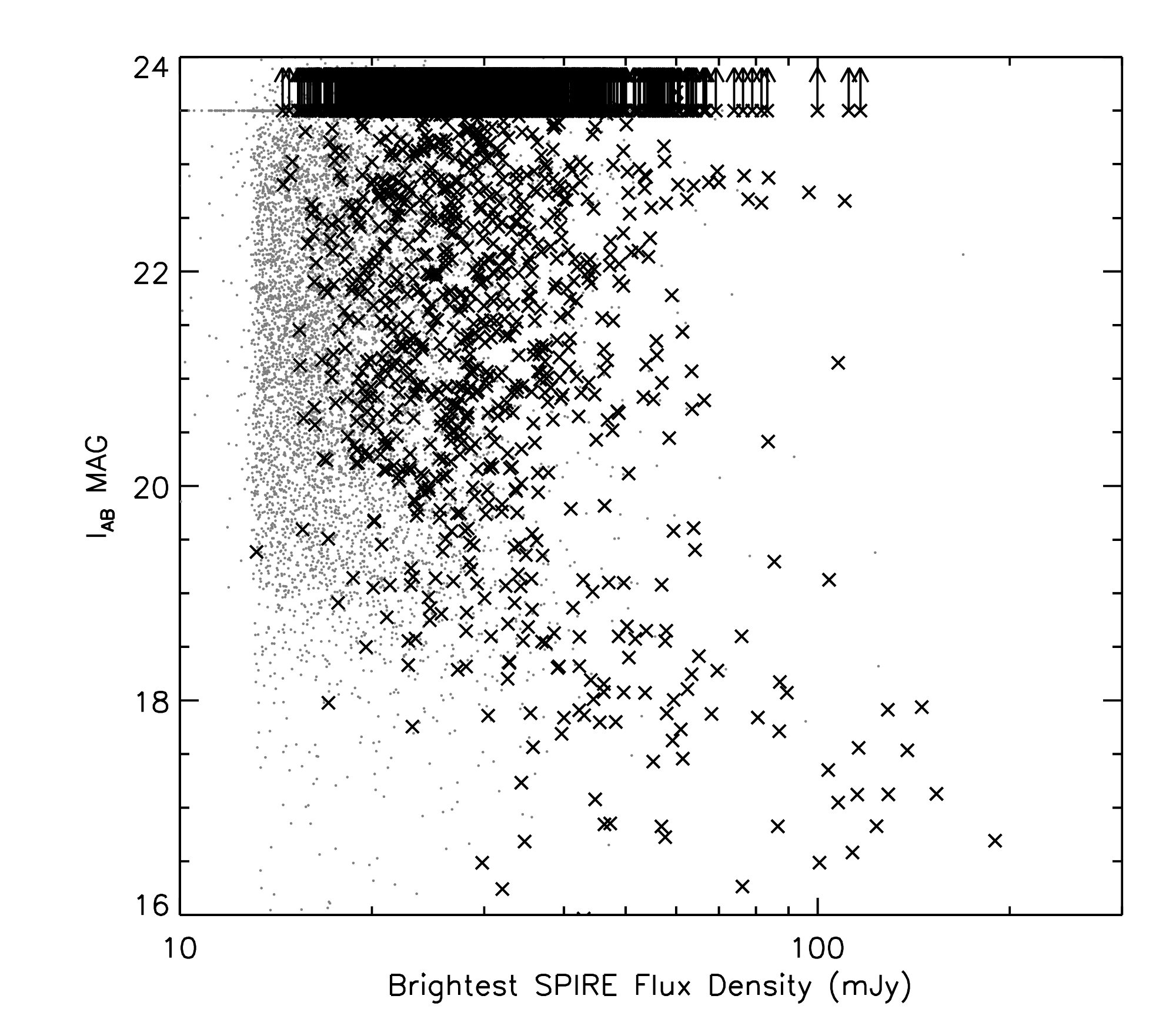}\\
  \caption{Optical $i$-band magnitude (AB) against {\sc Spire} flux
    density (the brightest flux density is taken from {\sc Spire}
    250\,\um, 350\,\um\ or 500\,\um) for sources in EN1 (similar
    results for other fields).  Black points are all significant {\sc
      Spire} sources used to fill our slitmasks.  The selection of
    these {\sc Spire} sources is not a function of optical magnitude.
  }
  \label{fig:optspire}
\end{figure}

Slit-masks for the Keck Low Resolution Imaging Spectrometer
\citep[LRIS;][]{oke95a} and DEep Imaging Multi-Ojbect Spectrograph
\citep[DEIMOS;][]{faber03a} were populated by a prioritization scheme,
whereby sources were graded in priority from 0 to 1000 by their {\sc
  Spire} photometry, radio detection or non-detection, and
24\,\um\ flux density.  No optical magnitude selection or
prioritization was made.  Note that there was no filtering of the
sample to remove low redshift targets or quasars \citep[as was often
  done for SMGs, e.g.][]{chapman05a}.  The $i$-band magnitudes of our
spectroscopic targets are plotted against the brightest {\sc Spire}
flux in Figure~\ref{fig:optspire}. Since our spectroscopic sampling is
exclusively driven by {\sc Spire} detectability and not by optical
magnitude, there is no clear relationship perceived in
Figure~\ref{fig:optspire}.  When there was no $i$-band or $z$-band
counterpart to center our slit on, we used the IRAC 3.6\,\um\ position
or radio VLA position, which are both good to $\sim$0.7\arcsec\ given
their relatively small beamsizes.

Sources detected at $>$3$\sigma$ in at least one out of the three
{\sc Spire} bands were given a priority $=$\,300, and sources detected in
all three {\sc Spire} bands at $>$3$\sigma$ were given a priority
$=$\,500.  Mask centers and orientations were chosen based on the
positions of rare, `red' 500\,\um-peaking sources ($S_{\rm 250}<S_{\rm
  350}<S_{\rm 500}$), thought to be the highest-redshift
{\sc Spire}-bright galaxies \citep{cox11a}.  These red sources were given
very high priority, $>$800, and if their multi-wavelength properties
were consistent with a high-$z$ source, e.g. $i_{\rm AB}>$22 and
$S_{\rm 24}<$500\,\uJy, priority was graded even higher, at 1000.  The
prioritization scheme is linear, such that a source with priority 1000
will be assigned a slit in favor of two $p$=300 sources, or one
$p$=300 source plus one $p=$500 source, however, a $p$=800 source
would be passed up in lieu of two $p$=500 sources.  While this scheme
could accidentally remove very high priority sources from our masks,
we adjusted mask position angles manually to ensure optimal spatial
sampling, and were only minimally affected by source overlap
(\simlt5\%\ of slits were conflicted).

The density of our sources, both low and high priority, is high enough
and comparatively uniform over the LHN, GOODS-N, EN1, UDS, CDFS and
COSMOS fields such that our mask coverages constitute a random
sampling, and completeness estimates are performed with respect to the
sky area probed by the masks alone (which total
$\sim$0.93\,deg$^2$).  Note that the centering of our masks around
high-priority `red' sources might give a high-redshift bias to our
sample, since we set out to find some of the rarest, high-$z$ HyLIRGs.
What we find (to be discussed later in the paper and more in C12) is
that the redshift distribution of `red' targets is {\it
  not} strongly biased towards high-$z$.  There are actually more
high-$z$ sources which are not red than are, leading us to
believe that the sky sampling is essentially random, despite our efforts to
detect more high-$z$ galaxies. The density of $>$3$\sigma$ {\sc Spire}
sources is also low enough so that our radio, 24\,\um\ or {\sc Spire}
color prioritization does not introduce statistically significant
selection biases into the sample; in other words, nearly all ($>$90\%)
the $>$3$\sigma$ {\sc Spire} sources within areas covered by slit-masks
were spectroscopically observed.

We assigned slits for 100\%\ of the very high priority `red' targets,
95\%\ of the high priority targets ($>$500), and 90\%\ of the lower
priority targets (300$<p<$500), filling any free space with additional
objects not discussed in this paper (radio sources or
24\,\um\ galaxies with insignificant {\sc Spire} fluxes).  Any
high-priority {\sc Spire} sources not observed were only excluded on
the basis of the mask configuration, where observing them would bump
another high-priority source off a slit.  However, the exclusion of
high-priority {\sc Spire} sources is rare enough that having a few
missing from our sample does not impact our completeness, particularly
since we find no significant bias towards high-redshift spectroscopic
identifications in the high-priority target sub-sample (see C12 for
more details on the $z>2$ sub-sample).

\begin{figure}
  \centering
  \includegraphics[width=0.99\columnwidth]{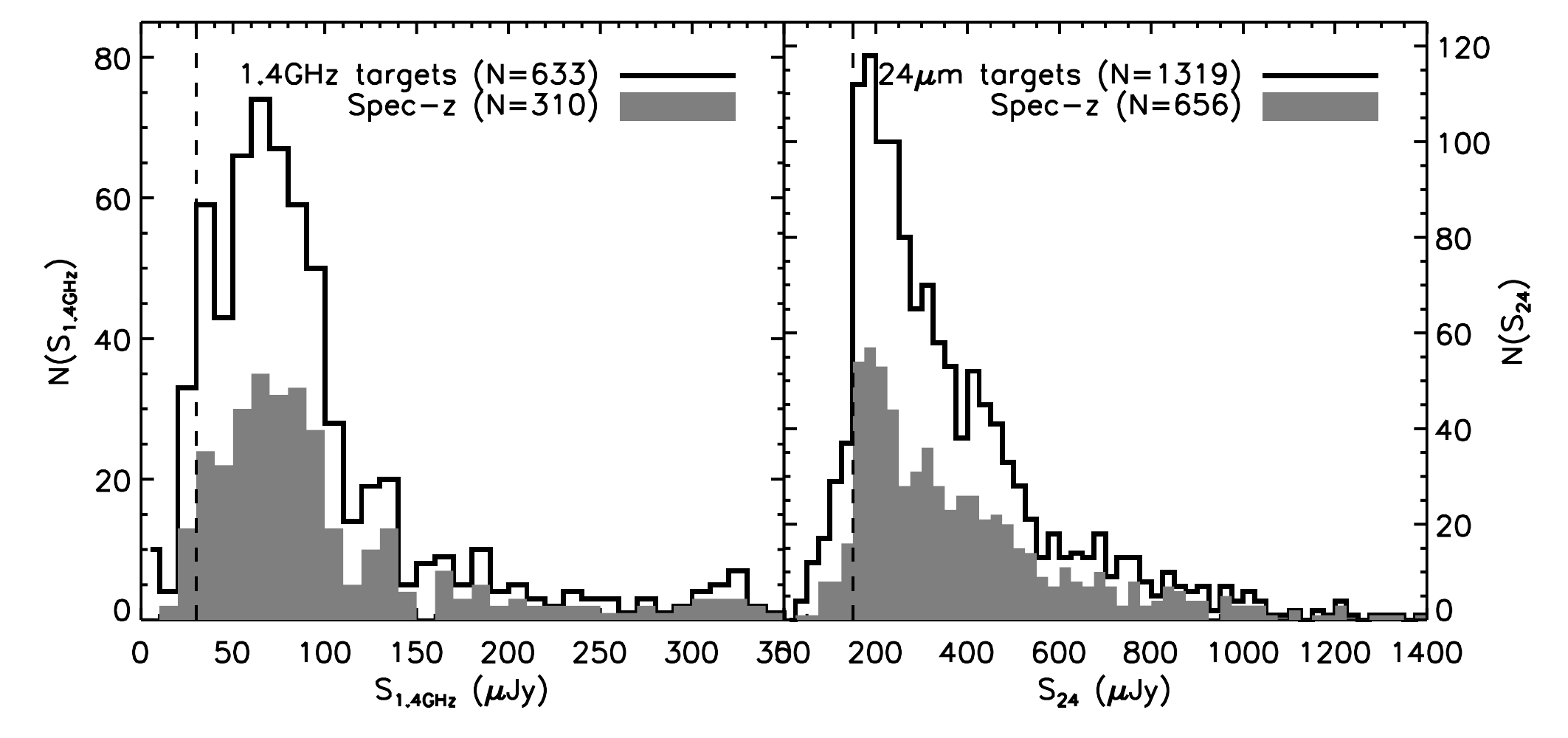}\\
  \includegraphics[width=0.99\columnwidth]{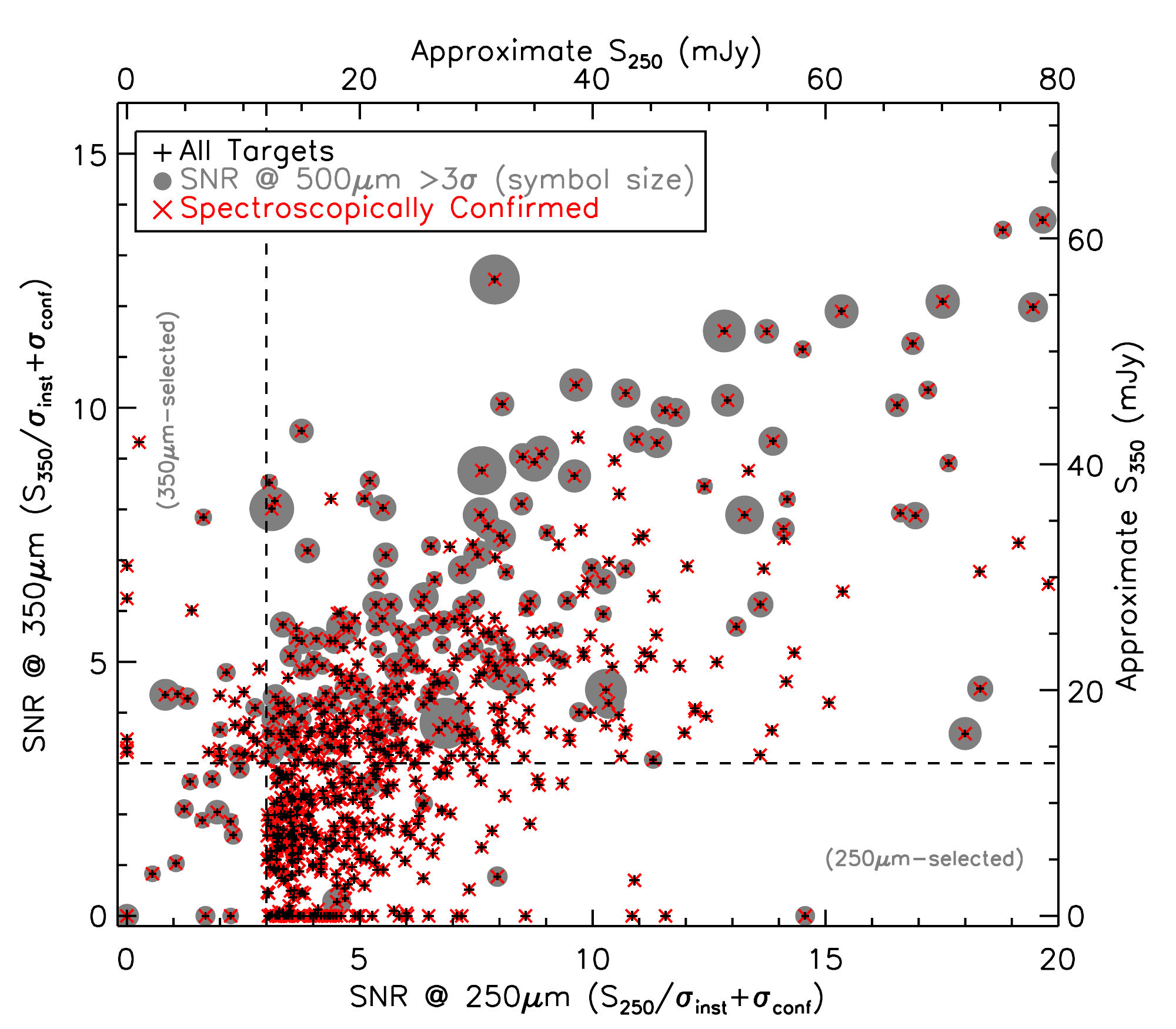}
  \caption{ 
    {\bf Top:} Distribution of 24\,\um\ and 1.4\,GHz radio flux density
    of our targets.  Of our sample, 1319 were primarily identified through
    24\,\um\ emission and 633 of were radio selected.  The gray
    distributions reflect the sources with confirmed spectroscopic
    redshifts (this number exceeds 767 since some sources are identified
    as both 24\,\um\ and radio sources).  The spectroscopic success rates
    for both samples is $\sim$55--60\%\ and does not correlate with flux
    density.  Dashed vertical lines represent average flux density
    limits of the radio data ($\sim$30\,\uJy) and the 24\,\um\ data
    ($\sim$150\,\uJy) in our fields.
    {\bf Bottom:} Each target in our survey must be detected at
    $>$3$\sigma$ in at least one of the three {\sc Spire} bands at 250,
    350, or 500\,\um.  Here we show the 250\,\um\ signal-to-noise ratio
    (SNR) against 350\,\um\ SNR.  The 500\,\um\ SNR scales with the size
    of the gray circles; if a target is under 3$\sigma$ at 500\,\um, it
    has no gray circle.  Spectroscopically confirmed sources are then
    marked with red crosses, showing no obvious correspondence between FIR
    flux density and spectroscopic success rates.  }
  \label{fig:spiredist}
\end{figure}

Our targets' distribution in radio flux density and in 24\,\um\ flux
density and {\sc Spire} signal-to-noise is shown in
Figure~\ref{fig:spiredist}.  Out of 1594 targets, 633 are 1.4\,GHz
identified, 1319 are 24\,\um\ identified, and 588 are both radio and
24\,\um\ identified.  The sources which are selected in both radio and
24\,\um\ maps come primarily from the LHN, COSMOS or CDF-S fields,
since EN1 and UDS lacks deep radio coverage and many sources in the
flanking fields of GOODS-N lack deep {\it Spitzer\/}
24\,\um\ coverage.

Throughout this paper, we refer to our sample as {\it Herschel} {\sc
  Spire}-selected galaxies (HSGs).  We prefer not to call them
submillimetre galaxies (SMGs) for sake of confusion with the
historical definition of SMG selected at 850\,\um--1.4\,mm.
We also choose not to use the term ULIRG for our sources since it
places the special qualification of a luminosity cut on the sample
($10^{12}$\,\lsun$<L<10^{13}$\lsun).  For the rest of the paper, any
reference to SMGs refers to a population selected at 850\,\um\ with
$S_{\rm 850}\ge$5\,mJy.  We use this strict definition of the SMG
population in order to draw comparisons between `classic SMGs' and
HSGs.

\section{Spectral Observations and Redshift Identification}\label{sec:ispire}

Optical spectroscopic observations were carried out at the W.M. Keck
Observatory using LRIS on Keck {\sc I} and DEIMOS on Keck {\sc II} in
2011 and 2012.

LRIS observations were carried out in adequate conditions on
2011-Feb-06 with $\sim$1\,\arcsec\ seeing and cloud cover, on
2012-Jan-26 and 2012-Jan-27 with $\sim$0.8--1.2\,\arcsec\ seeing with
minor to no cloud cover, and on 2012-Feb-27 with
$\sim$0.5-1.0\,\arcsec\ seeing and no cloud cover.  We used the
400/3400 grism for maximum wavelength coverage in the blue.  The 2011
observations used the 600/7500 grating in the red with a central
`multi-slit' wavelength of 6500\,\AA; the 2012-Jan observations used
the 400/8500 grating in the red with a central wavelength of
8400\,\AA, and the 2012-Feb observations used a central wavelength of
8000\,\AA.  All observations used clear filters for both red and blue
arms and the 560\,nm dichroic.  These setups give a
1.09\,\AA\ dispersion in the blue (e.g.~shortward of 5600\,\AA) and a
0.80\,\AA\ dispersion in the red.  Wavelength coverage for each source
varied by its position on the
5.5\arcmin$\times$7.8\,\arcmin\ slit-mask, the mean wavelength
coverage ranged from 2500--8200\,\AA\ in 2011 and from
2500\AA--1\,\um\ in 2012 but varied up to $\sim$1300\,\AA\ for sources
on each mask.  Due to the dichroic, some sources near the edge of the
slit-mask have gaps in wavelength coverage $\sim$800\,\AA\ wide in the
vicinity of 5600\,\AA.

We observed a total of 25 LRIS multi-slit masks, 13 of which were
observed in near-photometric conditions. LRIS data reduction,
including bias subtraction, flat fielding, wavelength calibration, and
sky subtraction were all performed using custom-built {\sc IDL}
routines.  Of 664 LRIS targets with $>$3$\sigma$ {\sc Spire}
detections, 268 were spectroscopically identified.  198 of the
identified sources (74\%) were confirmed in near-ideal conditions; we
add asterisks to the {\sc name} column of Table~\ref{tab_giant} to
distinguish these sources from sources confirmed in poorer weather.

DEIMOS observations were carried out in good conditions on 2011-May-28
and 2011-May-29 with $\sim$0.6--0.7\,\arcsec\ seeing, on 2011-Nov-28 in
average to cloudy conditions with $\sim$1.0--1.3\arcsec\ seeing, and
on 2012-Feb-16 and 2012-Feb-17 in very cloudy conditions with
1.0-3.0\,\arcsec\ seeing; we used the 600\,lines\,mm$^{-1}$ grating with a
7200\,\AA\ blaze angle (resulting in dispersion of 0.65\,\AA) and the
GG455 filter to block out higher-order light.  Wavelength coverage
varied with source position on the
5\,\arcmin$\times$16.7\,\arcmin\ slit-mask from 4400--9200\,\AA\ to
5200--9900\,\AA\ and averaged to 4850--9550\,\AA.

Sixteen of 29 DEIMOS multi-slit masks were observed in
near-photometric conditions, with integration times
$\sim$\,2700--4800\,s.  We used the DEEP2 DEIMOS data reduction
pipeline to reduce these data\footnote{The analysis pipeline used to
  reduce the DEIMOS data was developed at UC Berkeley by Michael
  Cooper with support from NSF grant AST-0071048}.  Of 930 DEIMOS HSG
targets, 499 were spectroscopically confirmed.  324 of those 499
(65\%) were confirmed in near-ideal conditions and are also marked
with asterisks in Table~\ref{tab_giant}.

Redshift identification was carried out through the identification of
multiple spectral signatures, primarily with the \oii\ doublet, \oiii,
\hb, \ha, \nii, Ca\,H\,\&\,K absorption and the Balmer break, \hg,
\lya, and the Lyman break (given in order of decreasing occurance in
the sample).  A minority of sources were identified by \ciii,
\civ\ and \heii\ emission (all sources identified via features in the
rest-frame ultraviolet are discussed in C12).  Full spectroscopic
details of all confirmed $z<2$ HSGs are given in Table~\ref{tab_giant}
(available online in full).  The reliability of redshift
identifications is also given in Table~\ref{tab_giant}, rated on a
scale of $q_{z}$=1--5, where 5 is best.  Sources with multiple feature
identifications have $q_{z}\ge3$, and sources with single line
identifications have $q_{z}=\,$1--2.  When visually inspecting the
spectra for our targets, redshifts were graded on a wider scale, with
some sources having $q_{z}$=0 (poor) or --1 (non-existent).  These
sources' potential redshifts are not reported in this paper due to
unreliability.

There are more HSGs with unconfirmed redshifts (826) than there are
with redshifts (767).  The vast majority of the unconfirmed sample is
unconfirmed due to poor weather including cloud-cover and poor seeing
(54\%\ of sources were observed in sub-optimal conditions).  The
average spectroscopic yield during good weather was 60$\pm$20\%\ while
the poor weather yeild was only 15$\pm$10\%.  The remaining
unconfirmed sources either very red faint continuum without
identifiable spectral features (20\%) or no continuum whatsoever
(80\%).  These sources are potentially misaligned on the slit (which
is based on an IRAC and $i-$ or $z-$band image when available) or are
too optically obscured to be detected in bright emission lines with a
$\sim$1--2 hr integration. They could also be at $z$\simgt3, thus
intrinsically much more difficult to detect optically.

Since we have very limited constraints on the spectroscopic failures,
we emphasise that our sample is spectroscopically incomplete.  We use
photometric redshifts and inferences on the redshift distribution
itself in \S~\ref{sec:nz} to estimate the spectroscopic completeness
as a function of redshift.

\subsection{Biases in Spectroscopic Confirmations}\label{sec:speccomp}

\begin{figure}
  \includegraphics[width=0.99\columnwidth]{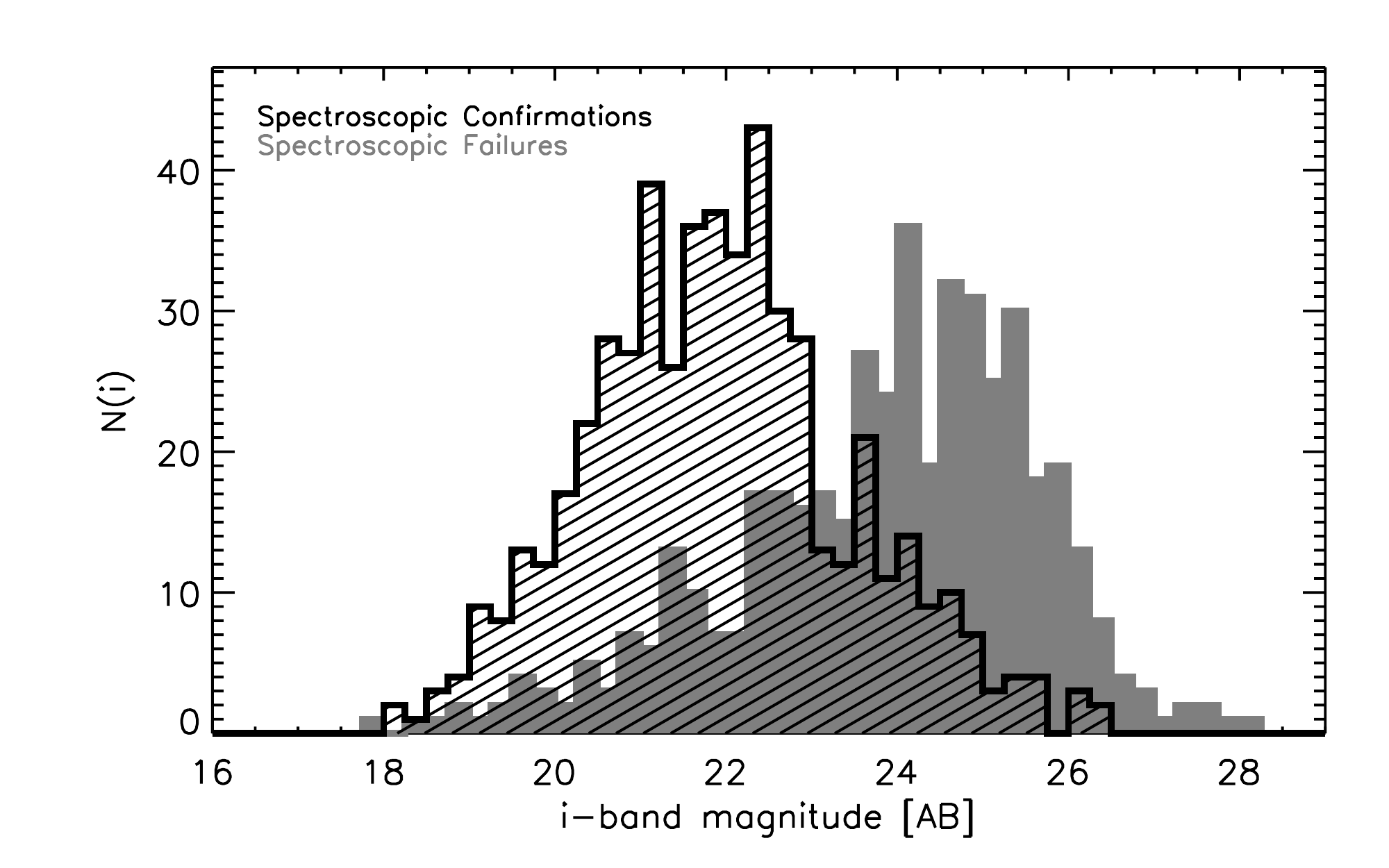}\\
  \includegraphics[width=0.99\columnwidth]{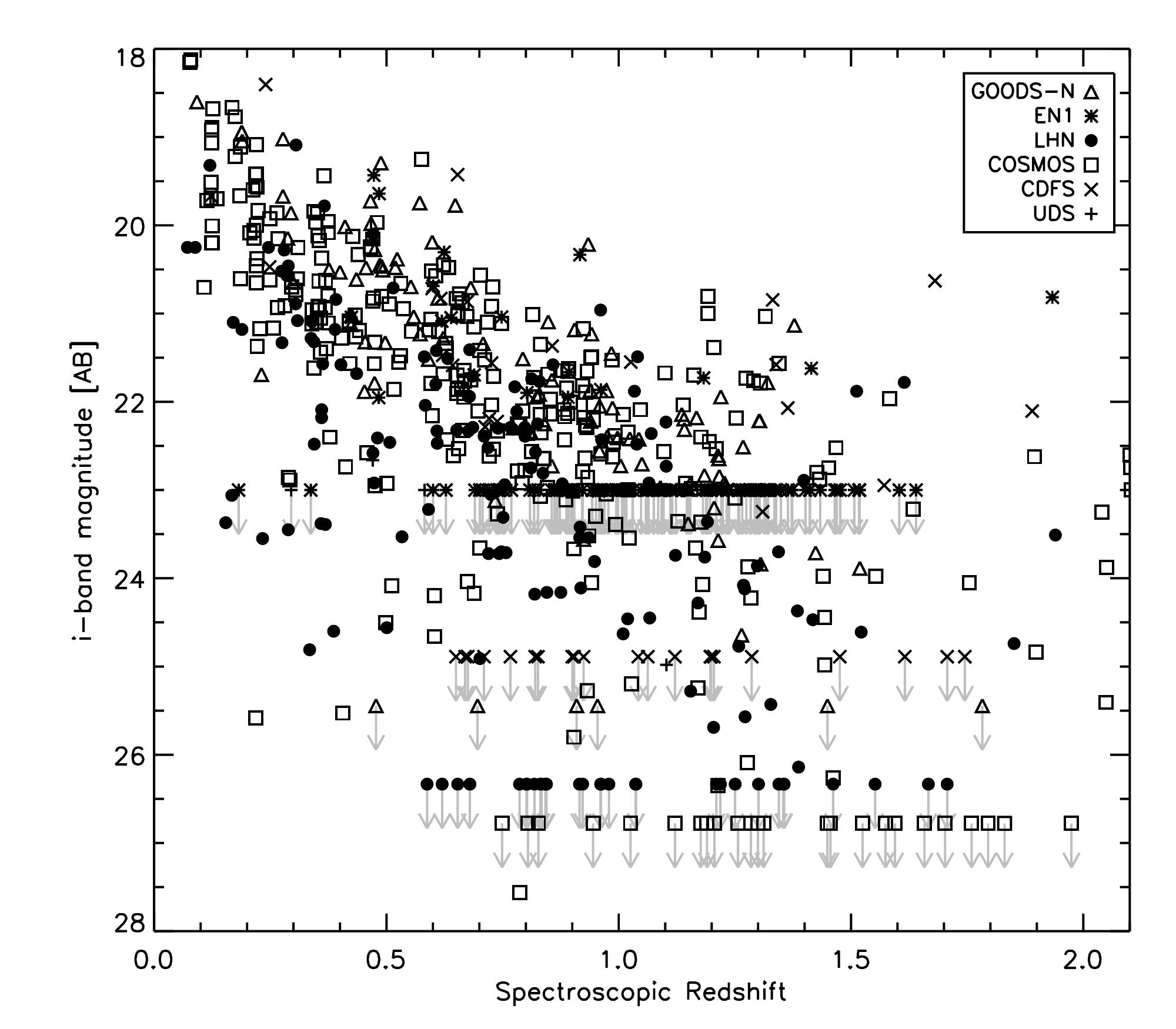}
  \caption{
    {\bf TOP:} Distribution in $i$-band magnitudes of spectroscopically
    confirmed (hashed area) and unconfirmed sources (filled gray area).
    There is clearly a bias in sources with spectroscopic confirmations
    being brighter at $i$-band magnitude, also much more likely to sit at
    lower redshifts.  However, note that for sources not included on this
    plot\,$-$\,those undetected in $i$-band\,$-$\,the spectroscopic
    success rate is 39\%, lower than the same fraction for sources with
    $i$-band detections, 54\%, although not as low as one might expect for
    very optically faint sources.  This is likely caused by many sources
    being confirmed through bright emission lines which can contribute
    minimally to broad-band photometry.
    {\bf BOTTOM:} The $i$-band magnitude against redshift for sources with
    confirmed redshifts.  At the bright end, $i<22$, sources show a clear
    trend with redshift, while many sources (marked as upper limits,
    according to the imaging depths in each field) have no available
    $i$-band photometry.  Clearly $i$-band magnitude strongly impacts the
    likelihood of measuring a spectroscopic redshift, a concept which is
    explored when measuring the spectroscopic completeness, shown in
    Figure~\ref{fig:speccomp}.}
  \label{fig:optz}
\end{figure}

Here we quantify the biases of our spectroscopic observations by
analyzing sources which failed to yield spectroscopic identifications.
This is arguably the most difficult completeness to quantify, since it
requires some knowledge of the redshift distribution of sources which
are: (a) the most optically obscured, (b) have featureless continuua,
or (c) have observed-frame emission lines outside of the wavelength
range of our observations (this redshift range is
$1.6$\simlt$z$\simlt$3.2$ for DEIMOS observations and
1.6\simlt$z$\simlt1.7 for LRIS observations).  Figure~\ref{fig:optz}
shows the optical $i$-band magnitude distributions for sources
spectroscopically confirmed and unconfirmed, and $i$-band magnitude
against redshift.  This makes it clear that optical $i$-band magnitude
need not be very bright for a spectroscopic identification based on
emission lines, and that optical magnitude itself does not constrain
redshift, given the number of sources at low-$z$ which are undetected
in the $i$-band.  However, it is clear that there is an overall trend
with redshift at brighter magnitudes $i<22$, with very few sources at
these magnitudes at $z>0.5$.  In that sense, it is also clear from
Figure~\ref{fig:optz} that optically bright sources are more likely to
be spectroscopically confirmed than those which are faint; the mean
$i$-band magnitude of spectroscopically confirmed sources is $i_{\rm
  AB}=22.1$, while for unconfirmed source it is $i_{\rm AB}=23.8$.

\begin{figure}
  \centering
  \includegraphics[width=0.99\columnwidth]{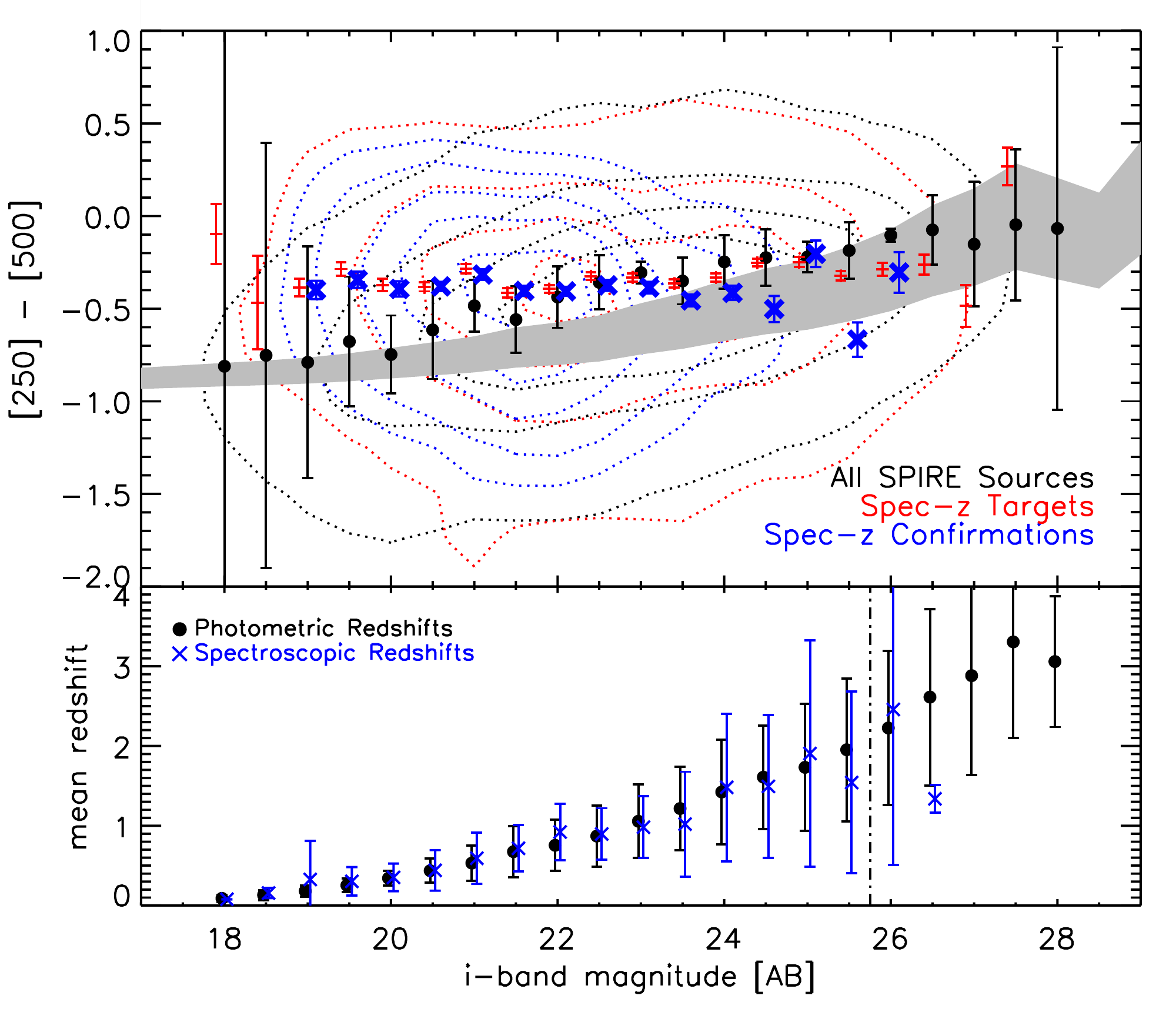}
  \caption{
    Relationship between $i$-band magnitude and {\sc Spire} color, [250]
    -- [500]=$-\log$\,(S$_{\rm 250}$/S$_{\rm 500}$).  {\sc Spire} color
    becomes redder with fainter magnitudes, which is expected given the
    redshift-$i$-band magnitude relationship seen in Figure~\ref{fig:optz}
    and an assumption of constant dust temperature from low to high
    redshifts.  The locus of all $>$3$\sigma$ {\sc Spire} sources in the
    upper panel consists of $\sim$5000 sources in the COSMOS field (black
    dotted contours and black points, the mean {\sc Spire} color for $i$
    magnitude bins).  Red dotted contours and points represent the
    distribution of spectroscopic targets ($\sim$1600 sources across all
    fields) and blue represent sources with spectroscopic confirmations,
    clearly skewed towards brighter $i$-band magnitudes.  In the bottom
    panel we compute the mean photometric and spectroscopic redshift for
    the same colored samples from the plot above, and see agreement down
    to $i_{\rm AB}\approx25.5$ (vertical dot-dashed line).  At top, the
    gray band represent the expected {\sc Spire} colors at the mean
    photometric redshift per $i-$band magnitude bin for dust temperatures
    ranging from 30\,K to 50\,K.  Within uncertainty, the full sample
    follows this expectation, while the spectroscopic sample deviates
    toward redder colors at brighter $i$-band magnitudes.
  }
  \label{fig:ispire}
\end{figure}

Prior studies of optical photometric properties of infrared starburst
galaxies find poor coorespondence between optical magnitude, IR
luminosity and redshift \citep[e.g.][]{chapman04a,chapman05a}.  This
makes it quite difficult to estimate the redshift distribution and IR
luminosities of the optically faint targets for which we fail to
measure redshifts.  Without redshift information, we must rely on
color information to infer if the optically fainter targets have a
significantly distinct redshift distribution (this also assumes {\sc
  Spire}-color varies with redshift and dust temperature is roughly
fixed).  Figure~\ref{fig:ispire} shows how {\sc Spire} color relates
to $i$-band magnitude for the entire parent galaxy sample of
$>$3$\sigma$ {\sc Spire} sources, spectroscopic targets, and also
spectroscopic confirmations.  This indicates that {\sc Spire} colors
should become redder with fainter $i$-band magnitude, assuming a
constant temperature and redshift range.  The mean
redshifts\,$-$\,photometric and spectroscopic\,$-$\,per $i$-band bin
are self-consistent, indicating no strong bias in the spectroscopic
redshift distribution with $i$ magnitude.  Note however that the
analysis as presented in Figure~\ref{fig:ispire} excludes sources (a)
without photometric redshifts, and (b) without $i$-band counterparts.
The variation of {\sc Spire} color with redshift, along with the
perceived bias of our spectroscopic sample towards redder colors at
bright magnitudes, is addressed further in \S~\ref{sec:spirecolor}.

Of the spectroscopically confirmed subsample, 65$\pm$15\%\ are
250\,\um-peaking (meaning $S_{\rm 250}>S_{\rm 350}>S_{\rm 500}$),
23$\pm$14\%\ are 350\,\um-peaking ($S_{\rm 350}>S_{\rm 250}$ and
$S_{\rm 350}>S_{\rm 500}$), and 13$\pm$9\%\ are 500\,\um-peaking
($S_{\rm 250}<S_{\rm 350}<S_{\rm 500}$).  These values are comparable
to the spectroscopic targets: 72$\pm$12\%\ were 250\um-peaking,
17$\pm$9\%\ were 350\um-peaking and 11$\pm$6\%\ were 500\um-peaking.
This leads us to conclude that successful spectroscopic identification
does not have a bias with respect to FIR SED shape.

\subsection{Redshift Distribution}\label{sec:nz}

\begin{figure}
  \centering
  \includegraphics[width=0.99\columnwidth]{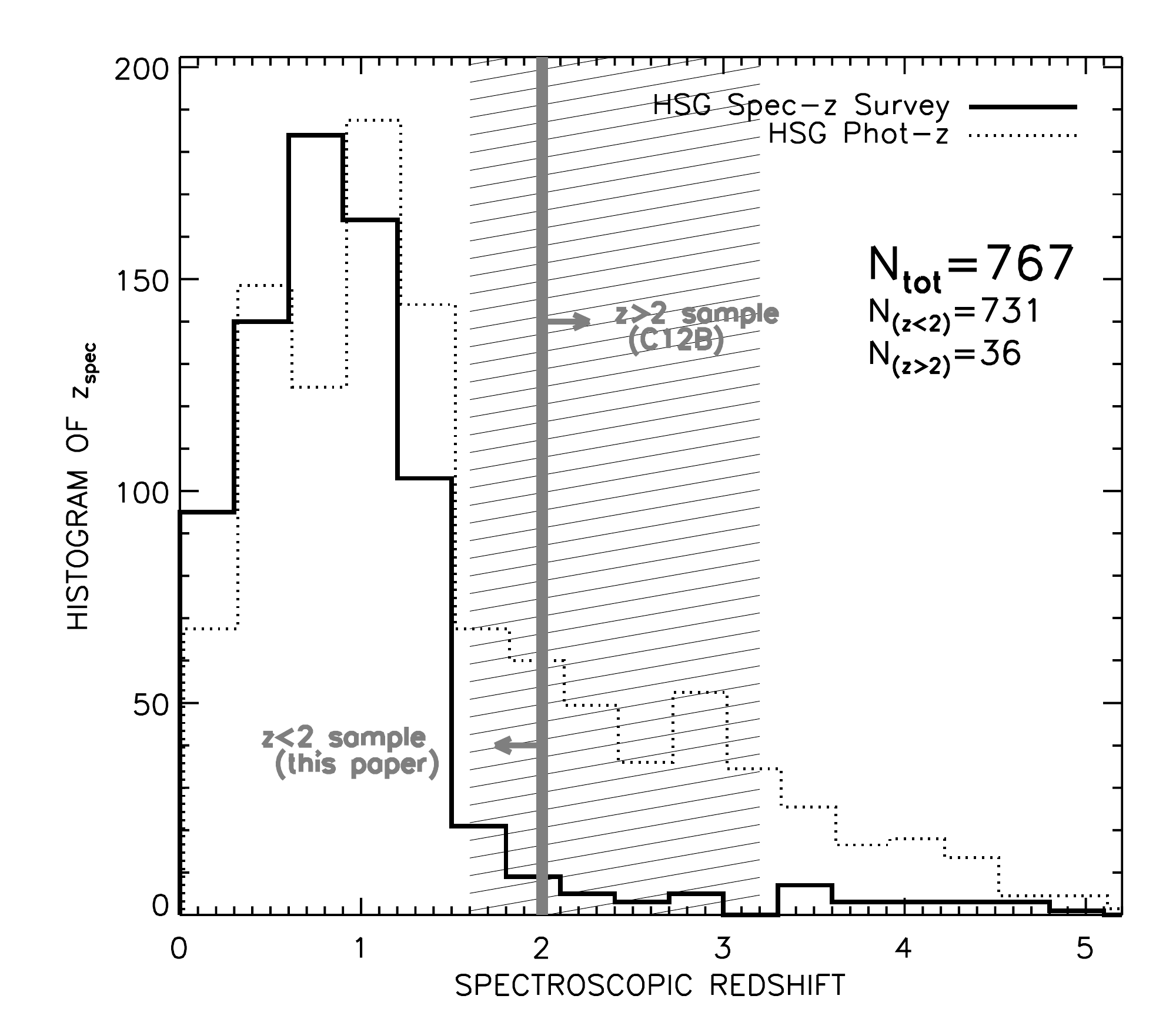}
  \caption{ The spectroscopic redshift distribution of HSGs. Since the
    majority of our 767 spectroscopically confirmed sources were
    observed with DEIMOS (66\%), the redshift range
    $z$=1.6--3.2 (hashed area) is sparsely populated (only two sources
    within this range are DEIMOS identified, both bright quasars).
    Distribution of photometric redshifts for the sample of
    spectroscopically targeted HSGs is shown as the dotted line,
    consistent at $z$\simlt1.5 but suggesting a significantly
    decreased spectroscopic yield at $z>$1.5.
  }
  \label{fig:nz}
\end{figure}

The redshift distribution of the 767 spectroscopically confirmed HSGs
in this survey is shown in Figure~\ref{fig:nz}.  Of 767, the vast
majority, 731, are at $z<2$. The distribution peaks at $z=0.85$ with a
tail of sources extending out to higher redshifts, discussed fully in
a separate accompanying paper, C12.  Also plotted is the distribution
in photometric redshifts for all spectroscopic targets. The
photometric redshift distribution peaks at the same epoch, but with a
much higher fraction of sources at $z$\simgt2.  The deficit in
spectroscopic redshifts at $z>2$ is caused by: (a) the DEIMOS `redshift
desert' (DEIMOS observations comprise 66\%\ of our sample and no strong
emission lines are visible within DEIMOS wavelength coverage from
$1.6<z<3.2$); and (b) by a decreasing spectroscopic completeness due
to enhanced obscuration in the rest-frame ultraviolet relative to
rest-frame optical.  The latter point relates to the very obscured
nature of infrared-luminous galaxies: the increased presence of dust
implies more significant extinction in the UV and optical, with more
substantial effects at bluer wavelengths.

\begin{figure}
  \centering \includegraphics[width=0.99\columnwidth]{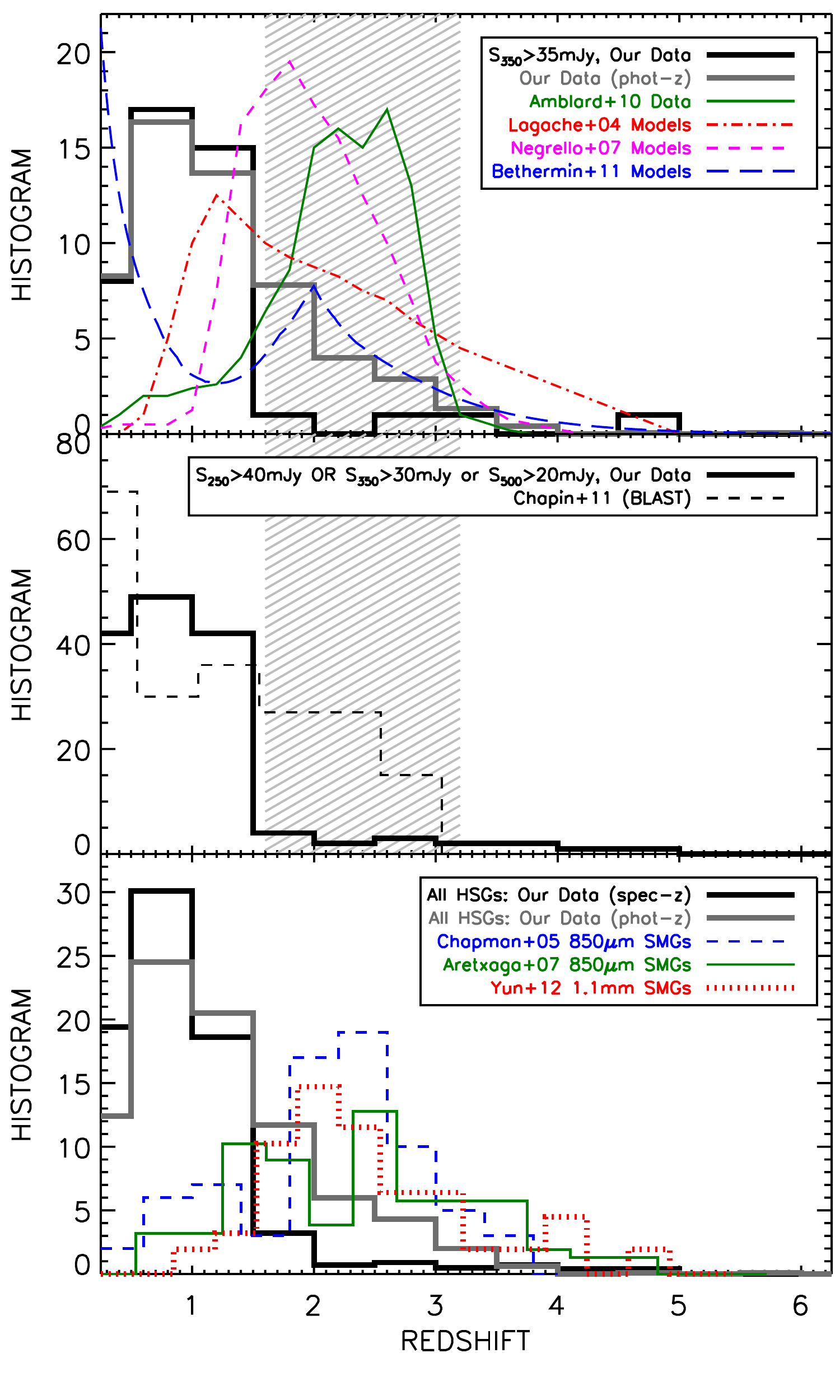}
  \caption{Comparison of our redshift distributions to those in the
    literature; at top, we compare to the \citet{amblard10a}
    distribution (green) for sources selected at 350\,\um\ above 35\,mJy
    (with $>$3$\sigma$ significant detections at both 250\,\um\ and
    500\,\um).  We also compare the full distribution of photometric
    redshifts of our targets (thick, solid gray) to the model
    distributions from \citet{lagache04a}, \citet{negrello07a}, and
    \citet{bethermin11a} (dot-dashed red, dashed magenta, and
    long-dashed blue respectively).  Neither observed distribution
    (spectroscopic and photometric) shows good agreement with the
    models, however the Lagache~\etal\ model (which peaks at $z\sim1$)
    agrees best with our data.  The middle panel compares to the
    redshift distribution of \citet{chapin11a} of for BLAST-bright
    galaxies in ECDF-S.  Given the limited number of sources in the
    BLAST analysis (and that they are all identified in ECDF-S, a small
    volume) we conclude that these two distributions are consistent
    within uncertainty.  At bottom we compare the HSG redshift
    distrubiton to those of 850\um--1.1mm selected SMGs, which peak at
    $z\approx$\,2.2--2.6 due to the longer wavelength selection.  The
    HSG distribution is taken from Figure~\ref{fig:nz} and scaled down
    10 times in total number for direct comparison.}
  \label{fig:nzcompare}
\end{figure}

Note that predicted redshift distributions for {\it Herschel\/}
sources have been studied in detail by \citet{amblard10a} and
\citet{bethermin11a}.  However, these are based on different {\sc
  Spire} selection methods and are thus not directly comparable to the
HSG sample discussed herein.  For instance, the Amblard
\etal\ distribution of 350\,\um-selected sources peaks at $z\sim2.2$,
as seen in the upper panel of Figure~\ref{fig:nzcompare}.  We extract
HSGs from our sample which would satisfy their detection criteria
($S_{350}>35$\,mJy, with 250\,\um\ and 500\,\um\ ${\rm SNR}>3$) for
comparison and find a statistically distinct distribution from Amblard
\etal

We also compare data with the model predictions of the expected {\sc
  Spire} distributions from \citet{lagache04a}, \citet{negrello07a}
and \citet{bethermin11a}.  While our data (both of the limited $S_{\rm
  350}>35$mJy sample and the distribution of all photometric
redshifts) are inconsistent with most models (particularly at the
$z>2$ end), our results are most consistent (tested via a
Kolmogorov-Smirnov statistic) with the predicted distribution of
\citet{lagache04a}; the number of sources we observe at $z<1$ is
notably different than the predictions from \citet{negrello07a} or
from the long-wavelength photometric-redshift based work of
\citet{amblard10a}.

In the middle panel of Figure~\ref{fig:nzcompare} we compare our
results to those of \citet{chapin11a} for BLAST-detected galaxies in
ECDF-S, a combination of photometric redshifts
\citep{dunlop10a,ivison10a} and spectroscopic redshifts
\citep{casey11a}.  Their selection is based on three different flux
cuts at 250, 350, and 500\,\um, and when applying the same selection
to our sample, we find an agreement between the median redshifts, both
of which have $\langle z\rangle\approx1$ but with differences at both
low and high-redshift. There is an additional peak in the
Chapin~\etal\ sample at $z<0.5$ which is not observed in our sample.
This could be due to cosmic variance, since the BLAST ECDF-S sample
probes a small volume in a single field, while our data sample
multiple deep fields over a larger sampling volume.  Given the DEIMOS
redshift desert, comparison cannot be drawn fairly beyond $z\sim1.6$.

Worth noting is the contrast of these redshift distributions with the
850\,\um-selected SMG redshift distribution \citep{chapman05a,yun12a},
which peaks at $z\sim2.2$ (see bottom panel of
Figure~\ref{fig:nzcompare}).  The peak in 850\,\um-selected galaxies
occurs at earlier epochs due to the selection wavelength: 850\,\um\ is
more sensitive to lower luminosity/colder sources at higher redshifts
and less sensitive to warmer sources at low redshift.  {\sc Spire}
selection is not constant in luminosity out to very high redshifts
like $\sim$1\,mm selection is, but it has the benefit of probing the
peak of the infrared SED at $z\sim$\,1--2, thus is not likely to miss
or select against sources at a given luminosity due to their warmer
SED shape.

\subsection{Comparison to photometric redshifts}

Photometric redshifts are far easier to obtain on larger samples of
galaxies than spectroscopic redshifts, and whenever the latter become
available, it is important to test the reliability of the former.
Many large statistical studies are now motivated exclusively by use of
large catalogs of photometric redshifts; here we explore possible
underlying biases which might persist in HSG photometric redshifts and
what might cause them.  Note that all photometric redshifts used in
this paper are calculated from ultraviolet through near-infrared
photometry and exclude any long-wavelength data $>$8\um\ (the catalogs
used in this analysis are described at the end of
\S~\ref{sec:selection}).

\begin{figure}
  \centering
  \includegraphics[width=0.99\columnwidth]{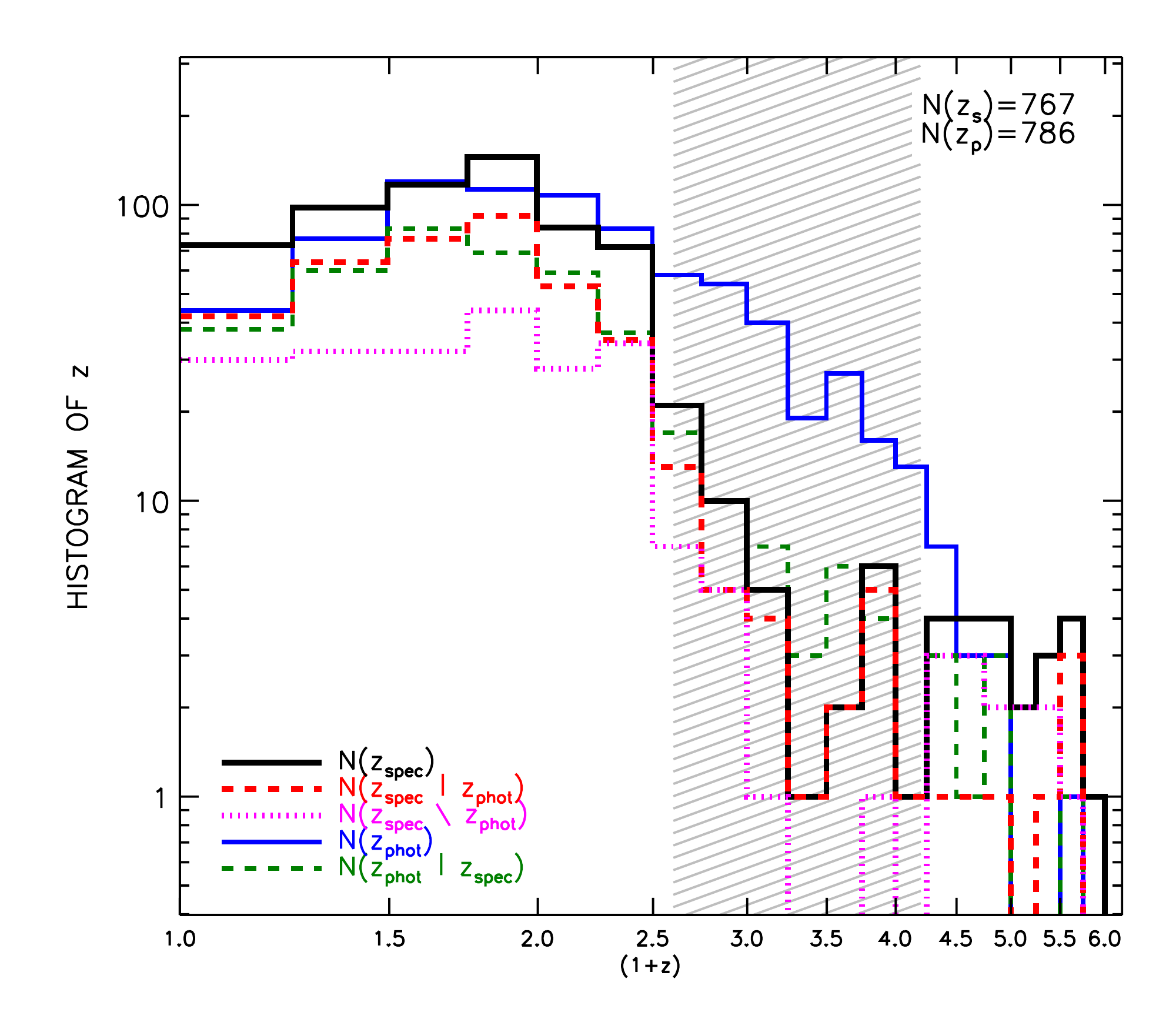}\\
  \includegraphics[width=0.99\columnwidth]{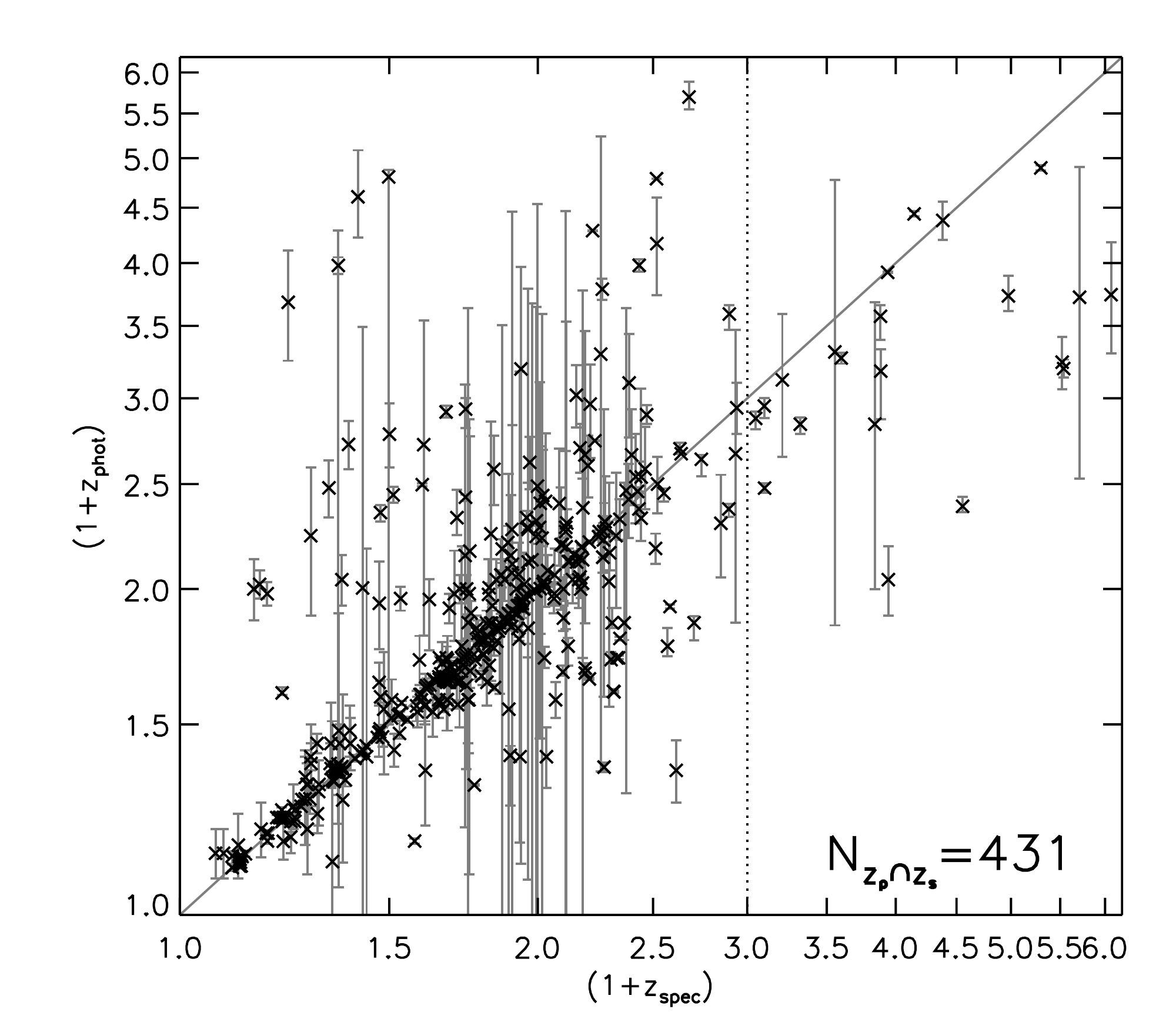}
  \caption{ {\bf TOP:} Redshift distributions for HSGs shown on a
    logarithmic scale: the full spectroscopic sample (solid black);
    spectroscopically confirmed sources with photometric redshifts
    ($z_{\rm spec}\,|\,z_{\rm phot}$, in other words ``$z_{\rm
      spec}~given~z_{\rm phot}$,'' dashed red); spectroscopically
    confirmed sources without photometric redshifts ($z_{\rm
      spec}~\setminus~z_{\rm phot}$, in other words ``$z_{\rm
      spec}~take-away~z_{\rm phot}$,'' dotted magenta); photometric
    redshifts (solid blue); and photometric redshifts for
    spectroscopically confirmed sources ($z_{\rm phot}|z_{\rm spec}$,
    in other words ``$z_{\rm phot}~given~z_{\rm spec}$,'' dashed
    green).  The DEIMOS redshift desert is shown as a hashed area.
    {\bf BOTTOM:} Spectroscopic versus photometric redshift for
    successfully identified targets \emph{which have} photometric
    redshifts (431 sources out of 767).  The photometric redshifts
    were taken from \citet{strazzullo10a} (Lockman Hole North),
    \citet{rowan-robinson08a} (ELAIS-N1, UDS), \citet{ilbert10a}
    (COSMOS) and \citet{cardamone10a} (ECDFS) with associated
    uncertainties on $z_{\rm phot}$ shown. The mean RMS scatter at is
    $\Delta z$/$(1+z_{\rm spec})$=0.29, notably worse than the
    expected photometric redshift accuracy of photometric redshifts of
    `normal' starburst galaxies.}
  \label{fig:photzspecz}
\end{figure}
Spectroscopic and photometric redshifts are plotted against one
another in Figure~\ref{fig:photzspecz}.  From this we measure that HSG
photometric redshifts have a mean rms scatter of $\sigma$=$\Delta
z$/$(1+z_{\rm spec})$=0.29, characteristically a factor 3--4 times
worse than the photometric redshifts for regular field galaxies
\citep[][measure $\Delta z$/$(1+z_{\rm spec})$=0.07 for galaxies with
  spectral confirmations in COSMOS]{ilbert10a}.  This might be
suprising given that the majority of sources in this comparison come
from fields with deep, multi-band photometry (e.g. COSMOS, LHN).  Why
are the photometric redshifts of infrared-bright galaxies
substantially worse than those of most field galaxies?

One might think that the disagreement originates from source blending
or mismatching due to the large beamsize of infrared observations.
However, we do not expect blending or mismatching to occur since the
photometric redshifts taken from the source catalogs are matched to
our spectroscopic targets' position within \simlt1\arcsec\ (i.e. even
if a {\sc spire} source's counterpart is mistakenly identified, we
would still expect the false counterpart's photometric redshift to
agree with its spectroscopic redshift).  Since mismatched counterparts
is not likely to be the source of the photometric redshift scatter,
the scatter is likely due to some other property of the
infrared-selected sample.

If the $z_{\rm phot}-z_{\rm spec}$ disagreement is not caused by
mismatched counterparts, then what intrinsic physical processes could
explain less reliable photometric redshifts in infrared-selected
galaxies?  Direct detection at far-infrared wavelengths implies two
things about a galaxy: (i) it has a significant dust reservoir which
has absorbed more energetic light and re-radiated it in the
far-infrared, and (ii) it has a high star-formation rate \citep[since
  FIR luminosity scales directly to star formation
  rate;][]{kennicutt98b}.  The first point highlights that the
galaxy's emission at rest-frame ultraviolet and optical
wavelengths--emitted by young, hot stars--is being obscured and
scattered by dust; as a result, infrared-selected galaxies are
optically fainter than their less-dusty counterparts of similar
redshifts and stellar masses.  The second point, that
infrared-galaxies have intrinsically higher star formation rates than
most `normal' field galaxies follows from the implied infrared
luminosities of the infrared-selected samples.  The vast majority of
field galaxies have star formation rates $<$10\sfr\ \citep[e.g. the
  vast array of 200K galaxies in COSMOS which are fit with stellar
  population templates from][]{bruzual03a}.  In contrast, most
infrared-selected galaxies satisfy $L_{\rm IR}>10^{11}$\lsun, which
implies star formation rates of $>17$\sfr\ or $L_{\rm
  IR}>10^{12}$\lsun\ which implies $>170$\sfr.  Higher star formation
rates translate to brighter rest-optical emission lines and higher
line-to-continuum ratios.

How might dust obscuration and enhanced line-to-continuum ratios
impact the reliability of photometric redshifts for HSGs?  The first
effect is straightforward: the fainter a galaxy is in the optical, the
more difficult it is to put a reliable constraint on its photometric
redshift.  Since HSGs are optically fainter than `normal' galaxies
which harbor less dust, this is one reason HSG photometric redshifts
are less reliable.  The second effect comes from the `contamination'
of bright emission lines of the optical broad band filters used to
compute the sources' photometric redshifts.  Although emission lines
can contaminate the broad band magnitudes for both normal and dusty
galaxies, dusty galaxies have comparably fainter continuum (or rather,
a more significant contribution from emission lines).  Also, despite
some efforts to account for emission lines in photometric redshift
code algorithms (by including them in the templates used for the
fitting), none of the existing optical templates in the literature
\citep[e.g.][]{calzetti94a,bruzual03a} include models with extremely
high star formation rates and dust content as exists in HSGs, hence
the poorer $z_{\rm phot}-z_{\rm spec}$ agreement.

The influence of dust obscuration on photometric redshift estimates is
not completely straightforward in that dust is not necessarily
expected to extinct optical flux uniformly across all wavelength
regimes.  Sources with high phot-$z$s and low spec-$z$s can be
explained by differential blue-to-red obscuration whereby the galaxies
dropout in several blue bands due to stronger absorption of higher
energy photons.  The handful of sources with low photometric redshifts
and high spectroscopic redshifts ($z>2$) are described in detail in
C12; these are thought to disagree due to differential obscuration of
resonant emission-line photons and continuum
\citep[e.g.][]{neufeld91a}, whereby emission lines are not extincted
as significantly as stellar emission continuum.  This is also highly
dependent on viewing geometry, but the observation holds that
photometric redshifts of dusty galaxies are significantly worse than
they are for normal field galaxies which contain less dust.  Caution
should be exercised when using large photometric datasets to quantify
the aggregate properties of infrared-selected samples.

\begin{figure}
  \centering
  \includegraphics[width=0.94\columnwidth]{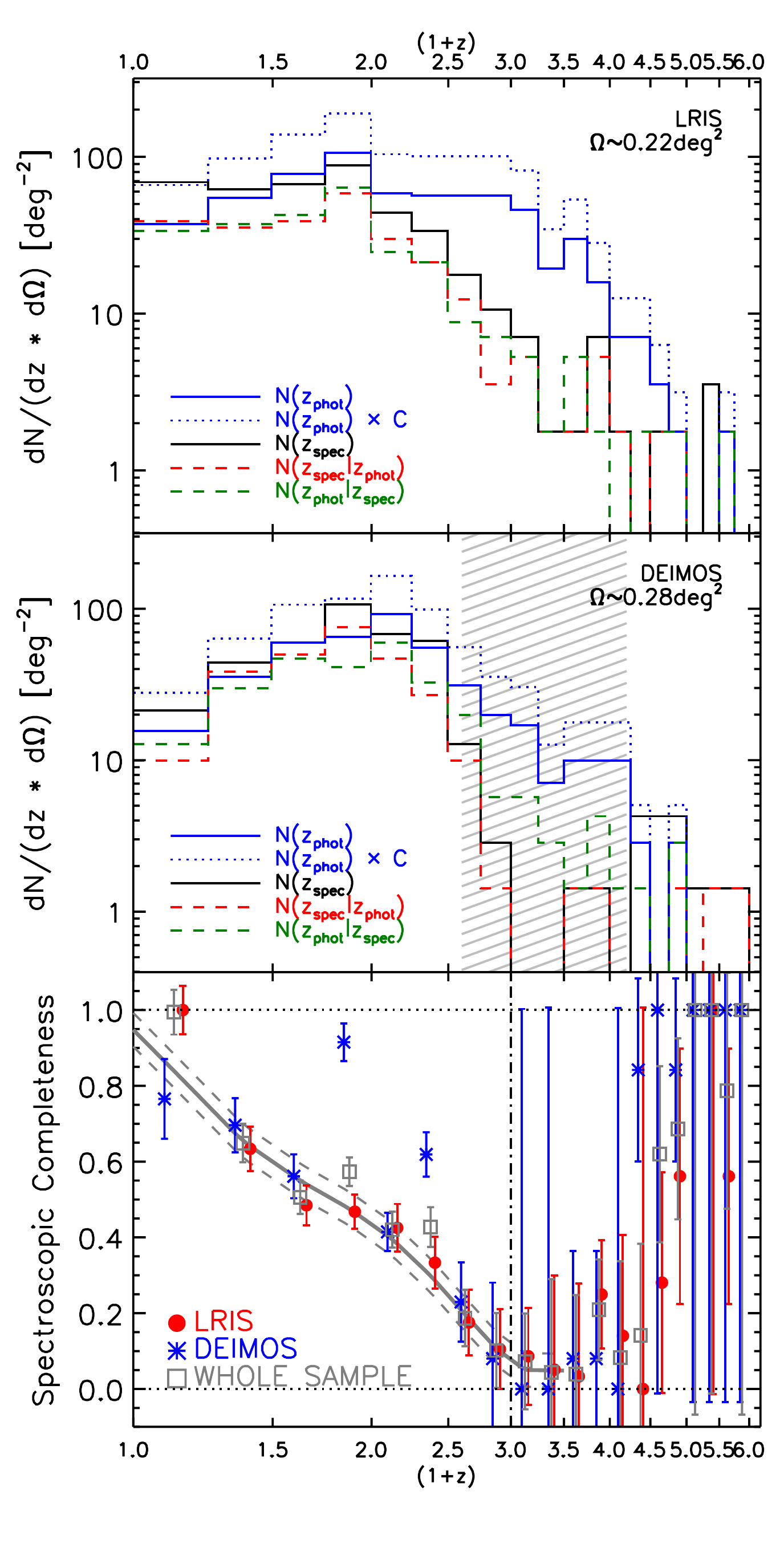}
  \caption{
    The top and middle panels show the redshift distributions of HSGs
    split into LRIS and DEIMOS samples.  DEIMOS redshift desert is marked
    in the middle panel.  Assuming the redshift distribution for sources
    without photometric redshifts resembles the distribution in
    photometric redshifts (see Figure~\ref{fig:photzspecz} top panel),
    then we can estimate the spectroscopic completeness of our survey,
    bottom panel, by dividing the spectroscopic redshift distributions (by
    instrument) by the photometric redshift distribution scaled up to
    account for sources without photometric redshifts.  From this, we
    determine that survey completeness cannot be constrained at $z>2$ (due
    to small number statistics) which is why that sample is discussed in a
    separate paper.  The best-fit curve at $z<2$ (plotted in gray with
    dashed lines showing uncertainty) is used later to estimate the total
    contribution of HSGs to the SFRD.  The two `outlier' high-completeness
    DEIMOS points are discussed in the text.}
  \label{fig:speccomp}
\end{figure}

\subsection{Spectroscopic Completeness}

With a lack of any better constraint on the redshift distribution of
unconfirmed HSGs, we use our targets' photometric redshifts (those
with and without spectroscopic redshifts) to constrain spectroscopic
completeness as a function of redshift.  Figure~\ref{fig:speccomp}
(top panel) contrasts the spectroscopic and photometric redshift
distributions by instrument, since LRIS and DEIMOS observations have
different wavelength coverage, thus different redshift completeness
levels.  Meauring spectroscopic completeness requires an estimate of
the underlying redshift distribution for the whole population,
including those without photometric redshifts.  The only constraint we
have on sources without photometric redshifts is from those
spectroscopically observed, shown in Figure~\ref{fig:photzspecz} at
top, consistent with the distribution in photometric redshifts.
Therefore, a scaling correction factor, $C$, is applied to the
distribution of photometric redshifts to account for the sources
without photometric redshifts.  Although this assumption might not be
entirely correct (i.e.~there might be a much higher fraction of $z>2$
sources which do not have photometric redshifts), the assumption that
the redshift distribution is weighted at lower redshifts effectively
places the lower limit of spectroscopic completeness at $z<2$.  If the
fraction of sources without photometric redshifts is higher at
high-$z$, then the spectroscopic completeness measurement at $z<2$
will increase.  The spectroscopic completeness is measured as a
function of redshift by dividing the distribution in spectroscopic
redshifts by that of the scaled up photometric redshifts
(i.e. $N(z_{\rm spec})/(C\times N(z_{\rm phot}))$\,).  This
spectroscopic completeness estimate is shown in the bottom panel of
Figure~\ref{fig:speccomp}.  As expected, it declines over $0<z<2$ and
then is unconstrained at $z>2$ due to limited samples.  The DEIMOS
completeness estimate has two `outlier' points at $z\sim0.8$ and
$z\sim1.3$; this is caused by the enhanced sensitivity of DEIMOS from
6500--7400\AA\ (where LRIS often has a gap in coverage), and
7600-9000\AA\ (where our LRIS observations were slightly more prone to
skyline contamination).  Since the completeness becomes unconstrained
at $z>2$, we choose $z=2$ as a natural boundary at which to divide the
sample, addressing the well-constrained $z<2$ population in this
paper, while discussing the full sample of spectroscopically
incomplete $z>2$ sources in C12.

\subsection{{\sc Spire} Color with Redshift}\label{sec:spirecolor}

\begin{figure}
  \centering
  \includegraphics[width=0.99\columnwidth]{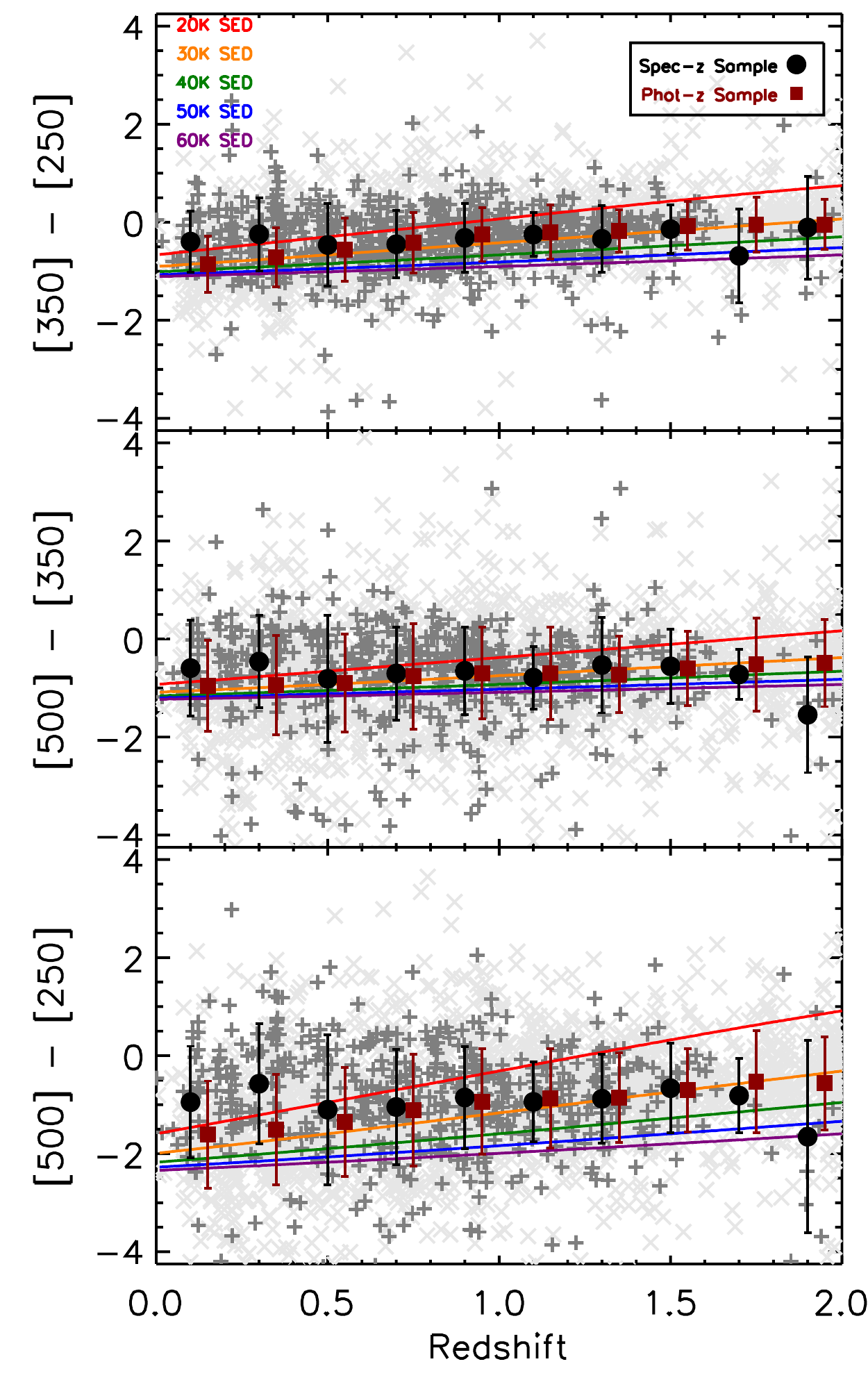}
  \caption{{\it Herschel}-{\sc Spire} color with redshift, where color
    is given as a magnitude difference between {\sc Spire} bands, such
    that high values correspond to red colors.  In light gray are {\sc
      Spire} colors of the entire photometric HSG sample from the COSMOS
    field; dark gray points are our spectroscopic sample.  The mean of
    the spectroscopic sample and photometric sample is shown as black
    circles and dark red squares, respectively, with associated
    uncertainties.  Overplotted are tracks of SEDs with fixed dust
    temperature, 20\,K (red), 30\,K (orange), 40\,K (green), 50\,K
    (blue), and 60\,K (purple).  We see no significant color evolution
    in the HSG sample with redshift; this is primarily due to selection
    properties of the sample.  }
  \label{fig:spirez}
\end{figure}
Recalling from the discussion in \S~\ref{sec:ispire}, our
spectroscopic HSGs sample the underlying distribution of all
{\sc Spire} sources well in {\sc Spire} color, despite a bias in
$i$-band magnitude (where brighter optical sources are more likely to
be spectroscopically confirmed).  Figure~\ref{fig:ispire} showed that
the mean redshift and {\sc Spire} color per $i$-band magnitude of our
spectroscopic HSGs is drawn from the parent sample of {\sc Spire}
sources without significant bias.  While this hinted at the evolution
of {\sc Spire} color with redshift, we can test this evolution
directly using our spectroscopic HSGs and larger photometric
redshift samples from COSMOS \citep{ilbert10a}.

Assuming that most infrared starbursts can be described with a 35\,K
dust SED \citep{chapman05a,rieke09a}, one would expect low redshift
sources to have blue {\sc Spire} colors and high redshift sources to
have red {\sc Spire} colors \citep[e.g.][]{cox11a,roseboom12a}.
Figure~\ref{fig:spirez} plots the [350]--[250], [500]--[350], and
[500]--[250] colors against redshift, where [250] is defined as a
250\,\um\ magnitude, i.e. $-2.5$log($S_{250}$).  We see no strong
trend in SED shape with spectroscopic redshift.  While it might be
expected that high-$z$ sources are significantly more `red' than
`blue' low-$z$ sources, this interpretation is simplistic since it
assumes no evolution in dust temperature, no relationship between dust
temperature and luminosity, and ignores the impact of population
selection effects.  Overplotted on Figure~\ref{fig:spirez} are the
color-redshift tracks for SEDs of fixed dust temperature systems from
20--60K; they evolve more strongly with redshift than our data,
particularly notable in [500]--[250] color.

The dispersion on the {\sc Spire} colors in both the spectroscopic and
photometric samples imply a wide range of dust temperatures.  This
highlights that far-infrared flux densities and colors cannot be used
exclusively as a proxy for redshift.  Notwithstanding the large
dispersion in {\sc Spire} colors, two main observations can be drawn
from Figure~\ref{fig:spirez}: (a) that the mean {\sc Spire} colors do
not redden significantly with redshift as might be expected; and (b)
at low-$z$ ($z<0.5$) our spectroscopic sample is redder, i.e. cooler,
than the photometric redshift sample.

Both of these observations can be understood by investigating how {\sc
  Spire} selection works in the $L_{\rm IR}-T_{\rm dust}$ plane,
discussed fully in \S~\ref{sec:lfirtd}.  Dust temperature increases
with luminosity (which is partially a selection effect against
high-$z$ cold systems and partially thought to be a real correlation).
Since the highest-redshift sources in our sample are significantly
more luminous than our lower redshift sources, the correlation between
$L_{\rm IR}$ and $T_{\rm dust}$ translates to a correlation between
$z$ and $T_{\rm dust}$.  This results in roughly constant {\sc Spire}
colors with redshift.  At $z<0.5$, we note that our spectroscopic
survey is limited in the number of very luminous (\simgt10$^{12}$\lsun)
galaxies it can detect simply due to sky area probed
($\approx1deg^2$); since (in \S~\ref{sec:lfirtd}) we will observe that
more luminous HSGs are hotter, it is natural that a sample of HSGs
with photometric redshifts (covering a much larger area than our
spectroscopic survey) will probe naturally warmer HSGs.  This gives
rise to the `redder' $z<0.5$ spectroscopic sample.  Although the two
redshift bins at $1.6<z<2.0$ seem bluer than at any other redshift, we
note that the number of sources in these redshift bins is
significantly fewer (and the observation is not the same for the much
larger sample of sources with photometric redshifts in that range).

\begin{figure*}
  \centering
  \includegraphics[width=0.99\columnwidth]{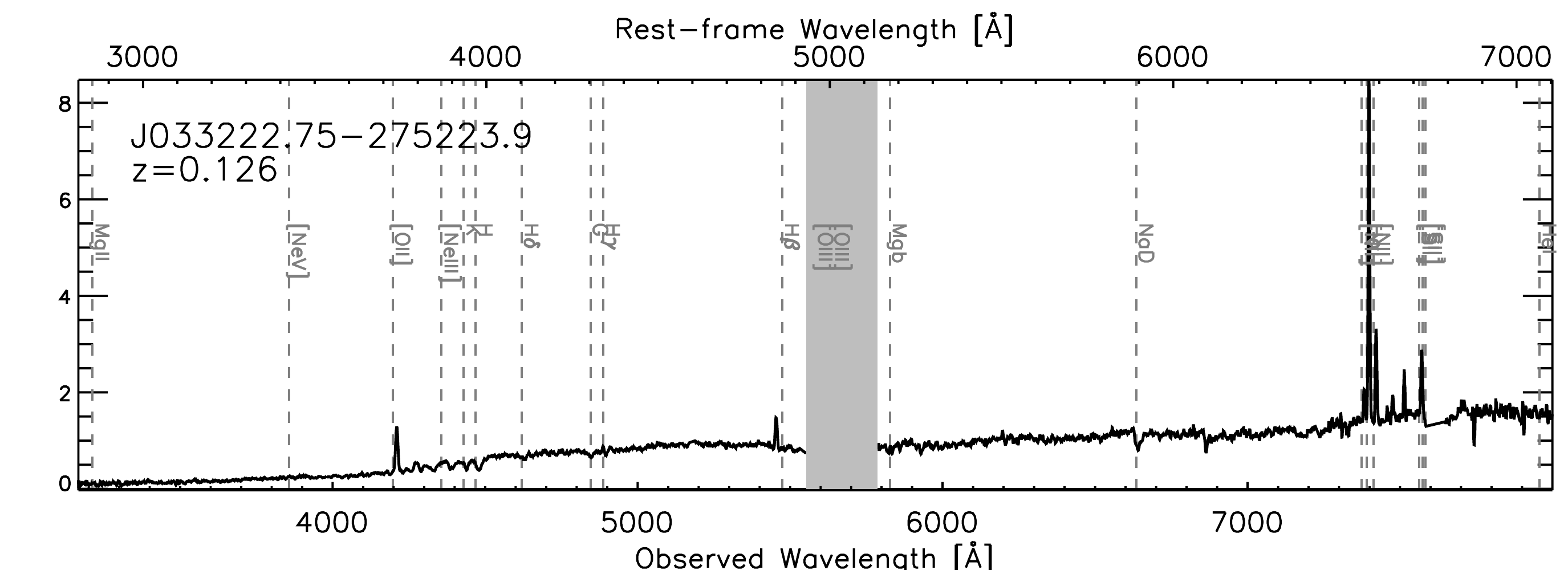}
  \includegraphics[width=0.99\columnwidth]{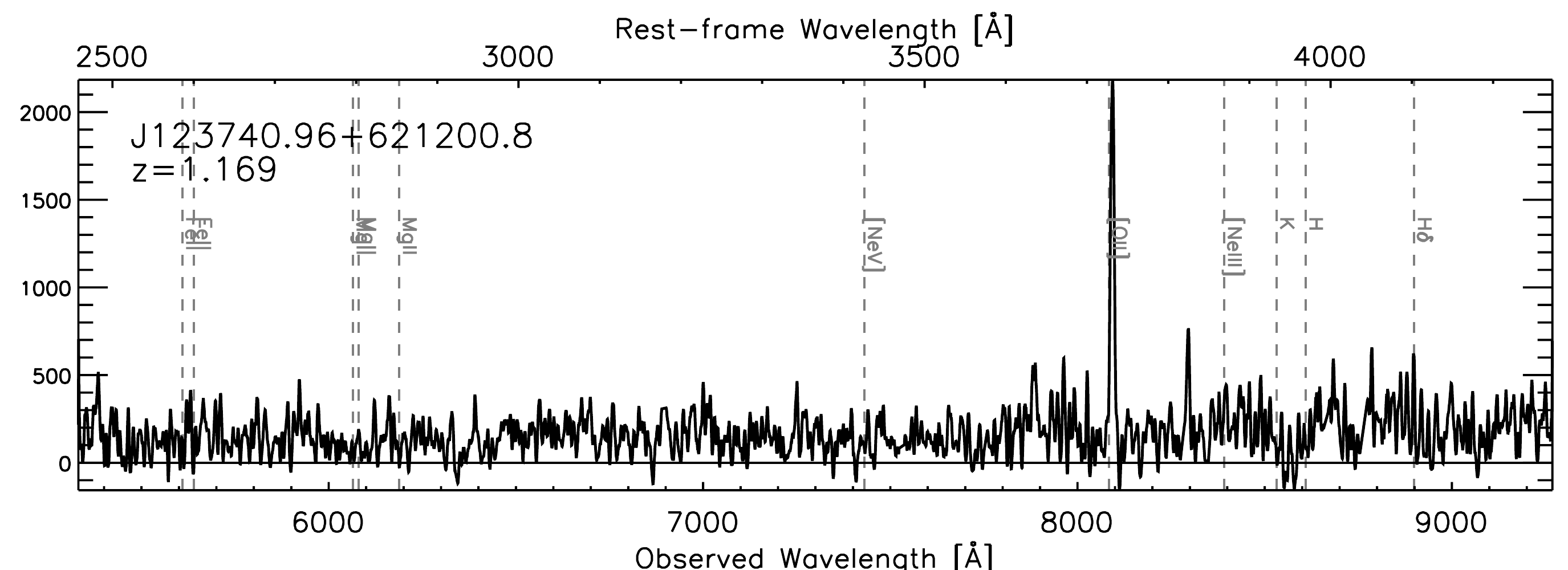}\\
  \includegraphics[width=0.99\columnwidth]{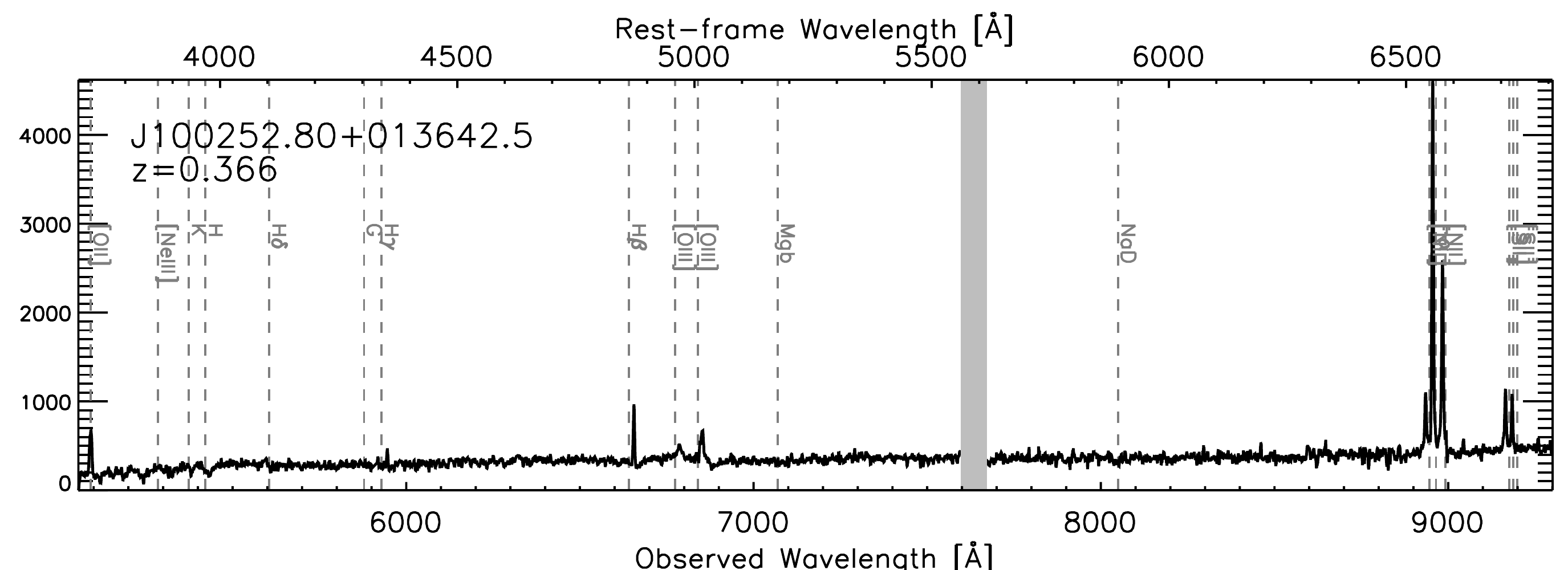}
  \includegraphics[width=0.99\columnwidth]{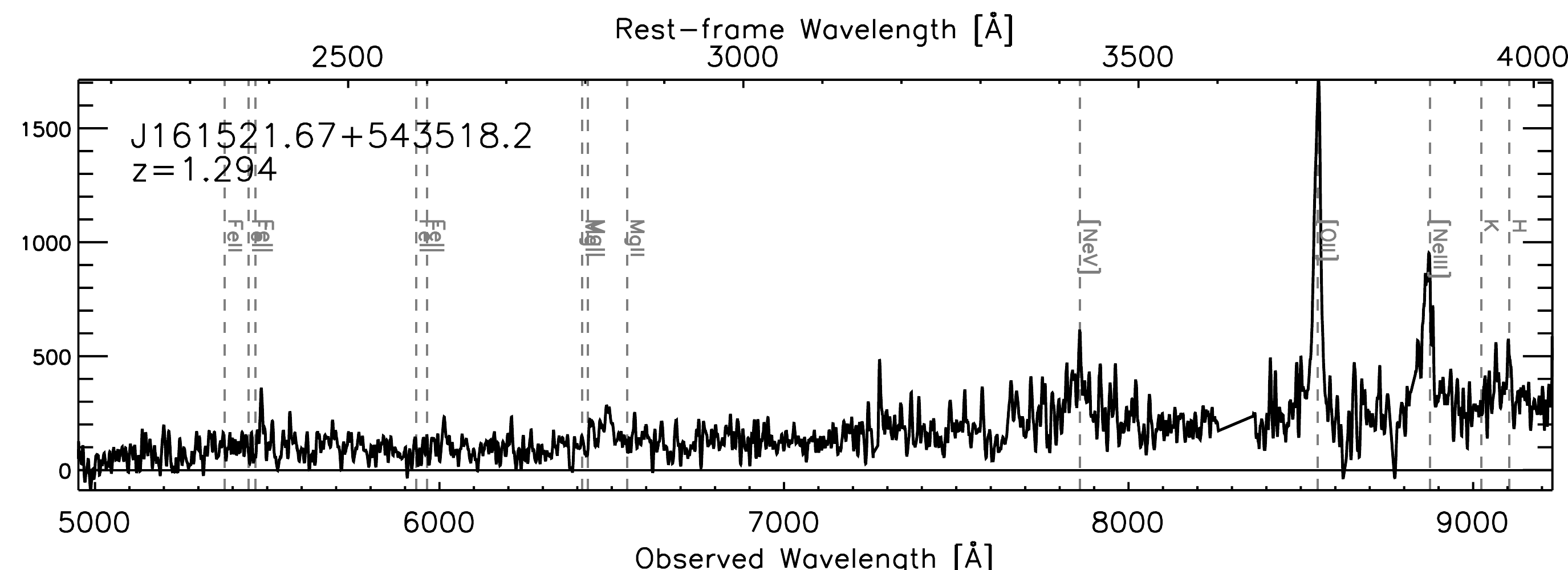}\\
  \includegraphics[width=0.99\columnwidth]{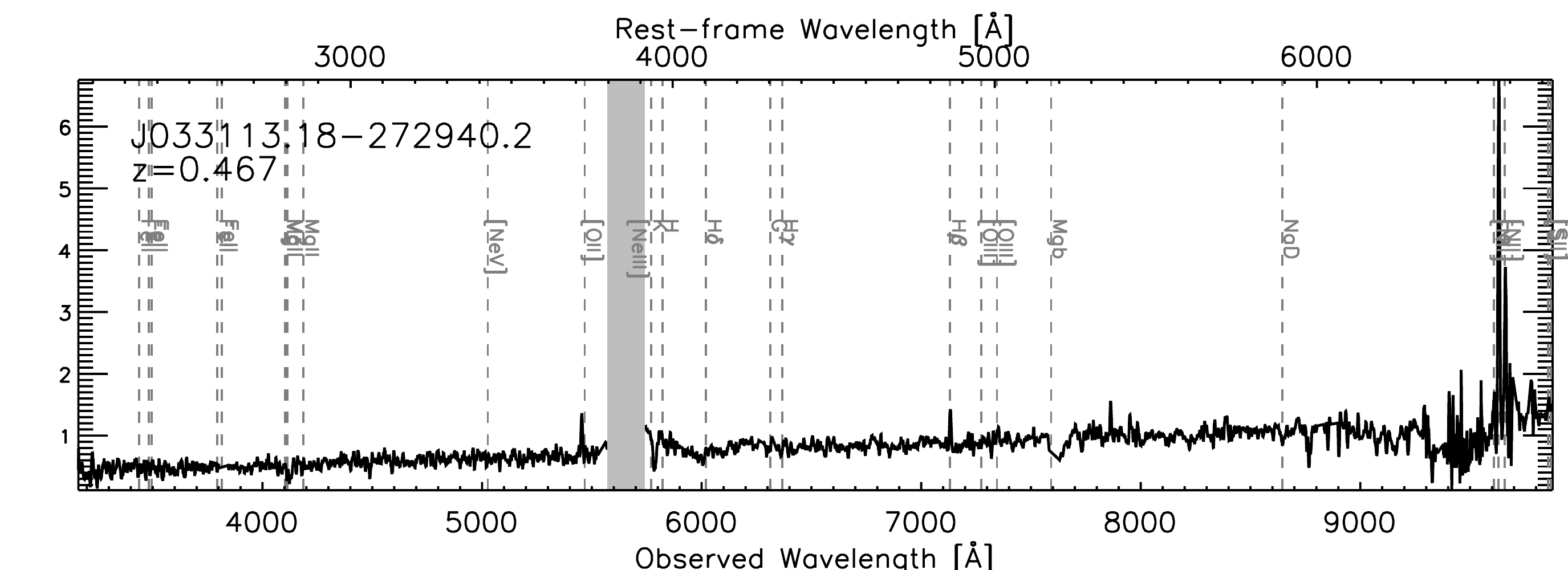}
  \includegraphics[width=0.99\columnwidth]{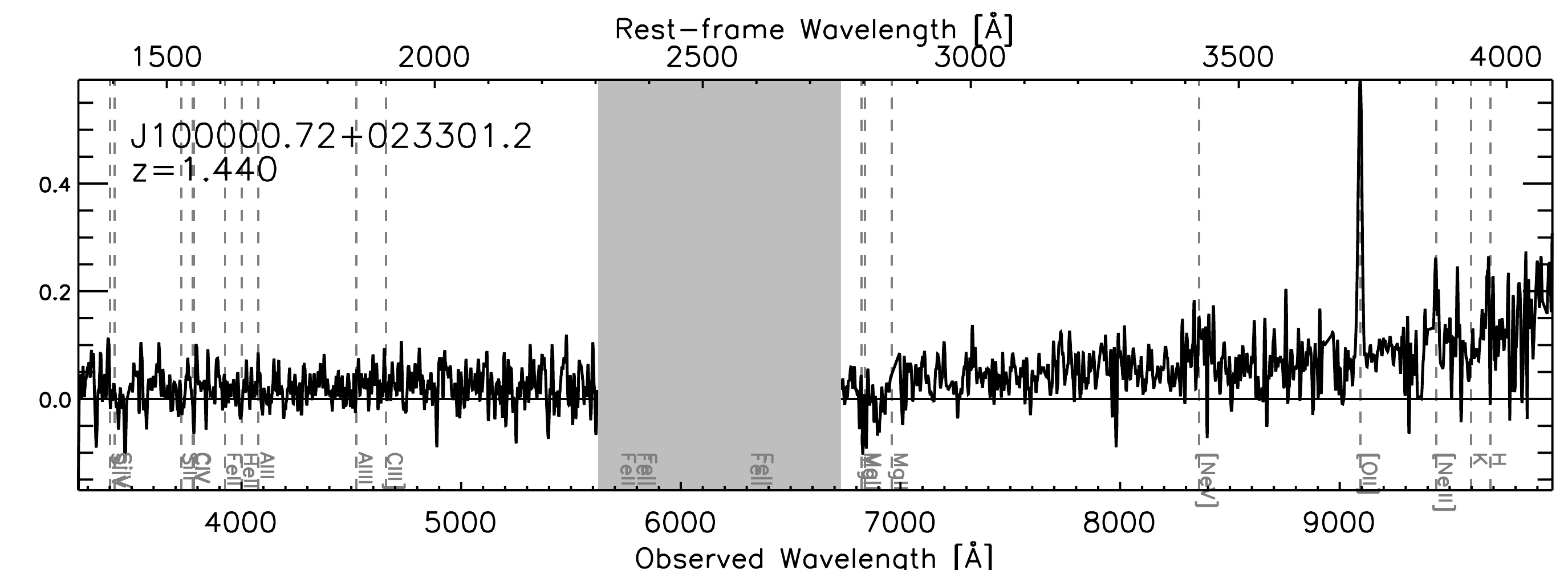}\\
  \includegraphics[width=0.99\columnwidth]{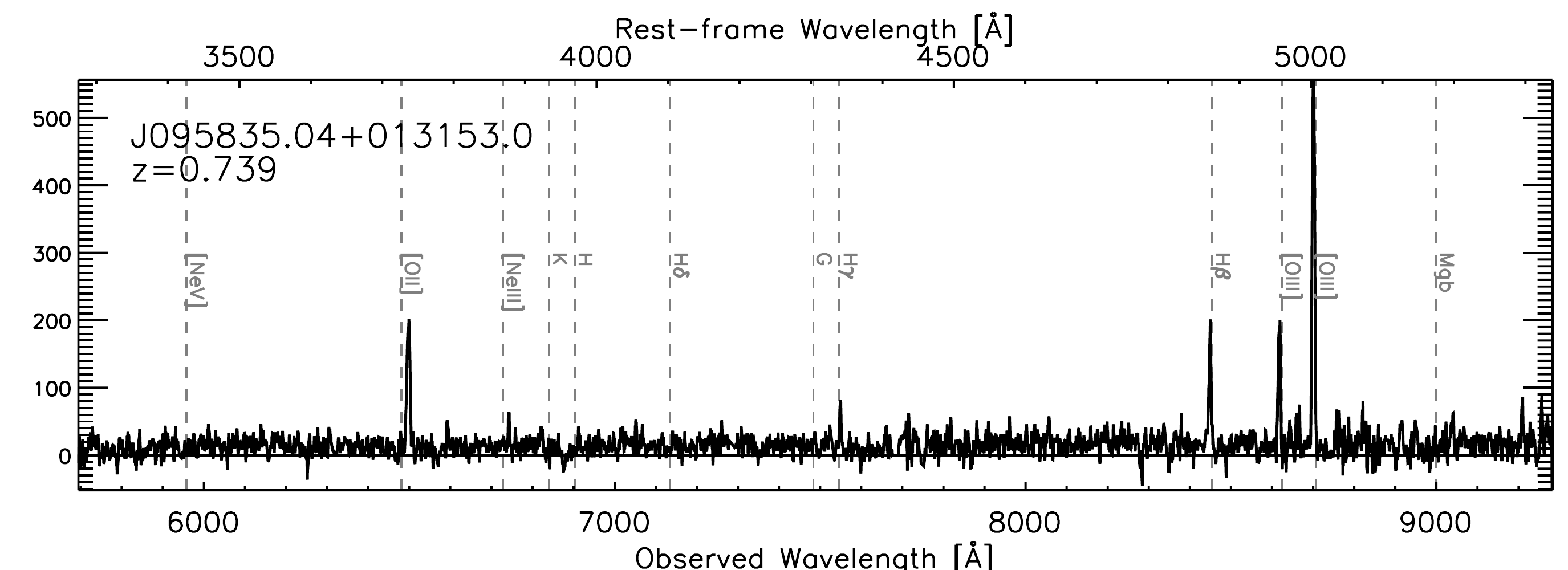}
  \includegraphics[width=0.99\columnwidth]{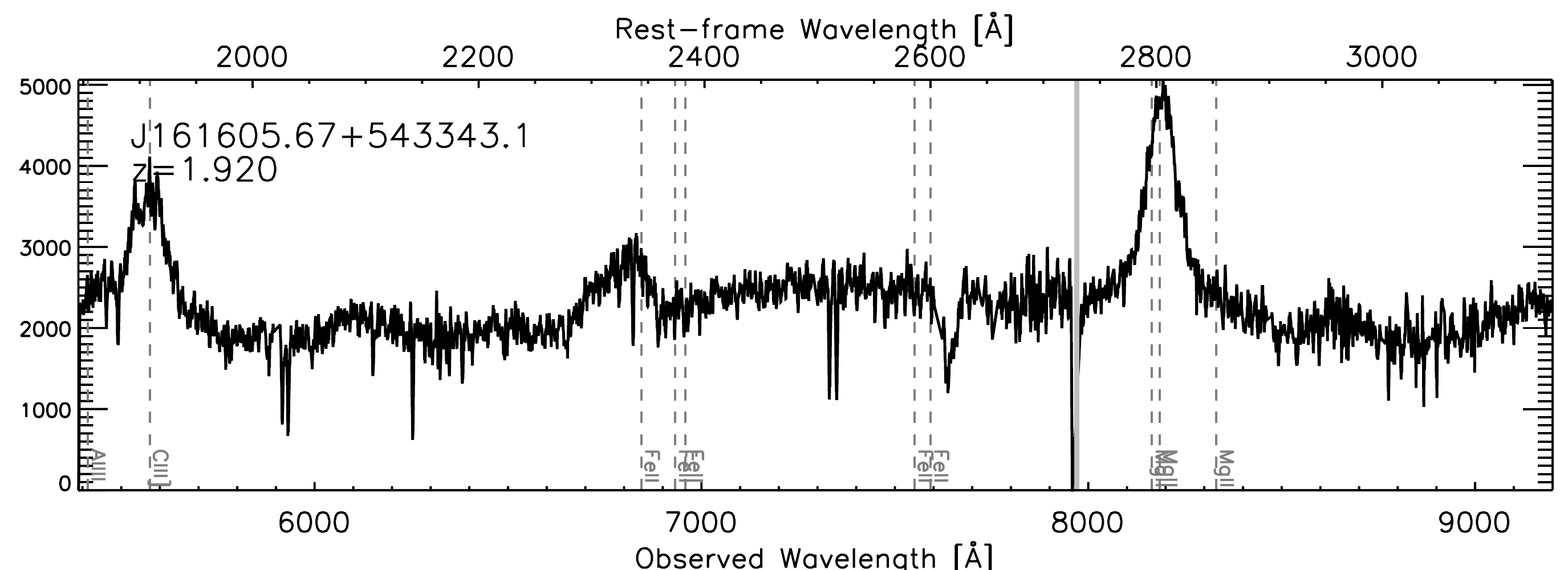}\\
  \includegraphics[width=0.99\columnwidth]{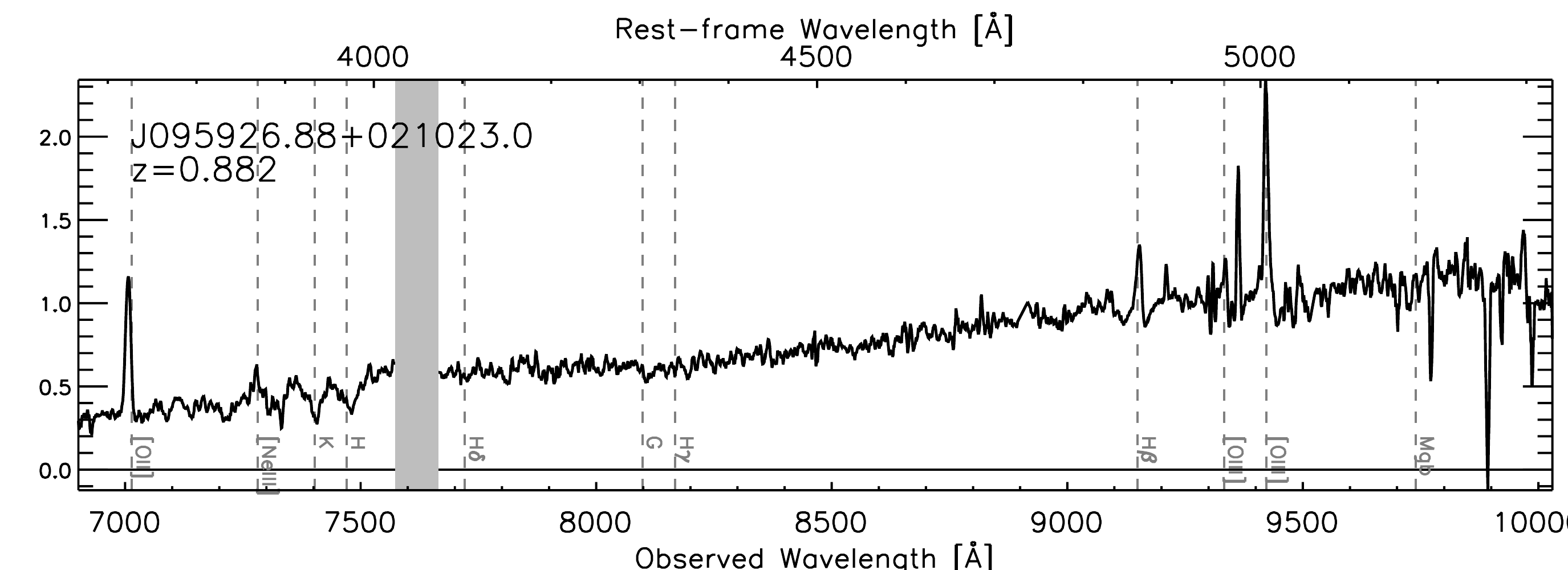}
  \includegraphics[width=0.99\columnwidth]{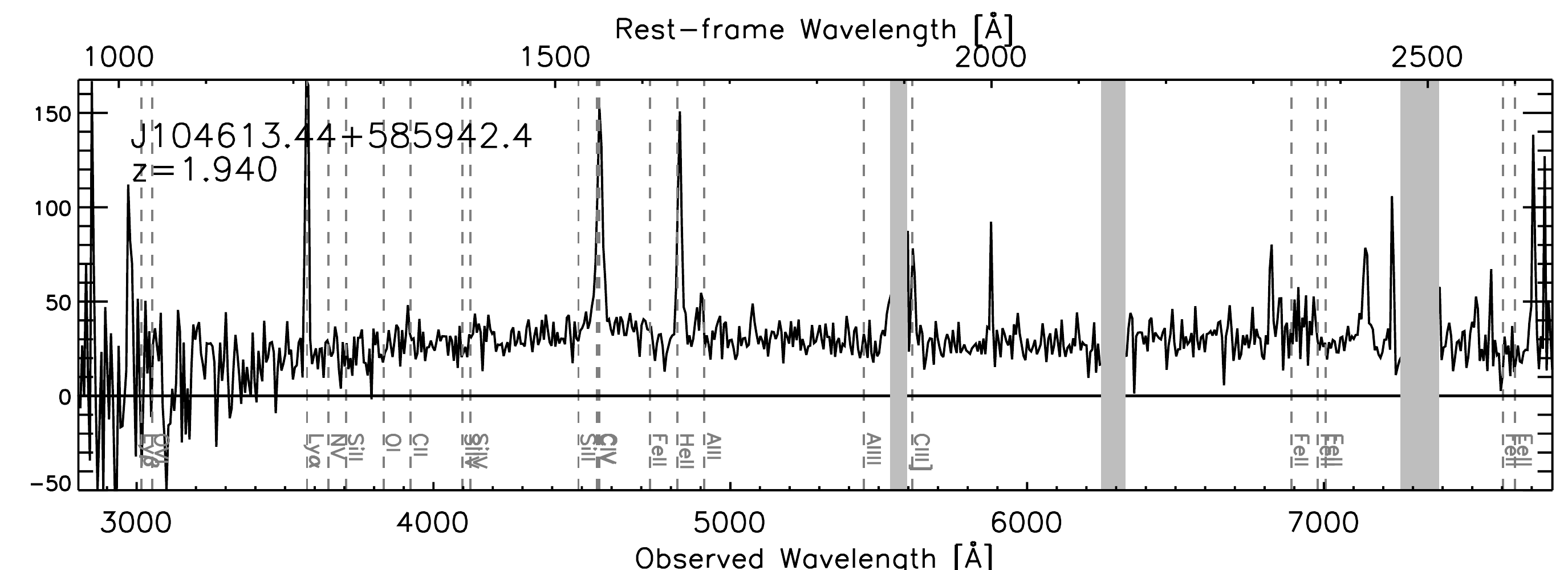}\\
  \caption{ Ten sample spectra of {\it Herschel}-{\sc Spire} selected
    galaxies ranging in redshift from $z=0.126$ to $z=1.940$. These
    sample spectra span a range of spectral types and include both
    LRIS and DEIMOS observations.  The flux scales are in arbitrary,
    uncalibrated flux units.  Spectra of the $z>2$ HSGs are given in
    full in C12. Individual spectra for sources given in
    Table~\ref{tab_giant} are available on request. }
  \label{fig:keckspec}
\end{figure*}

\section{Results}

\subsection{Summary of Spectra and Redshifts}

Table~\ref{tab_giant} lists the positions, redshifts, {\sc Spire} flux
densities, 24\,\um\ and radio flux densities, luminosities and dust
temperatures for the full sample of $z<2$ spectroscopically confirmed
HSGs. Sample spectra of a variety of spectroscopically-confirmed
targets of the $z<2$ confirmed HSG sample are shown in
Figure~\ref{fig:keckspec}.  Of 1594 targets with $>$3$\sigma$ {\sc
  Spire} detections, 767 have spectroscopically identified redshifts
from these data, 731 of which are at $z<2$. By field, there are 288 in
COSMOS, 163 in LHN, 119 in GOODS-N, 139 in EN1, 51 in CDFS, and 8 in
the UDS.  The total sky area probed by this redshift survey is
$\sim$0.93\,deg$^2$.

Besides the selection and spectroscopic biases mentioned in the
previous section, the spectroscopic yield for sources varied strongly
as a function of observing conditions and integration time, anywhere
from 15\%\ to 80\%.  However, unlike the selection and spectroscopic
biases, this variation in yield is much easier to correct.  While the full
0.93\,deg$^2$, 767 source sample is used in most analysis, only the
sources surveyed under photometric conditions to full survey depth,
i.e. 13 LRIS masks and 16 DEIMOS masks, covering an area of
0.43\,deg$^2$, are used to measure quantities which rely on accurate
measurements of source density (e.g. the integrated infrared
luminosity function and star formation rate density).

\subsection{Infrared SED Fitting}\label{sec:seds}

We note that best-fit SED template fits from \citet{chary01a},
\citet{dale01a}, \citet{dale02a}, \citet{siebenmorgen07a} or
\citet{draine07a} recover $L_{\rm IR}$ well, however we choose not to
fit our data to these complex models given the limitations of our data
and the variation in SED shapes.  As discussed in \citet{casey12a} at
length, the templates should be used with caution when the number of
free parameters in the templates exceeds the number of data points;
this is certainly the case for those sources which have three to five
photometric FIR data points.  The dust temperature range of these SED
templates is restricted to the 20$<T_{\rm dust}<$60\,K range and quantized.  For
these reasons, we opt to use the functional fits in lieu of FIR SED
templates.

Measuring infrared luminosities and dust temperature for {\sc Spire}
galaxies is performed by fitting a modified blackbody SED extrapolated
over the rest-frame FIR to existing photometric data.  In addition to
the {\sc Spire} flux densities, we use photometry from {\it
  Herschel}-PACS at 100\,\um, and 160\,\um\ in Lockman Hole North and
COSMOS, and {\it Spitzer}-MIPS at 24\,\um, and 70\,\um\ where
available.

We fit the data for each galaxy to a modified blackbody fit of general
opacity coadded to a mid-infrared power law, where the flux density,
$S_{\nu}$ at rest-frame frequency $\nu$ is represented by
\begin{eqnarray}
S_{\nu} & \propto(1-e^{-\tau(\nu)})B_{\nu}(T) \nonumber \\
 & = N_{\rm bb}\frac{(1-e^{-\tau(\nu)})\nu^{3}}{e^{h\nu/kT}-1} + N_{\rm pl}\nu^{-\alpha}\,e^{-(\nu/\nu_{\rm c})^{2}}
\end{eqnarray}
where $\tau(\nu)$=($\nu/\nu_{0}$)$^{\beta}$ is the optical depth
\citep{draine06a}, $\beta$ is the emissivity, the physical dust
temperature is $T$, and $\nu$=$c$/$\lambda_{0}$
($\lambda_{0}=$\,200\,\um) is the frequency where optical depth is
unity \citep[see][for full details on the SED fitting
  method]{casey12a}.  Note that the physical dust temperature, $T$,
used in this equation is not the same as the dust temperature used for
the remainder of the paper; we choose to measure a ``peak-SED'' dust
temperature, called $T_{\rm dust}$, which is the dust temperature
measured from the peak of the SED via Wien's law.  $T_{\rm dust}$ is
less dependent on model parameters than $T$, so is more easily
compared to dust temperatures measured via other means \citep[see][,
  Figure 2 for more detail]{casey12a}.  $N_{\rm bb}$ and $N_{\rm pl}$
are the coefficients of the modified black body and power law terms,
where $N_{\rm pl}$ is set such that the two functions are equal at the
frequency, $\nu_{\rm c}$, where the gradient of the black body is
equal to the slope of the power law, $\alpha$.  We allow $\alpha$ to
vary with the range $0.5<\alpha<5.5$ for sources with
24\,\um\ measurements (all sources except those in GOODS-N).  If we
have no mid-infrared measurements for the source, we fix $\alpha=2.0$,
which is the mean value for the rest of the {\sc Spire} galaxies in
our sample and a common value used throughout the literature
\citep[see][]{younger09a,magnelli10a,hilton12a}.  Since our
observations cover a relatively narrow wavelength range without
measurements on the Rayleigh-Jeans tail in the millimeter (from
$\approx$850\,\um--2mm), we cannot constrain $\beta$ in a meaningful
way (except for very low-redshift sources where the {\sc Spire} points
sit on the Rayleigh-Jeans tail).  For consistency in our fitting
technique, we decided to fix the emissivity to $\beta=1.5$ \citep[this
  is a commonly chosen value of $\beta$ in the literature,
  e.g.][]{chapman05a,pope05a,younger09a}.  There is very little change
in derived $L_{\rm IR}$ or $T_{\rm dust}$ by fixing $\beta$ within the
1--2 range, so we determine this procedure to be reasonable.

\begin{figure*}
  \centering
  \includegraphics[width=1.95\columnwidth]{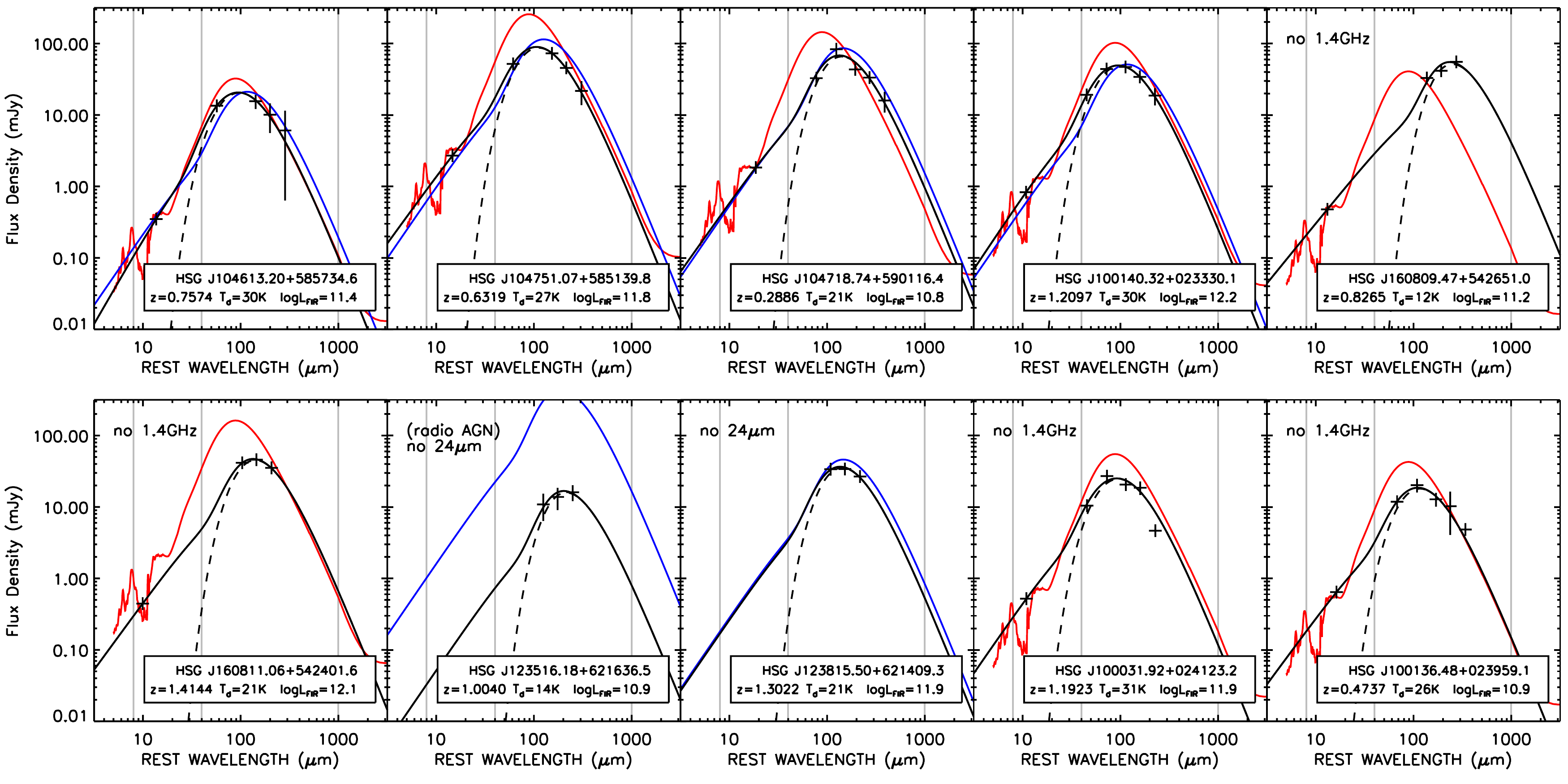}\\
  \caption{ Ten example FIR SED fits to {\sc Spire}-selected
    starbursts: three from LHN; three from COSMOS; two from EN1; and
    two from GOODS-N.  The FIR data points ({\sc Spire}, and PACS,
    where available) are shown with 1$\sigma$ uncertainties, and our
    best-fit modified blackbody SED is shown in black (see
    \S~\ref{sec:seds} for details on SED fitting).  The underlying
    cold-dust blackbody dominating the FIR portion $>$40\,\um\ of our
    SED is shown as a dashed line. For sources which have {\it
      Spitzer}-MIPS coverage, we overplot the mean composite SED of
    SMGs from \citet{pope08a} normalised to 24\,\um\ ($red$).  Sources
    with 1.4\,GHz VLA coverage have another SED overplotted ($blue$):
    a modified blackbody infered from assuming the radio/FIR
    correlation holds \citep[with $q_{\rm IR}$ as given
      in][]{ivison10b} with $T_{\rm dust}$ best fit to the data.  Gray
    vertical lines mark rest-frame 8\,\um, 40\,\um, and 1000\,\um.  We
    note that the cold-dust blackbody only dominates the SED in the
    40--1000\,\um\ wavelength range, although IR luminosity is
    computed across the range 8--1000\,\um, using the combination of
    IR modified blackbody and mid-infrared power law.  }
  \label{fig:sedpanels}
\end{figure*}

Figure~\ref{fig:sedpanels} illustrates ten example SED fits using our
method, contrasting them with both radio-implied FIR SEDs (from the
radio/FIR correlation) and extrapolations from 24\,\um\ assuming a
nominal template for high-$z$ ULIRGs, like the \citet{pope08a}
composite SED.

\begin{figure}
  \centering \includegraphics[width=0.99\columnwidth]{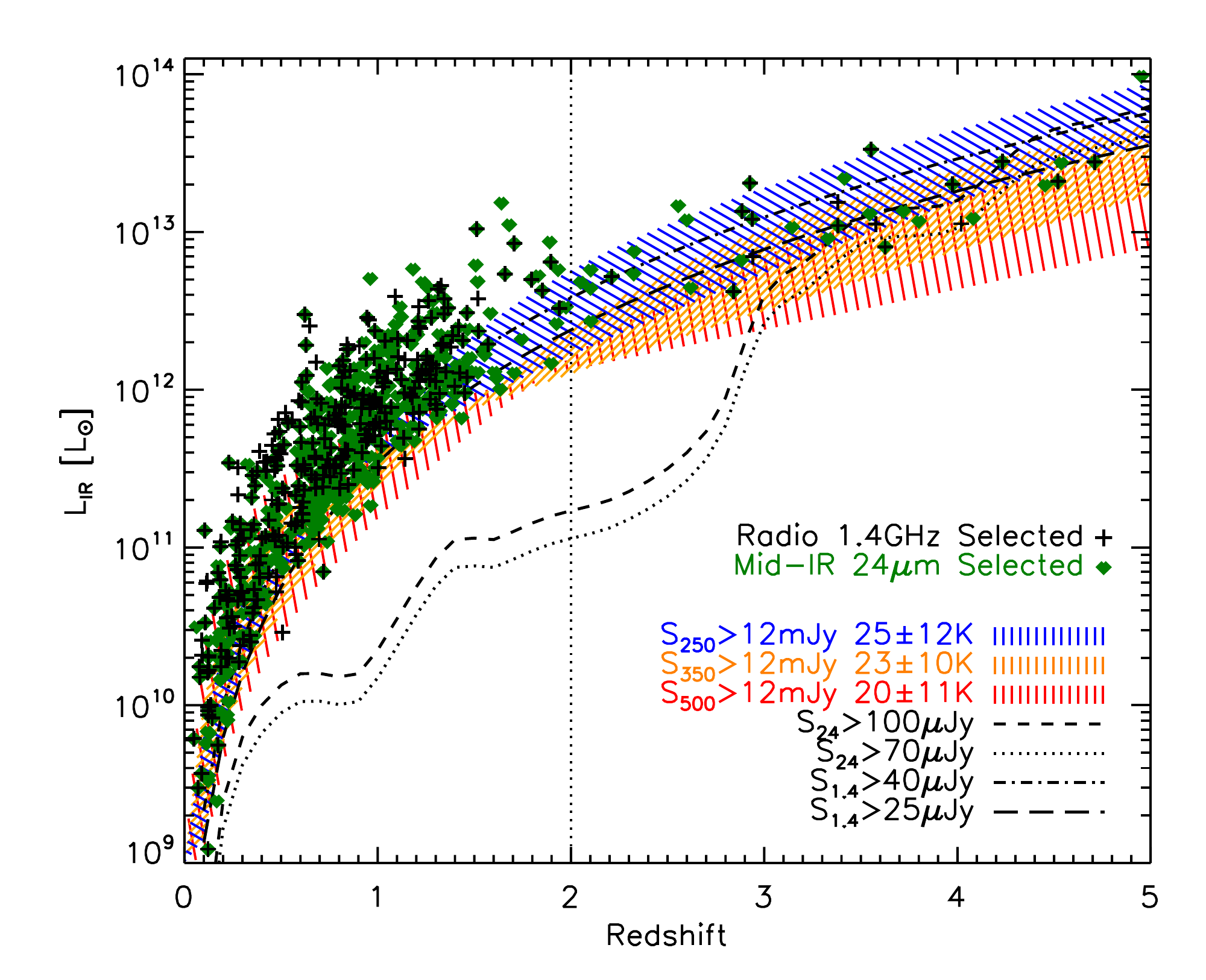}
  \caption{The 8--1000\um\ integrated infrared luminosity against
    redshift for {\it Herschel}-{\sc Spire} spectroscopically identified
    sources.  The lower luminosity limits at the selection wavelengths
    are marked, both for the prior source catalogs at 24\,\um\ ($S_{\rm
      24}>$100\uJy, dashed line) and 1.4\,GHz ($S_{1.4}>$25 or 40\,\uJy,
    long dashed and dot-dashed lines), and {\sc Spire} wavelengths
    (colored hashed regions).  The detection limits at the {\sc Spire}
    bands change with SED shape; we measure the dust temperatures of
    galaxies (via Wien's law) selected in each filter to determine the
    range of peak-SED dust temperatures to integrate between to
    calculate the IR luminosity limits at 250\,\um\ (blue),
    350\,\um\ (yellow), and 500\,\um\ (red).  Sources selected as radio
    sources are marked as crosses and 24\,\um\ sources are marked as
    green diamonds.  }
  \label{fig:lirz}
\end{figure}

Table~\ref{tab_giant} lists the 8--1000\,\um\ IR luminosities and dust
temperatures for the whole sample, and Figure~\ref{fig:lirz} shows how
the sources fall in luminosity versus redshift with respect to the
various selection criteria.  Note that the {\sc Spire} selection
criteria are not fixed in infrared luminosity since luminosity depends
both on dust temperature and flux density.  Similarly, the radio and
24\um\ detection limits are not absolute, since survey depths at both
wavelengths vary between fields.

Figure~\ref{fig:lirz} shows all spectroscopic HSGs in $L_{\rm
  IR}$--$z$ space.  Despite the fact that our sources are identified
initially by 24\,\um\ or radio source observations, the detection
threshold in $L_{\rm IR}$--$z$ space corresponds well with the {\sc
  Spire} $>$3$\sigma$ luminosity limit alone (at $z<3$, where {\it
  Spitzer\/} is effectively more sensitive than {\sc Spire}).

\subsection{Aggregate Infrared SEDs}\label{sec:netsed}

While individual galaxies in our sample have at most five flux
density measurements in the FIR, some with very low signal-to-noise,
we can combine the measurements from many sources to infer the
aggregate infrared properties of our sample in greater detail.

\begin{figure}
  \centering
  \includegraphics[width=0.99\columnwidth]{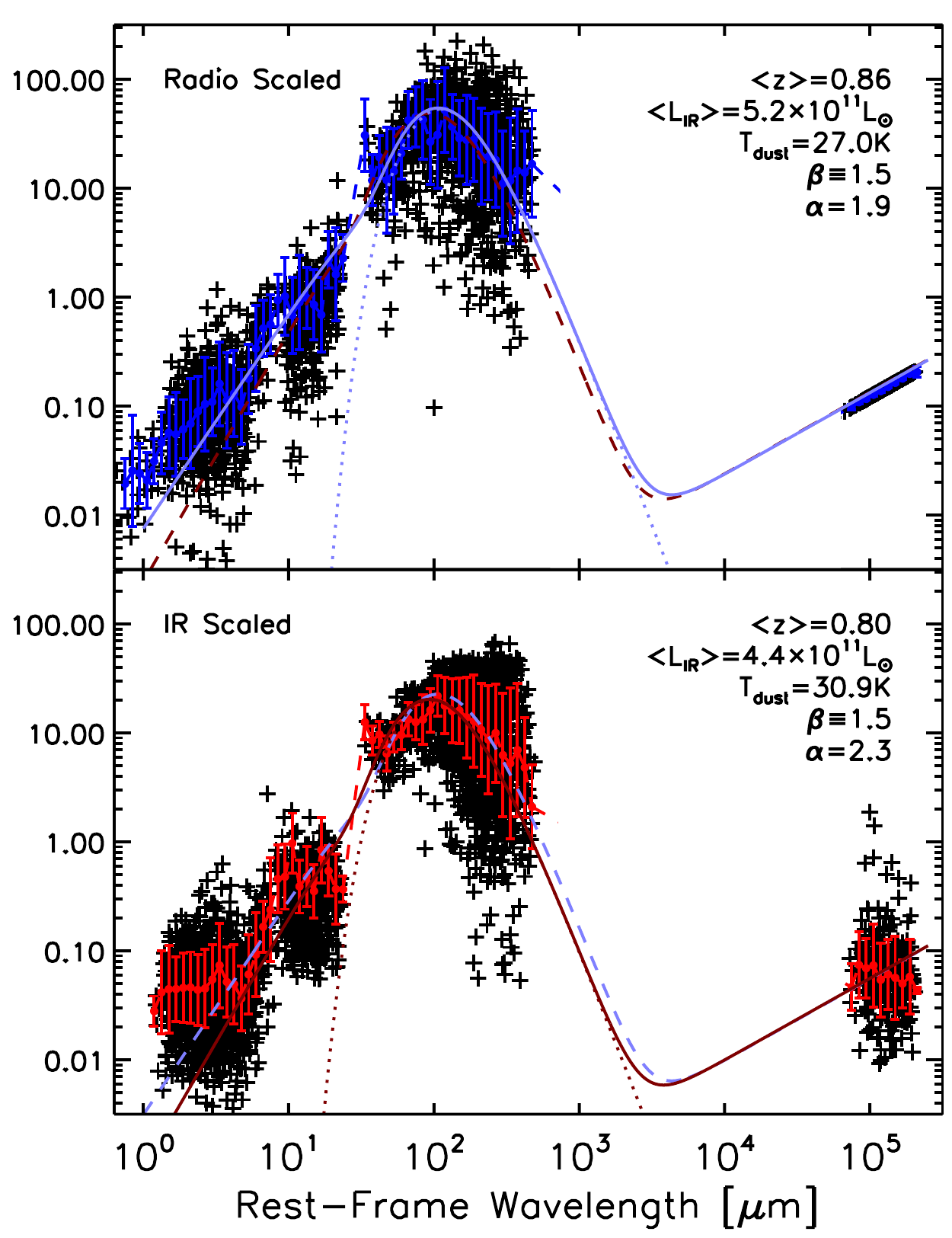}
  \caption{
    Composite infrared to radio spectral energy distributions of HSGs
    scaled to their radio flux densities (top; 331 radio-detected HSGs)
    and integrated infrared luminosities (bottom; all 767 HSGs).  The best
    fit SED shapes (light blue, top, dark red, bottom) are remarkably
    similar between the two samples, with best-fit dust temperatures of
    27\,K and 30\,K and mid-infrared slopes of 1.9 and 2.1, respectively.
  }
  \label{fig:netsedradio}
\end{figure}

Figure~\ref{fig:netsedradio} shows two composite near-infrared through
radio SEDs for the entire $0<z<2$ HSG sample.  The first comprises the
331 sources which are radio-detected at $z<2$: their flux densities
are scaled so that the $K$-corrected radio flux density equals the
mean of the sample, 133\,\uJy, at the mean redshift of the sample,
$\langle z\rangle=0.89$.  The second composite SED is constructed of
all 731 $z<2$ sources, with flux densities re-normalized to the mean
infrared luminosity, 4.1$\times$10$^{11}$\lsun, and redshift, $\langle
z\rangle=0.80$, of the sample.  The two SEDs are remarkably similar,
with dust temperatures of 27\,K and 30\,K, and mid-infrared slopes of
$\alpha=1.9$ and 2.1, respectively.  The most noticable difference
lies in the near-infrared, where radio normalization seems to
artificially wash out the stellar mass bump.  The other observation to
make is the spread in the radio flux densities in the IR scaled SED;
although this is large ($\sim$2\,dex, most falling within $\sim$1\,dex
scatter), we note that this could very well be due to individual
source variation in the synchrotron slope or bright radio AGN.  Since
the mean value of the radio flux densities falls within $\pm$0.2\,dex
of the expected radio flux density (from the FIR radio correlation),
we deduce agreement.  The overall agreement between the two radio
scaled and IR scaled SEDs re-enforces that: (a) the FIR radio
correlation seems to hold in this sample within uncertainty; and that
(b) radio selection is not clearly biased in integrated infrared
properties in comparison to 24\,\um\ selected HSGs.

\begin{figure}
  \centering
  \includegraphics[width=0.99\columnwidth]{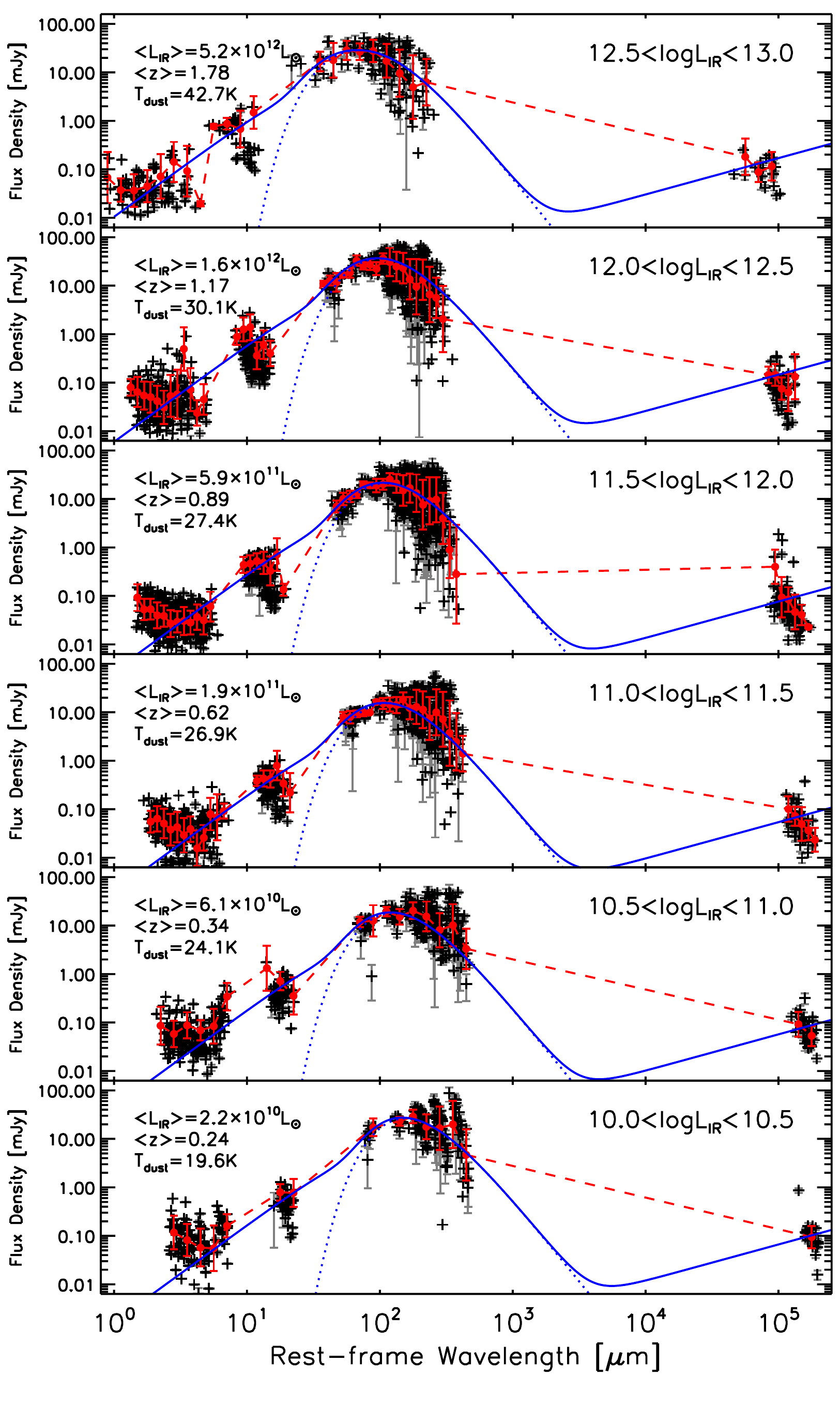}
  \caption{
    Composite infrared to radio spectral energy distributions of HSGs in
    six luminosity bins, from $L_{\rm IR}=$10$^{10}$\,\lsun\ to
    10$^{13}$\,\lsun.  Flux densities are renormalized first to the mean
    luminosity and redshift of each bin.  Infrared dust SEDs are fit to
    binned data (red points) from rest-frame 8\,\um\ to 1000\,\um, with
    fixed $\alpha=2.0$ and $\beta=1.5$.  The radio portion of the SEDs are
    generated by assuming the FIR/radio correlation holds; overall, these
    radio SEDs agree with radio data.  From the SED fits, we measure a
    steady increase in dust temperature with luminosity, from $T=$19\,K at
    2$\times$10$^{10}$\,\lsun\ at $\langle z\rangle=0.23$ to $T=46$\,K at
    5.2$\times$10$^{12}$\,\lsun\ at $\langle z\rangle=1.81$.  We also see
    evidence for the evolution of near- to mid-infrared properties, most
    prominent in the highest luminosity bin, with detection of the
    $\approx$10\,\um\ Si absorption feature (from a 24\um\ flux density
    deficit of sources around $z\sim1.4$).
  }
  \label{fig:netsed}
\end{figure}

To test for luminosity and redshift evolution of infrared SED type,
Figure~\ref{fig:netsed} illustrates the near-infrared through radio
SEDs of all $0<z<2$ sources split into several infrared luminosity
bins.  We measure a steady increase in dust temperature with
luminosity (and also consequently redshift) from $\approx$\,20\,K at
10$^{10}$\,\lsun, to $\approx$\,30\,K at 10$^{12}$\,\lsun, and a jump to
45\,K at 10$^{12.5-13.0}$\,\lsun.  Understanding whether or not this
increase in dust temperature is due to redshift evolution, luminosity,
or selection effects is difficult, due to the redshift
dependence of luminosity; this issue is discussed more in context of
the $L_{\rm IR}-T_{\rm dust}$ plane in the next section.

An interesting conclusion can be drawn from the composite SEDs about
the evolution of near- and mid-infrared properties.  At lower
luminosities (up to 10$^{12}$\,\lsun), the near-infrared data
$<$10\,\um\ increases towards shorter wavelengths, indicative of
emission from old stars (i.e.  the stellar mass bump) with a local
minimum at $\sim$5--6\,\um.  In the highest luminosity bin, the
prominence of the stellar bump disappears, as the emission seems to be
dominated by a power law.
Furthermore, the rest-frame mid-IR is unremarkable at low
luminosities, largely consistent with the expected continuum flux
density from our SED fit, while the highest luminosity bin shows a
strong absorption feature at $\approx$10\,\um, thought to be
9.7\um\ Si absorption, seen in some of the most luminous local
infrared galaxies, including Arp\,220 \citep{charmandaris97a}.  This
feature is detected in this composite SED only due to sources at the
corresponding redshift ($z\sim1.4$) having a 24\um-flux density
deficit relative to sources at redshifts above and below $z\sim1.4$.

\subsection{The Temperature-Luminosity Plane}\label{sec:lfirtd}

\begin{figure*}
  \includegraphics[width=0.99\columnwidth]{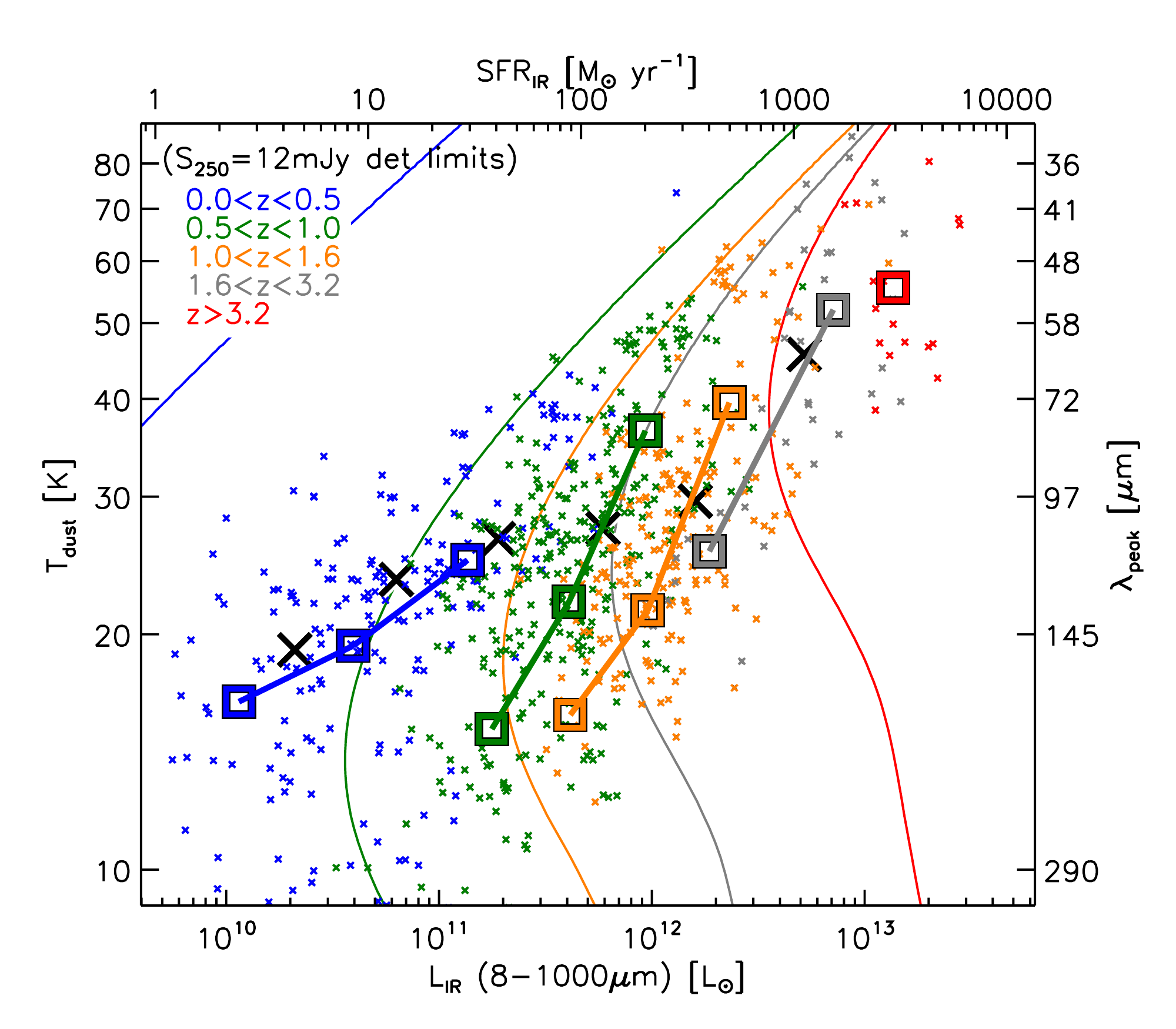}
  \includegraphics[width=0.99\columnwidth]{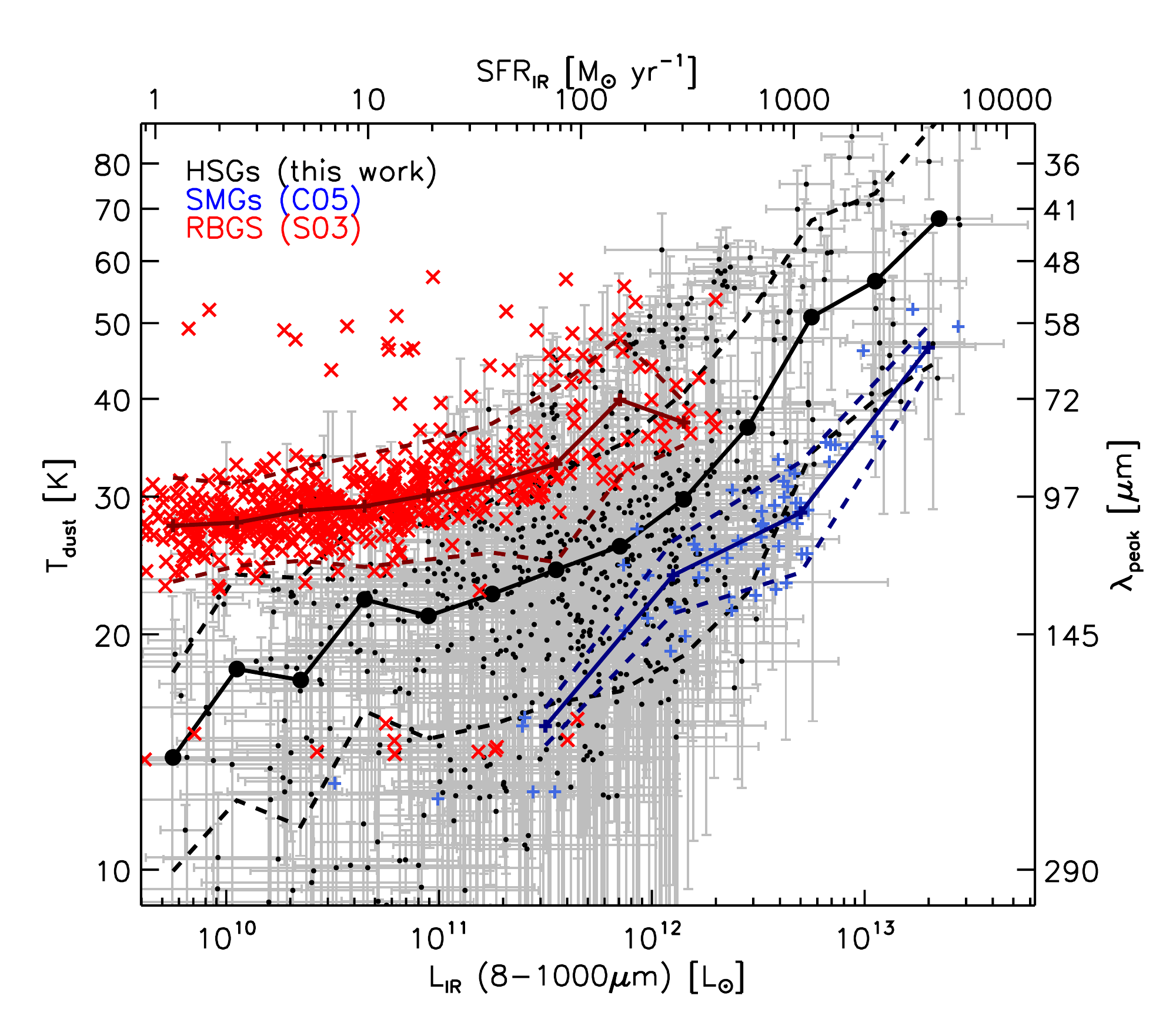}
  \caption{ $L_{\rm IR}$ against $T_{\rm dust}$ as an indicator of the
    variation of SED type in our sample.  Dust temperature is
    estimated via Wien's Law (inversely proportional to peak of the
    SED).  Colors on the left denote redshift bins, and the mean
    $L_{\rm IR}-T_{\rm dust}$ relation for each redshift slice is
    shown as large squares.  This relation changes with increasing
    redshift, however this can be attributable to the selection bias
    of {\sc Spire} selection: the 250\,\um\ detection limits at fixed
    redshifts are shown as solid lines, whose color corresponds to the
    lower redshift limit of the bin.  At a fixed temperature, only
    sources to the right of the line (i.e. at higher luminosities) are
    detectable.  The mean luminosities and temperatures for the
    composite SEDs in Figure~\ref{fig:netsed} are large crosses.
    At right, we illustrate the $L_{\rm IR}-T_{\rm dust}$ relation for
    {\sc Spire} galaxies ($black$) in comparison to local {\it IRAS\/}
    galaxies, from the RBGS and GOALS samples
    \citep[$red$;][]{sanders03a,armus09a}, and SMGs
    \citep[$blue$;][]{chapman05a,pope06a}.  We refit SEDs for all
    samples consistently so that both dust temperatures and
    luminosities are directly comparable.  This demonstrates that our
    {\sc Spire} galaxy sample is statistically colder than local {\it
      IRAS\/} galaxies of similar luminosities but warmer than SMGs of
    similar luminosities.  Although we observe a different slope to
    the $L_{\rm IR}-T_{\rm dust}$ relation than is seen locally, our
    sample is not large enough to measure evolution, since luminosity
    is largely a function of redshift \citep[e.g.][]{seymour10a}.  }
  \label{fig:lfirtd}
\end{figure*}

Figure \ref{fig:lfirtd} shows our sample in $L_{\rm IR}-T_{\rm dust}$
space.  Overall, dust temperature increases with luminosity, which is
partially attributable to our selection and partially to a real
physical effect.  The selection effect (that more luminous sources are
hotter) stems from the nature of selection at wavelengths on the
Rayleigh-Jeans tail, where hot-dust SEDs are selected against (at
comparable IR luminosities).  This dust-temperature selection effect
was much more pronounced in the SMG population selected at
850\,\um\ \citep[see][]{blain04a,chapman04a,casey09a,chapin09a}, than
it is in {\sc Spire} samples, which probe the dust SED closer to its
peak and even beyond, at $z$\simgt3.  The solid lines in the left
panel of Figure~\ref{fig:lfirtd} represent {\sc Spire} detection
limits, or contours of fixed 250\,\um\ flux density and redshift.
When split into different redshift bins, the $L_{\rm IR}-T_{\rm dust}$
relation appears to evolve, although this is primarily attributable to
luminosity-limit selection effects.  In order to probe intrinsic
redshift evolution of sources in $L_{\rm IR}-T_{\rm dust}$, a larger
dynamic range of luminosities is needed over narrow redshift ranges
(e.g. expanding to $>10^{12-13}$\lsun sources at $0.0<z<0.5$ would
help greatly).  In contrast to 850\,\um\ selection,
250\,\um\ selection is relatively unbiased with dust temperature
\citep{casey11a}.


There has been work which argues that the selection biases in $L_{\rm
  IR}-T_{\rm dust}$ space are minor compared to physical effects
\citep{sajina07a}.  The argument is that there is more intense dust
heating in more extreme star forming environments.  Our data support
this, since the dust temperature bias effects are less prominant using
{\sc Spire} selection.  Although selection effects play a role in our
perceived relation, we fit dust temperature as a function of $L_{\rm
  IR}$ and find a significant correlation for {\sc Spire} selected
samples of $\langle T_{\rm dust}\rangle=0.47^{+0.09}_{-0.06}\,(L_{\rm
  FIR})^{0.144\pm0.006}~[K\,L_{\odot}^{-1}]$.  The slope of this
correlation, 0.14, is steeper than is seen in the local {\it IRAS\/}
sample.  We highlight that the difference between the two samples can
be attributed to two factors.  First, the dust temperature bias of the
{\it IRAS\/} sample, selected at 60\,\um\ with $S_{\rm 60}>5.24$\,Jy,
which selects against low dust temperature sources (see
Figure~\ref{fig:lfirtd}, right). Second, the local sample is a
single-redshift snapshot at $z=0$, while our sample is spread over a
wide epoch range, $0<z<2$ (with 36 sources at $2<z<5$).  Fitting the
whole sample together washes out any evolution in the $L_{\rm
  IR}-T_{\rm dust}$ relation due to the strong redshift dependence of
luminosity \citep[e.g.][]{seymour10a}.  Eventually, much larger
samples will make it possible to probe the $L_{\rm IR}-T_{\rm dust}$
relation as a function of redshift.

\subsection{Comparison to SMGs and `missing' warm-dust ULIRGs}\label{sec:tdust}

Submillimetre Galaxies (SMGs) selected at 0.85--1.3\,mm by SCUBA,
LABOCA, AzTEC, or MAMBO have dominated the studies of high-$z$
starbursts for the past decade.  While {\it Herschel}-selected
galaxies overlap with the SMG population \citep{roseboom12a}, the
correspondence between source detectability at 250--500\,\um\ and
$\sim$850\,\um\ is not obvious.  Although {\it Herschel\/} detects
galaxies at their SED peak at $z\sim1-2$ (and is more sensitive to
lower luminosities at $z<2$ than SCUBA), the submm has the advantage
of preferentially detecting higher-redshift galaxies and detecting
them to lower luminosity limits than {\it Herschel}.

\begin{figure*}
  \centering
  \includegraphics[width=0.98\columnwidth]{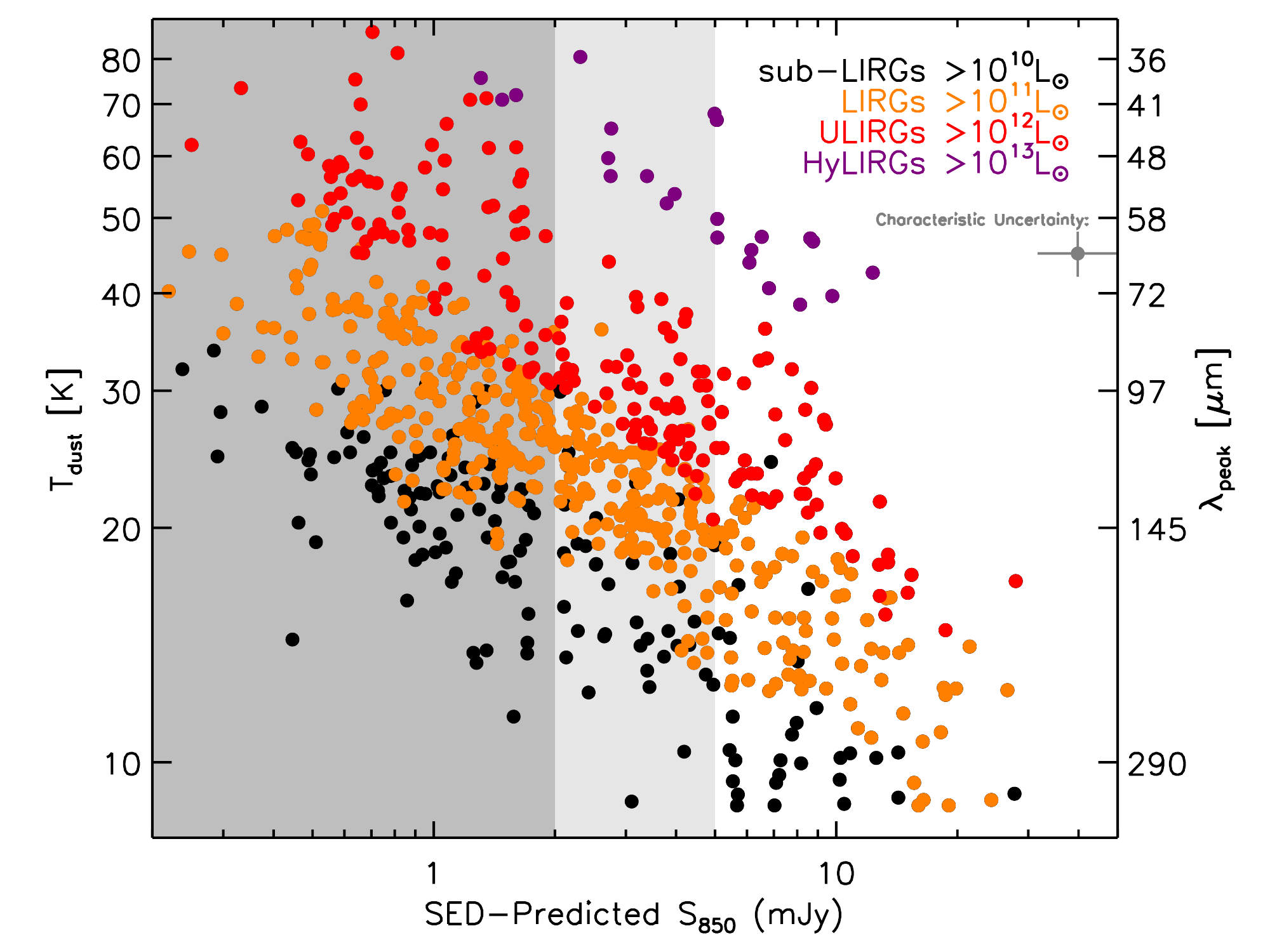}
  \includegraphics[width=0.84\columnwidth]{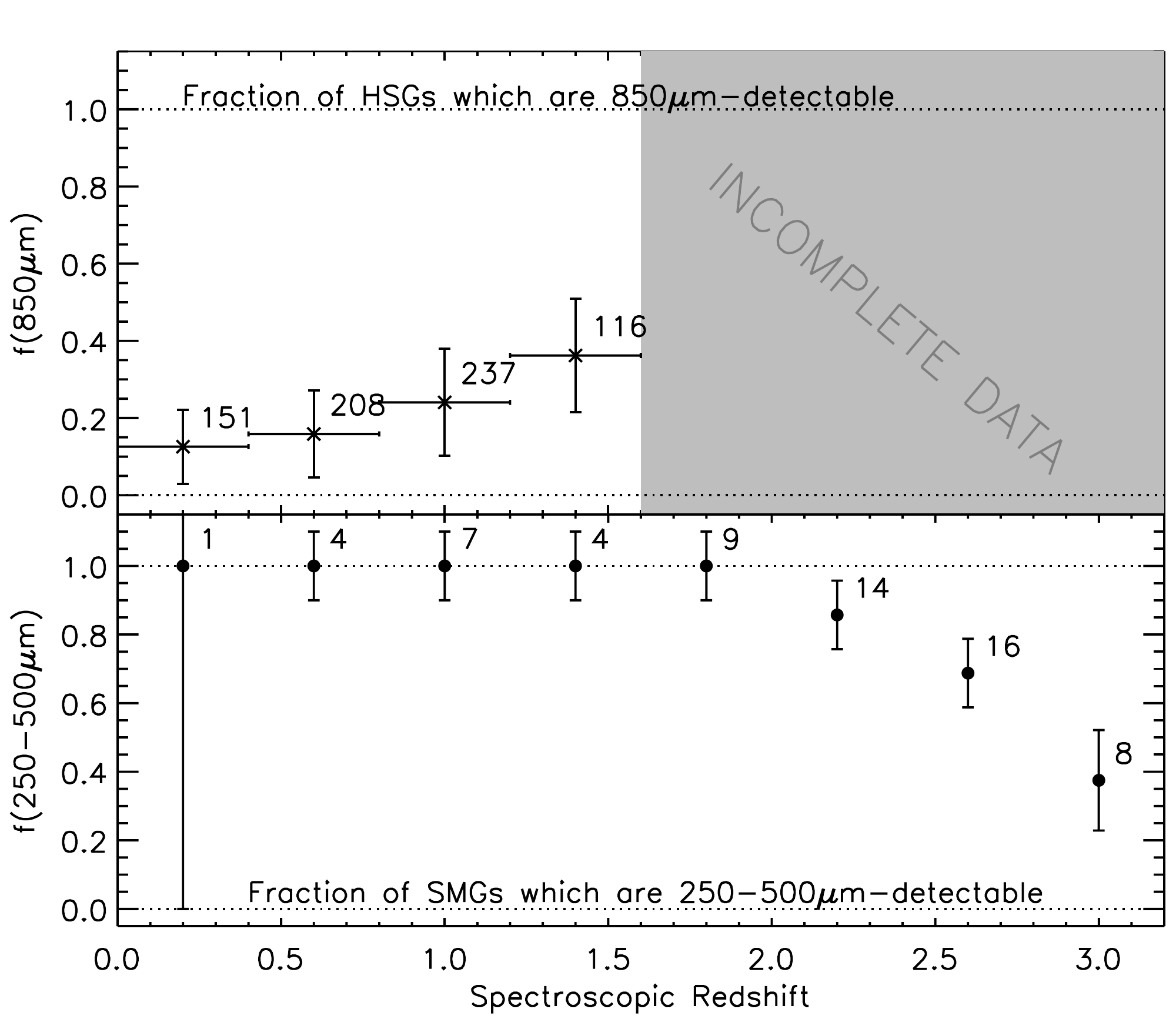}
  \caption{Predicted 850\,\um\ flux densities for our {\sc Spire} sample
    against the fitted SED dust temperatures.  The characteristic
    uncertainty in either measurement is shown in the upper right.  Note
    that the classic SMG selection criteria, $S_{850}>5$\,mJy would
    exclude 79$^{+4}_{-8}$\%\ of {\sc Spire} sources from detection in
    traditional 850\,\um-based SMG surveys: 83$^{+4}_{-7}$\%\ of HSGs at
    $z<1$ and 72$^{+5}_{-11}$\%\ of HSGs at $1<z<2$.  This is
    significantly higher than previous estimates to the `missing'
    fraction of warm-dust ULIRGs at $z>1$ missed by SMG surveys. At
    right, we show the fraction of {\sc Spire} galaxies which are
    850\um-detectable and 850\um-selected SMGs which are {\sc
      Spire}-detectable, as a function of redshift.  Here {\sc Spire}
    detectability is defined as $S_{\rm 250}$, $S_{\rm 350}$ or $S_{\rm
      500}>12$\,mJy, and qualification as an SMG is $S_{\rm
      850}>5$\,mJy. This illustrates that 60-80\%\ of {\sc Spire}
    galaxies are undetectable at 850\,\um\ out to $z\sim1.6$, and that
    SMGs are only expected to drop out of the {\sc Spire} bands at $z>2$
    for 20-60\%\ of sources.  The numbers next to each point represents
    the number of galaxies used in each bin.  There are too few {\sc
      Spire} galaxies to reliably determine this statistic at $z>1.6$
    (gray shaded region).}
  \label{fig:s850td}
\end{figure*}

Unfortunately, very few of our {\sc Spire} galaxies were directly
observed with SCUBA, LABOCA, AzTEC or MAMBO on account of small survey
areas for those instruments, so it is difficult to assess population
overlap directly.  However, using the SED fits from \S~\ref{sec:seds},
we can estimate 850\,\um\ flux densities for every HSG.  These
SED-predicted 850\,\um\ flux densities span a wide range, from
$\sim$0.2--30\,mJy. Figure~\ref{fig:s850td} shows the relation between
fitted dust temperature and extrapolated 850\,\um\ flux density for
{\sc Spire} galaxies; there is a clear relation between dust
temperature and 850\,\um-detectability.

We infer that 79$^{+4}_{-8}$\%\ of all {\sc Spire} galaxies are
\emph{undetectable} at 850\,\um\ at a flux cut-off of $S_{\rm
  850}<5$\,mJy.  Even considering a submillimeter detection
threshold as low as 2\,mJy \citep[the lowest 3$\sigma$ detection limit
  for SCUBA, which had a confusion noise of $\sim$0.7\,mJy at
  850\,\um][]{blain99a}, 47$^{+27}_{-14}$\%\ of {\sc Spire} galaxies
would be undetectable at 850\,\um.  While this is a large fraction, it
could be that most of these are at low redshift, and therefore low
luminosities and low SFRs.  The `missing' fraction as a function of
redshift is plotted on the right panel of Figure~\ref{fig:s850td}.
Indeed the fraction of HSGs not detectable at 850\um\ is high at
$z<1$, 83$^{+4}_{-7}$\%, yet even at higher redshifts, $1<z<2$, the
fraction is substantial: 72$^{+5}_{-11}$\%\ of HSGs have $S_{\rm
  850}<5$\,mJy.  

The fact that 850\,\um\ selection misses a large fraction of
infrared-bright starbursts is not new, but has been difficult to
measure directly or estimate in the past.  This dust temperature
selection effect was studied in detail in the pre-{\it Herschel\/} era
\citep{blain04a,chapman04a,casey09b,casey11a,casey11b} where
conservative estimates of the `missing' warm-dust ULIRG population
(often refered to as `OFRGs' or Submillimeter-Faint Radio Galaxies,
`SFRGs') were on the order of 10--20\%, and later works went on to
estimate the missing fraction at \simgt\,30\%\ using photometric data
and limited spectroscopic data from {\it Herschel\/}
\citep{chapman10a,magdis10a}.

While we note that roughly half of {\sc Spire} galaxies are below the
detection threshold at 850\,\um, we can also investigate the converse
statistic.  How many SMGs are not detectable by {\sc Spire}?  We use
the original sample of SMGs described in \citet{chapman05a}, along
with assumed IR luminosities estimated via the FIR/radio correlation
to fit SEDs and extrapolate to {\sc Spire} fluxes.  We then use these
hypothetical fluxes to calculate a {\sc Spire} detectability, in other
words, we determine what fraction of the SMGs would pass our
$>$3$\sigma$ selection criteria.  The percentage of SMGs which are
250--500\,\um\ detectable is shown at bottom on the right panel of
Figure~\ref{fig:s850td}.  At $z<2$, all SMGs are detected with {\sc
  Spire}, but above $z>2$, anywhere from 20\%\ to 60\%\ of SMGs are {\sc
  Spire} dropouts.

We also investigate this fraction using real {\sc Spire} coverage of
SMG fields together with LHN data presented in \citet{chapman10a} and
{\sc Spire} coverage around SMGs from \citet{chapman05a}.  The
statistic is consistent, roughly 1/3 of SMGs are not detected in {\sc
  Spire}, and nearly all at $z>2.5$.  Since the sample size of SMGs
with overlapping {\sc Spire} data is about half the size of the full
SMG sample, we use the extrapolated flux densities in
Figure~\ref{fig:s850td}.

\subsection{The FIR/Radio Correlation}

\begin{figure}
  \centering
  \includegraphics[width=0.99\columnwidth]{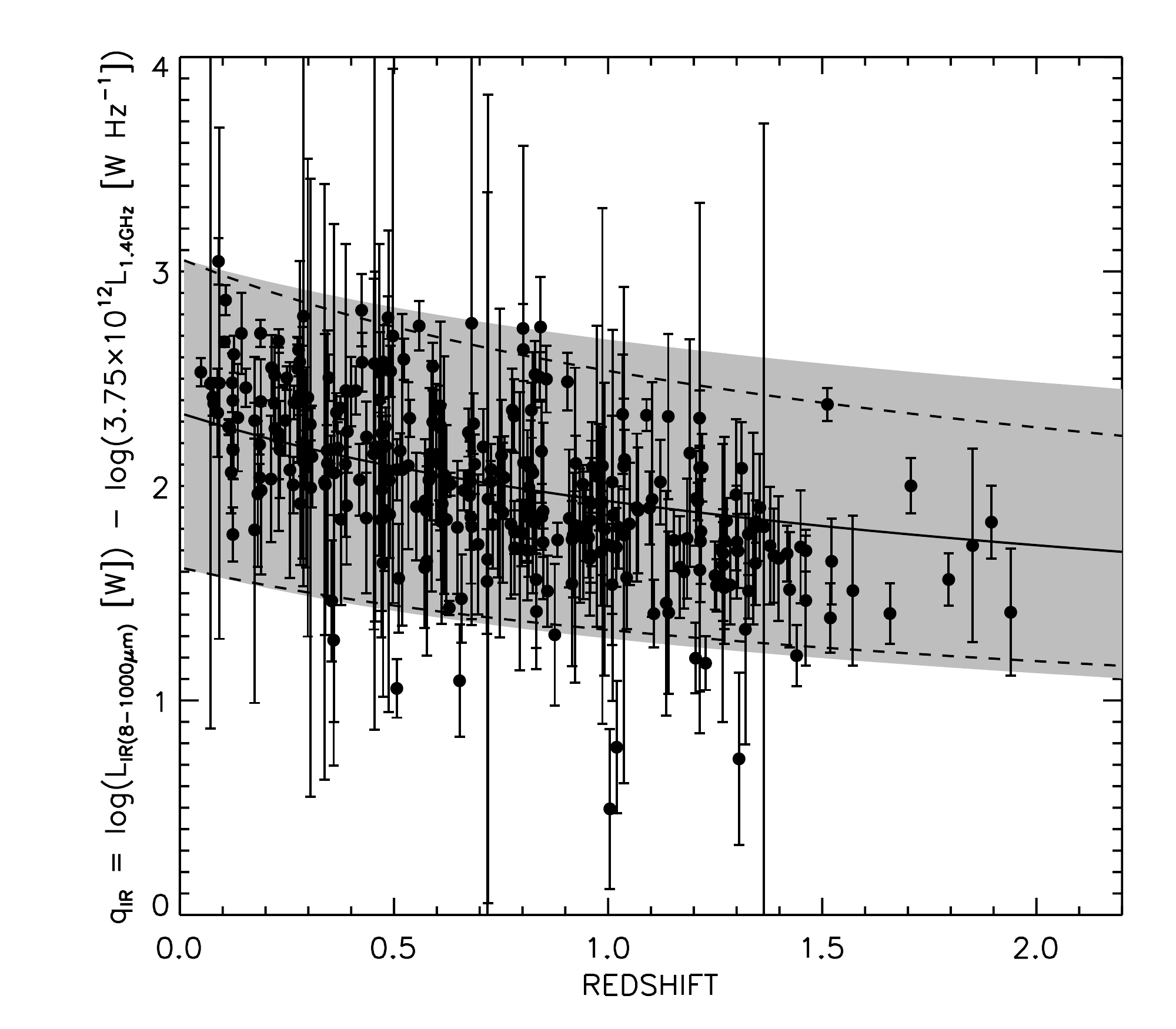}
  \caption{Ratio between FIR luminosity and radio luminosity in our
    sample, $q_{\rm IR}$, against redshift.  We
    compare against the moderate evolution measured by
    \citet{ivison10a} $\propto(1+z)^{-0.26\pm0.07}$ (gray
    band).  We measure slightly stronger evolution, $q_{\rm
      IR}\propto(1+z)^{-0.30}$.  }
  \label{fig:firradio}
\end{figure}

Since IR luminosities were derived independently of radio luminosity,
we measure the FIR/radio ratio, $q_{\rm IR}$, for radio-detected HSGs
in order to assess the FIR/radio correlation for starbursts
\citep[see][]{helou85a,condon92a,ivison10a,ivison10b} for HSGs.  We
use the bolometric $q_{\rm IR}$ ratio between 8--1000\,\um\ flux and
1.4\,GHz radio flux as defined in \citet{ivison10a}, with IR
luminosities measured from rest-frame 8\,\um\ to 1000\um.
Figure~\ref{fig:firradio} shows the $q_{\rm IR}$ ratio evolving with
redshift for synchrotron slope $\alpha=0.75$, where
$S_{\nu}\propto\nu^{-\alpha}$.  Our sample is consistent with the
moderate evolution measured by \citet{ivison10a,ivison10b} and
\citet{magnelli10a}.  We measure $q_{\rm IR}\propto(1+z)^{\gamma}$
where $\gamma=$--0.30$\pm$0.02 using only the $0<z<2$ HSG sample (an
independent assessment of the $z>2$ sample is given in C12).
\citet{ivison10b} measured $\gamma=$--0.04$\pm$0.03 at $z=$\,0--2, but
found that their $z<0.5$ samples contaminated this measurement since
they were few in number and not well matched in luminosity to the
higher redshift sources;
they measure $\gamma=$--0.26$\pm$0.07 for the $0.5<z<2.0$ sample,
which is consistent with our finding, $\gamma=$--0.30.  Either with or
without our $z<0.5$ sample, we measure $\gamma=$--0.30, likely because
our sample is dominated (68\%) by sources at $0.5<z<1.5$.  Note that
the evolution we measure is much more pronounced if the original
luminosity limits defining $S_{\rm FIR}$ are used, 40--120\um: $q_{\rm
  FIR}\propto(1+z)^{-0.59\pm0.04}$.  Note also that the significantly
limiting factor of this measurement is that our sample is partially
radio-selected and is not a luminosity-matched sample.

\section{Discussion}\label{sec:discussion}

The characterization of galaxies by {\it Herschel} is the largest
advance in understanding star formation in the infrared since the
discovery of SMGs by SCUBA.  This paper has presented a spectroscopic
survey of 1594 {\sc Spire} sources, 767 have identified spectroscopic
redshifts from $0<z<5$, and 731 at $z<2$.  Securing spectroscopic
redshifts is itself very valuable for follow-up studies of this
population, from their metalicities, stellar populations,
morphologies and evolutionary histories to their molecular gas
properties and dust content. However, arguably the impact of this
population on the cosmic star formation rate density (SFRD) is the
most relevant computation for galaxy formation and evolution studies
in general.  The luminosity function has only ever been measured in
integrated IR luminosity ($L_{\rm IR}(8-1000\mu\!m)$) for limited
populations of SMGs with known incompleteness
\citep{chapman05a,wardlow11a} or from extrapolations to the IR from
      {\it Spitzer} \citep{le-floch05a,caputi07a,magnelli11a} making
      use of model templates or explicitly assuming SEDs from
      libraries \citep{le-borgne09a,bethermin11a,marsden11a}.  This
      section presents the integrated IR luminosity function of {\it
        Herschel\/} galaxies, a discussion of the completeness of our
      samples, the implications for the cosmic SFRD and comparisons to
      these previous studies.

\subsection{Luminosity Function}\label{sec:lf}

\begin{figure*}
  \centering
  \includegraphics[width=1.65\columnwidth]{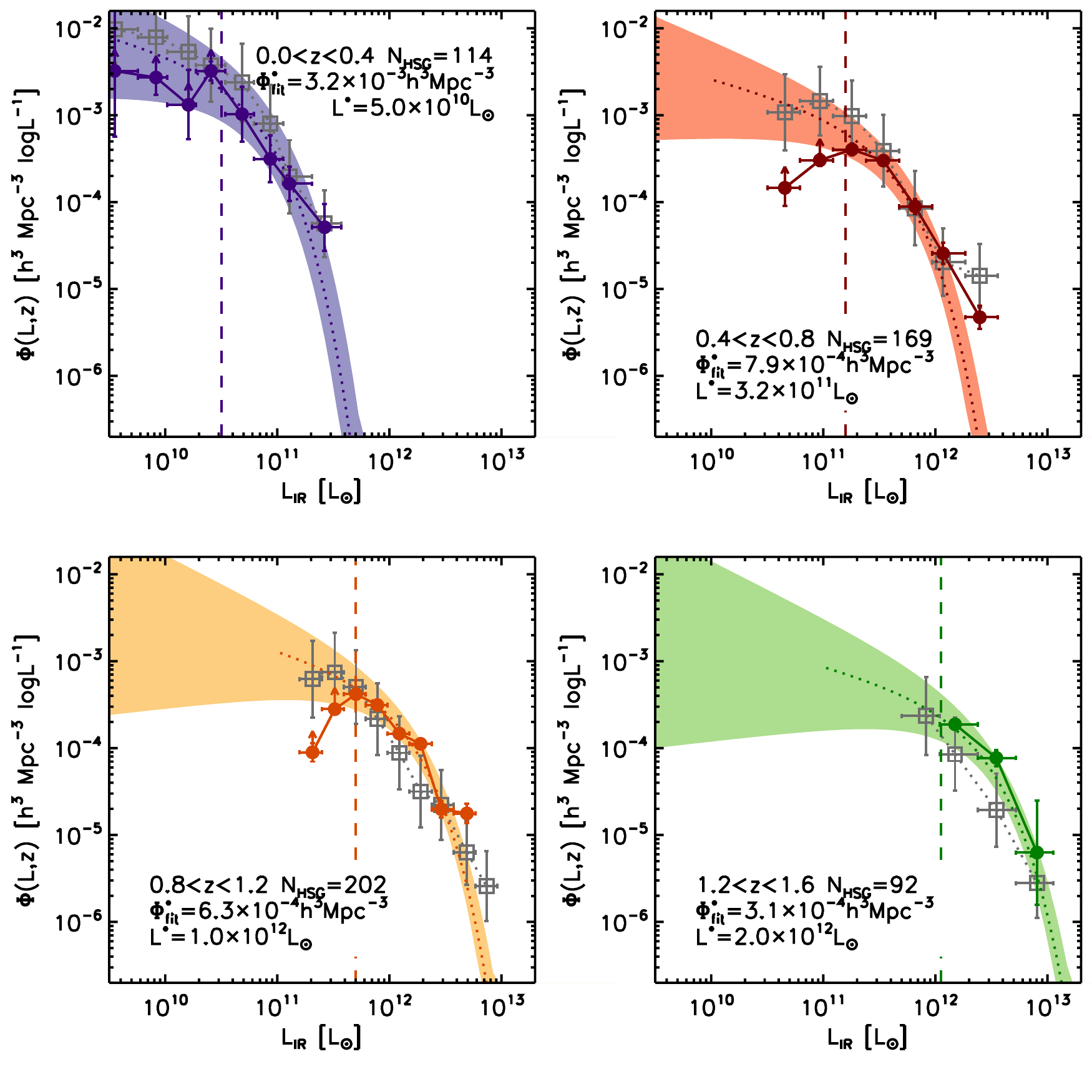}
  \caption{
    Evolving luminosity functions for our HSG spectroscopic sample.  The
    redshift bins are $0.0<z<0.4$ ($purple$), $0.4<z<0.8$ ($red$),
    $0.8<z<1.2$ ($yellow$) and $1.2<z<1.6$ ($green$).  We also include the
    estimated luminosity functions from photometric redshifts in COSMOS
    \citep{ilbert10a} as gray boxes and associated uncertainties.  The
    best-fit Schechter functions are shown as dotted colored lines, with
    associated uncertainties (shaded regions) based on uncertainty in
    $\alpha$, $\Phi^{*}$ and $L^{*}$.  The lower
    luminosity completeness limits of each bin are shown as dashed
    vertical lines.  Below these luminosities, our {\sc Spire} samples are
    incomplete (in terms of {\sc Spire} detectability).  Note that these
    luminosity functions are corrected for spectroscopic incompleteness
    (as detailed in Figure~\ref{fig:speccomp}).
  }
  \label{fig:lf}
\end{figure*}

We compute the integrated IR (8--1000\,\um) luminosity functions (LFs)
of {\sc Spire}-detected galaxies in four redshift bins where the
spectroscopic completeness is constrained and the number of
identifications is $\sim$100 per bin: 0.0$<z<$0.4, 0.4$<z<$0.8,
0.8$<z<$1.2, and 1.2$<z<$1.6.  We use the $1/V_{\rm max}$ accessible
volume technique where
\begin{equation}
\Phi_{z}(L)\Delta L = \sum\limits_{i=0}^n \frac{1}{V_{i}(L)}
\end{equation}
$\Phi_{z}(L)$ is given in units of $h^{3}\, $Mpc$^{-3}\, $log$L^{-1}$
and $\Phi_{z}(L)\Delta L$ is the number density of sources with
luminosities between $L$ and $L+\Delta L$.  Here $V_{i}(L)$ is the
comoving volume within which the $i$th source is detectable in our
survey.  This accessible volume is determined by constructing a
detection limit in $L_{\rm IR}$--$z$ space for each source.  As was
evident in Figure~\ref{fig:lirz}, the {\sc Spire} detection limit is
highly dependent on dust temperature (i.e. 40\,K galaxies will have a
different $L_{\rm IR}$--$z$ detection threshold than 20\,K galaxies).
The luminosity limit per source is determined by assuming an SED (of
type given by Eq~1) with fitted $T_{\rm dust}$, and the 3$\sigma$
detection bounds at 250\,\um, 350\,\um, and 500\,\um, where the lowest
luminosity limit with redshift is adopted for the accessible volume
calculation.  The measured luminosity function, after correction for
spectroscopic incompleteness (as discussed earlier in the text and
shown in Figure~\ref{fig:speccomp}), is given in Table~\ref{tab:lf}
and shown in Figure~\ref{fig:lf} as filled, colored circles.

\begin{deluxetable}{cccc}
  \renewcommand\baselinestretch{1.0}
  \tablewidth{0pt}
  \tablecaption{IR Luminosity Function from the {\it Herschel}-{\sc Spire} spectroscopic sample}
  \scriptsize
  \tablehead{
    \multicolumn{2}{c}{\underline{\it $0.0<z<0.4$} N=116} & 
    \multicolumn{2}{c}{\underline{\it 0.4$<z<$0.8} N=168} \\
    log$L_{\rm IR}$ & log$\Phi(L_{\rm IR})$ & 
    log$L_{\rm IR}$ & log$\Phi(L_{\rm IR})$ \\
    (\lsun) & {\footnotesize (h$^3$\,Mpc$^{-3}$\,log$L^{-1}$)} & 
    (\lsun) & {\footnotesize (h$^3$\,Mpc$^{-3}$\,log$L^{-1}$)}
  }
  \startdata
  9.55   & ($-$2.49$\pm$0.76)  & 10.66 & ($-$3.84$\pm$0.18) \\
  9.92   & ($-$2.57$\pm$0.19)  & 10.97 & ($-$3.52$\pm$0.06) \\
  10.21  & ($-$2.88$\pm$0.36)  & 11.26 &  $-$3.39$\pm$0.06 \\
  10.41  & ($-$2.49$\pm$0.11)  & 11.54 &  $-$3.52$\pm$0.05 \\
  10.69  &  $-$2.99$\pm$0.32   & 11.82 &  $-$4.05$\pm$0.09 \\
  10.94  &  $-$3.51$\pm$0.28   & 12.07 &  $-$4.59$\pm$0.12 \\
  11.11  &  $-$3.79$\pm$0.19   & 12.40 &  $-$5.32$\pm$0.13 \\
  11.42  &  $-$4.29$\pm$0.27   & & \\
  \hline
  log$\Phi^{\star}$= & $-$2.5  & log$\Phi^{\star}$= & $-$3.1 \\
  log$L^{\star}$=    & 10.7    & log$L^{\star}$=    & 11.5 \\
  $\alpha\equiv$     & $-$0.35 & $\alpha\equiv$     & $-$0.35 \\
  \hline\hline
  & & & \\
  \multicolumn{2}{c}{\underline{\it 0.8$<z<$1.2} N=202} & 
  \multicolumn{2}{c}{\underline{\it 1.2$<z<$1.6} N=92} \\
  log$L_{\rm IR}$ & log$\Phi(L_{\rm IR})$ &
  log$L_{\rm IR}$ & log$\Phi(L_{\rm IR})$ \\
  (\lsun) & {\footnotesize (h$^3$\,Mpc$^{-3}$\,log$L^{-1}$)} & 
  (\lsun) & {\footnotesize (h$^3$\,Mpc$^{-3}$\,log$L^{-1}$)} \\
  \hline
  11.31 & ($-$4.05$\pm$0.11) & 12.18 & $-$3.72$\pm$0.05 \\
  11.51 & ($-$3.55$\pm$0.06) & 12.54 & $-$4.11$\pm$0.10 \\
  11.70 & ($-$3.38$\pm$0.03) & 12.91 & $-$5.20$\pm$0.54 \\
  11.89 &  $-$3.50$\pm$0.06 & & \\
  12.09 &  $-$3.83$\pm$0.06 & & \\
  12.28 &  $-$3.95$\pm$0.05 & & \\
  12.46 &  $-$4.70$\pm$0.10 & & \\
  12.69 &  $-$4.75$\pm$0.11 & & \\
  \hline
  log$\Phi^{\star}$= & $-$3.2  & log$\Phi^{\star}$= & $-$3.5 \\
  log$L^{\star}$=    & 12.0    & log$L^{\star}$=    & 12.3 \\
  $\alpha\equiv$     & $-$0.35 & $\alpha\equiv$     & $-$0.35 \\
  \hline\hline
  \enddata
  \label{tab:lf}
  \tablecomments{The luminosities in this table are integrated within
    8--1000\,\um.  They represent the central luminosity of sources in
    each bin, in other words, the bin for luminosity $L$ covers sources
    with luminosities between $L-\Delta L/2$ and $L+\Delta L/2$.  The
    luminosity functions in this table are based exclusively on our {\sc
      Spire} galaxy spectroscopic sample.  We compute the luminosity
    function in units of h$^3$\,Mpc$^{-3}$\,log$L^{-1}$.  }
\end{deluxetable}

As a check, we also compute the luminosity functions for the same XID
selection using the photometric redshifts in COSMOS \citep{ilbert10a}
and the same luminosity limit technique (for $\sim$6000 sources over
$\sim$2.2\,deg$^{2}$).  The photometric redshift results are shown on
Figure~\ref{fig:lf} as open gray squares.  Within uncertainties, the
spectroscopic and photometric luminosity functions agree at each
epoch.

Figure~\ref{fig:lf} clearly demonstrates evolution in the luminosity
function with redshift, with ultraluminous ($>$10$^{12}$\lsun) and
hyperluminous ($>$10$^{13}$\lsun) galaxies becoming far more abundant
with increasing $z$ than in the local Universe.  At each of the four
epochs, we fit Schechter functions of the form
$\Phi=\Phi^{*}x^{\alpha}e^{-x}$, where $x=L/L^{*}$.  The faint end of
the luminosity function, dominated by the $x^{\alpha}$ power law, is
largely unprobed by our dataset since $L^{*}$, the turnover
luminosity, is approximately equal to the lower luminosity limit of
{\sc Spire} in each bin.  Without constraints, we decided to fix
$\alpha=$--0.35, our measurement from the local RBGS sample
\citep{sanders03a} and the $0<z<0.4$ photometric redshift sample,
which is also consistent with prior local and high-$z$ measurements
\citep{le-floch05a,caputi07a}.  We add a large margin of uncertianty
on the slope $\alpha$, since its impact on integrated star formation
density increases with redshift ($\sigma_{\alpha}=$0.1 at $z\approx0$,
increasing to 0.5 at $z\approx1.4$).  At the top end of the luminosity
function, we note slight excesses at $0.0<z<0.4$, $0.4<z<0.8$ and
$0.8<z<1.2$, which, in other samples, has been attributed to AGN
contributions to infrared luminosity; while a double power law
luminosity function might provide an alternate fitting method, the
difference in this work is inconsequential considering the uncertainty
on the luminosity function points themselves.

The parameters $\Phi^{*}$ and $L^{*}$ are then varied to provide an
optimal fit to data at each epoch.  We measure evolution in both
parameters, as given in Table~\ref{tab:lf}.  However, it should be
stressed that the measurement of these parameters' values is highly
degenerate with variation in the faint end slope, $\alpha$ and with
each other.  Much larger data sets, particularly those which constrain
the faint end slope via stacking techniques, are needed for detailed
analysis on the evolution of $\Phi^{*}$ and $L^{*}$.  This detailed
work is being done with larger samples from COSMOS (M. Vaccari, E. Le
Floc'h, private communication).

\subsection{Star Formation Rate Density}

To estimate the contribution of {\sc Spire}-detected sources to the
cosmic star formation rate density (SFRD), our measured luminosty
functions are converted to SFR functions via the \citet{kennicutt98a}
prescription assuming a non-evolving Salpeter IMF.  Note that there is
some recent discussion, particularly in relation to very high
luminosity sources at $z>2$, that the Kennicutt conversion from
infrared luminosity to SFR might not hold due to IMF variation
\citep{swinbank08a}; however, in this paper we use it for consistency
with previous work on the infrared SFRD
\citep{le-floch05a,hopkins06a,caputi07a,magnelli11a}.

\begin{figure*}
  \includegraphics[width=0.99\columnwidth]{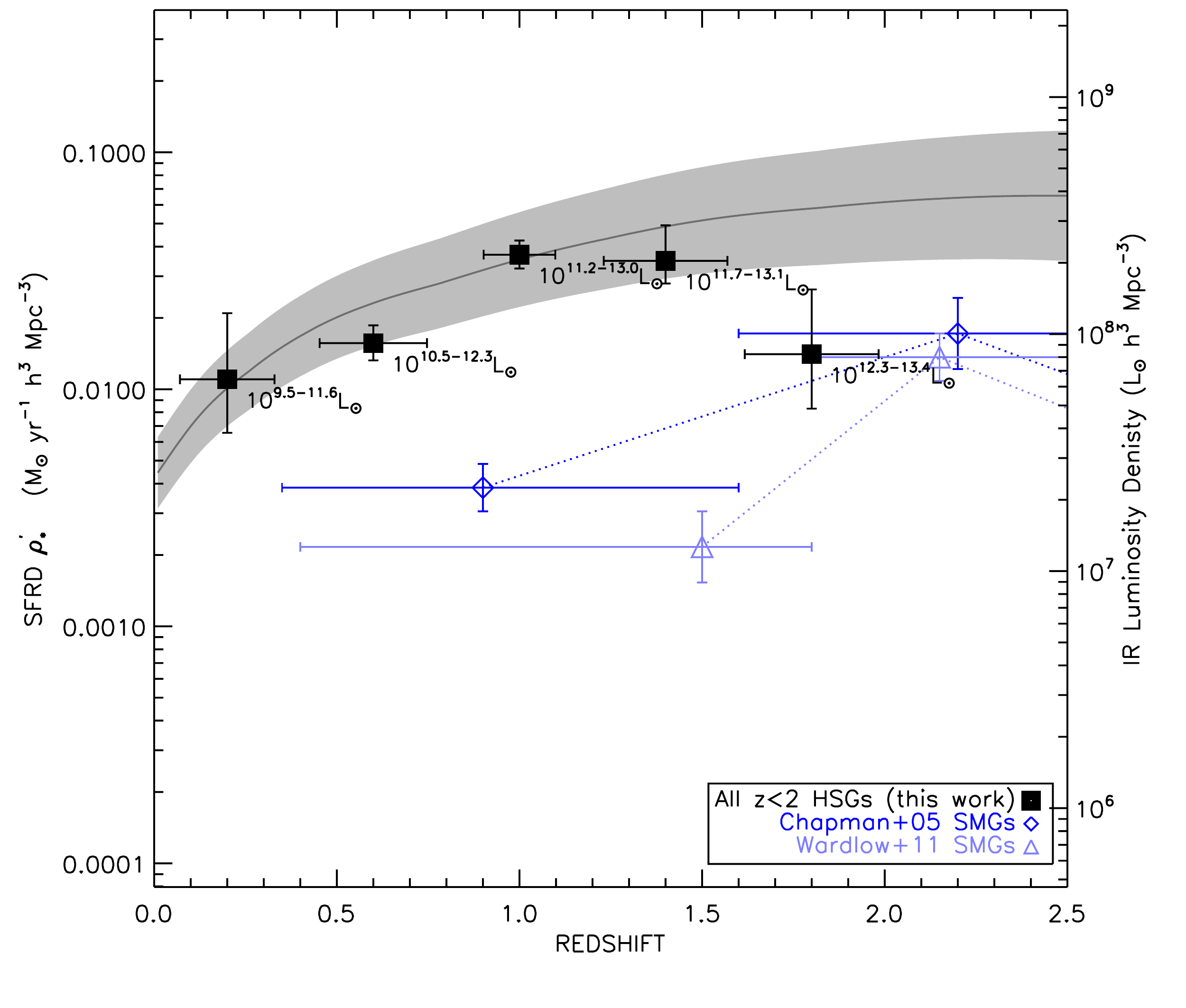}
  \includegraphics[width=0.99\columnwidth]{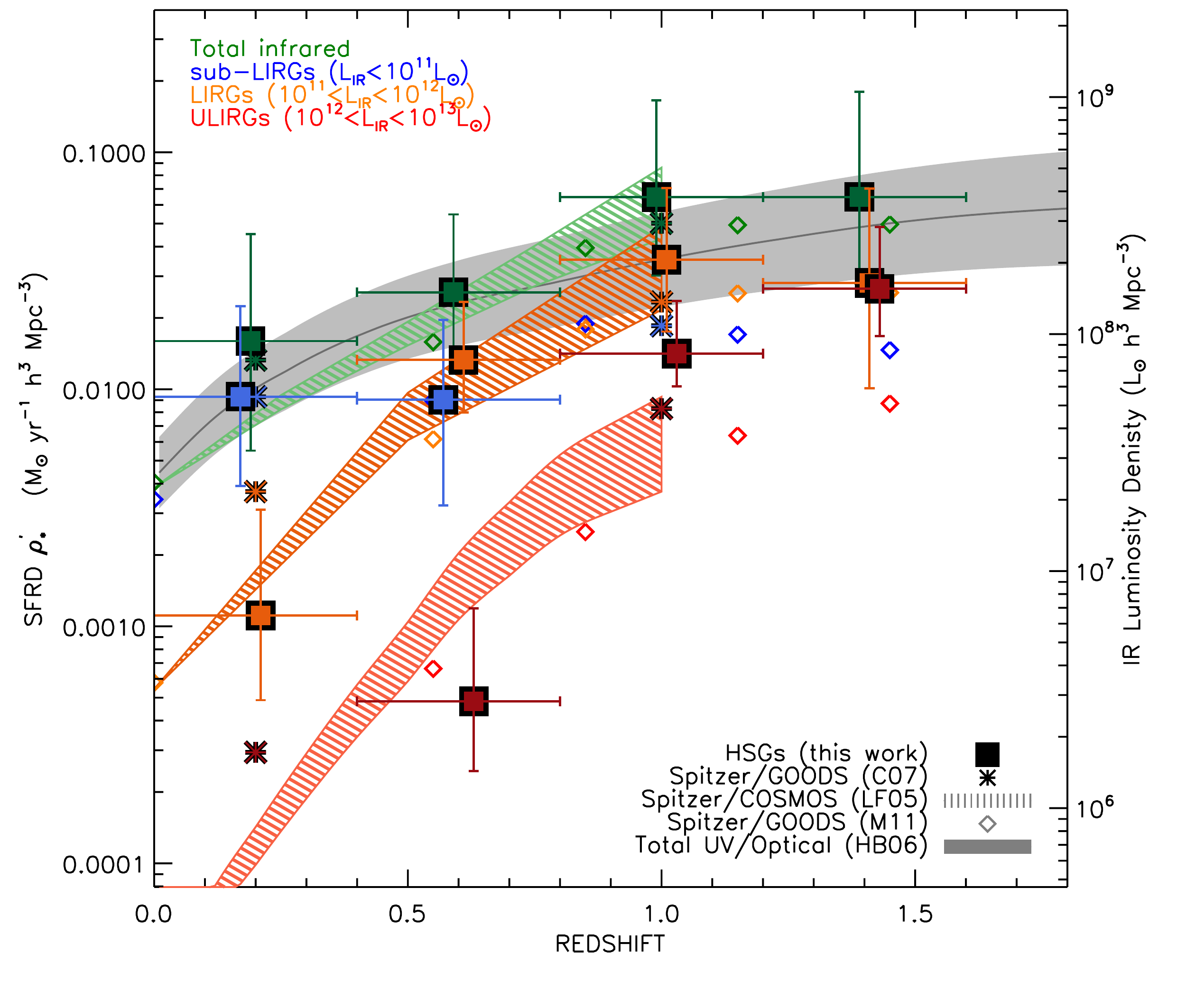}
  \caption{Contribution of {\sc Spire} galaxies to the cosmic star
    formation rate density from $z=0$ to $z=2$.  We compare against
    the compilation in Hopkins \&\ Beacom 2006 of optical/UV SFRD
    estimates which are dust-corrected (large grey area, representing
    the associated uncertainty of UV/optical estimates). We also
    overplot the SFRD points estimated by \citet{chapman05a} and
    \citet{wardlow11a} for SMGs which are substantially lower by
    $\sim$10$\times$ than the optical/UV dust-corrected SFRD and HSGs
    at $z<$1.5.  At left, we measure the SFRD from this spectroscopic
    sample by converting the measured luminosity functions to star
    formation rate functions, then integrating.  At right, we use the
    best-fit Schechter functions in Figure~\ref{fig:lf} to extrapolate
    over luminosities not directly probed by our survey, to measure
    the total infrared (green), sub-LIRG (blue), LIRG (orange) and
    ULIRG (red) contributions to the SFRD (colors are consistent
    across different data sets).  The right y-axes give the value of
    the IR luminosity density which translates directly to SFRD via
    the \citet{kennicutt98a} scaling.  We gather similar computations
    from the literature for comparison: \citet{le-floch05a} {\it
      Spitzer} samples in COSMOS (hashed regions), \citet{caputi07a}
    {\it Spitzer} samples in GOODS (asterisks), and
    \citet{magnelli11a} {\it Spitzer} samples in GOODS (diamonds). }
  \label{fig:sfrd}
\end{figure*}

The SFRD contribution from HSGs is shown in Figure~\ref{fig:sfrd}.
The left panel shows the raw conversion of the luminosity function
data points to SFRD contribution, including all sources above and
below the completeness luminosity limit (dashed lines on
Figure~\ref{fig:lf}), and without extrapolation to lower or higher
luminosities.  Although we excluded $1.6<z<2.0$ sources from the
luminosity function calculation due to small numbers, we bring them
back for this computation.  The measured HSG SFRD appears to increase
from $z=0$ to $z=1$, and then fall sharply at $z=1.5$; however, this
is due entirely to the luminosity limits of the sample at the given
epochs.  For example, the highest redshift bin, $z\approx1.8$, has a
lower luminosity limit of log$L_{\rm IR}=12.3$, significantly higher
than the lower redshift bins.  Similar luminosity limit restrictions
exist for the literature values of the SFRD found from SMGs, in
\citet{chapman05a}, \citet{wall08a} and \citet{wardlow11a}.

Although SMGs are thought of as a very rare class of galaxy
in this context, it is important to point out that this depends
entirely on the adopted definition of `SMG' and associated luminosity
limits.  The $\sim$1\,dex difference between the SFRD contributions of
850\um-selected SMGs and 250--500\,\um-selected HSGs at $z\sim1$ is due
to the effective luminosity limits of {\sc Spire} or 850\um\ selection
(where HSG selection probes luminosities $\sim$5 times fainter) and
also due to the dust temperature bias of 850\um-selection (another
factor of $\sim$2).

Drawing a fair comparison with other samples requires interpolation in
the luminosity function across equal luminosity bins via our best-fit
Schechter functions.  Several literature sources have estimated the
total integrated star formation rate density from infrared sources
using a variety of datasets: {\it Spitzer}-24\,\um-identified sources
in COSMOS \citep{le-floch05a}; {\it Spitzer}-24\,\um\ sources in both
GOODS fields \citep{caputi07a,magnelli11a}; and {\it Akari}-selected
sources \citep{goto10a}.  We also draw on the full optical and
ultraviolet data compilation of \citet*{hopkins06a}, which is
extinction corrected to account for infrared emission in
optically-bright galaxies.  On the right panel of
Figure~\ref{fig:sfrd}, we show the breakdown of the SFRD in 1\,dex
luminosity bins, $<$10$^{11}$\lsun\ (sub-LIRGs),
10$^{11}<z<$10$^{12}$\lsun\ (LIRGs),
10$^{12}<z<$10$^{13}$\lsun\ (ULIRGs), and total integrated infrared.
On a whole, we observe the same trends as other literature works: the
total integrated infrared SFRD contribution is comparable to the total
optical and ultraviolet contribution (i.e.~just as much star formation
is obscured as is unobscured) LIRGs dominate at $z\approx1$, and
ULIRGs become increasingly important at $z>1$.  However, a few subtle
differences stand out.  For example, both \citet{le-floch05a} and
\citet{caputi07a} find higher contributions from ULIRGs at $z<0.5$
than is measured in the \citet{magnelli11a} sample or in our sample.
Since both prior samples were 24\um-selected, this could be indicative
of the disagreement between 24\um\ and total intregrated luminosity.
At $z>1$, the \citet{magnelli11a} sample shows a deficit in ULIRGs
with respect to our sample \citep[also seen in the models
  of][]{bethermin11a}.  While this deficit could be due to SED
assumptions, it could also be due to cosmic variance and the size of
the GOODS fields used in Magnelli \etal\ \citep[e.g., as is known to be
  a problem in CDFS, which has a submm deficit, see][]{weis09b}.

The uncertainties on the SFRD measurements are dominated by the
uncertainty in the faint-end slope of the luminosity function, which
is not well constrained.  At higher redshifts $L^{\star}$ increases;
this impact of the faint-end slope uncertainty then grows, since
sub-LIRGs and LIRGs are not constrained with our data beyond $z>$1.2.
This is not only true for our data set, but any data set (selected via
24\,\um\ or radio with a {\sc Pacs} or {\sc Spire} detection) which does
not employ stacking analysis to constrain faint sources.  Further work
is needed on stacking 24\,\um- and radio-selected samples of galaxies to
probe fainter luminosities out to high redshifts over large areas of
sky less prone to cosmic variance.

It is clear from this work that the importance of infrared-luminous
star formation in the context of cosmic star formation is high, and
equally high as rest-frame UV/optical estimates, even though prior
surveys of SMGs seem to imply that infrared starbursts are rare even
at $z\sim$\,1--2.  {\it Herschel}-{\sc Spire} has enabled us to
directly constrain the far-infrared unlike previous work in the area,
extrapolating from 24\um.  Our observations confirm the importance of dust
obscured star formation across the first 10 billion years of the
Universe's history.

\section{Conclusions}

This paper has presented a large spectroscopic survey of galaxies
selected in the {\it Herschel}-{\sc Spire} 250--500\,\um\ bands and
followed up spectroscopically on Keck LRIS and DEIMOS.  Out of 1594
spectroscopic targets spanning 0.93\,deg$^2$ in multiple deep legacy
fields, 767 sources had identifiable redshifts from their rest-frame
UV or optical spectra, 731 of which are at $z<2$.  We present the
following conclusions:
\begin{itemize}
\item The redshift distribution of {\sc Spire}-selected
  spectroscopically-confirmed galaxies peaks at $z=0.85$ and has a
  tail of sources extending out to $z\approx5$.  The vast majority of
  this spectroscopic sample (731/767\,$\approx$\,96\%) is at $z<2$,
  mostly spectrally identified by \oii, \oiii, \hb, or \ha\ emission.
\item Our spectroscopic sample excludes sources without 24\,\um\ or
  radio 1.4\,GHz counterparts.  A negligible fraction of {\sc
    Spire}-sources in deep legacy fields are thought to drop out in
  the mid-IR and radio at $z<2$, meaning our {\sc Spire} targeting is
  close to complete at $z$\simlt2.
\item Only $\sim$50\%\ of spectroscopic targets yield redshifts (767
  with identifications, 826 without).  Under photometric observing
  conditions, $\sim$60\%\ yield identifications.  We constrain the
  spectroscopic completeness at $z<2$ to 20--80\%, depending on
  redshift, using large catalogs of photometric redshifts.  The
  sources without spectroscopic identifications either were observed
  in poor weather conditions, have no optical counterpart (i.e. no
  source picked up on the slit after 1--2\,hrs integration), or have
  spectra not easily identified with a single redshift (e.g. continuum
  without emission or absorption features).
\item We measure the accuracy of HSG optical/near-infrared photometric
  redshifts at $\Delta z$/($1+z_{\rm spec}$)=0.29, a factor of
  $\sim$3--4 times worse than photometric redshifts for normal field
  galaxies.  We determine that the lack of reliability in photometric
  redshifts for HSGs is due to more significant optical obscuration
  and higher overall star formation rates (thus higher
  line-to-continuum ratios).  We caution that future analysis of
  aggregate properties of infrared-selected samples like HSGs might be
  biased by only making use of large photometric redshift catalogs.
\item We observe a correlation between dust temperature and infrared
  luminosity which is partially a selection effect and partially
  thought to be physically real.  The selection effect is caused by
  the dependence of {\sc Spire} flux density on dust temperature and
  the physical motivation comes from the local observation that more
  luminous galaxies have more compact, clumpy and therefore hot dust.
\item {\sc Spire} color does not evolve with redshift in our
  spectroscopic sample, due to the redshift dependence of luminosity
  and the luminosity dependence on dust temperature (a combination of
  selection effects and physical mechanisms).
\item We infer the aggregate properties of HSGs by combining the
  near-infrared to radio measurements by selection method and
  luminosity.  We find little difference between the bulk infrared
  properties of radio-selected galaxies and 24\,\um-selected galaxies,
  and good overall agreement with the FIR/radio correlation.
  Consistent with our results from individual sources, we see an
  increase in dust temperature with luminosity in the composite IR
  SEDs.  At luminosities $>$10$^{12.5}$\lsun, we detect the
  $\approx$10\,\um\ Si absorption feature in the composite, resembling
  the SED of the local ULIRG Arp\,220.
\item By extrapolating our infrared SED fits to 850\um, we determine
  that only 79$^{+4}_{-8}$\%\ of HSGs would be detectable as `classic'
  SMGs ($S_{\rm 850}>5$\,mJy).  This fraction is highest at low
  redshift, 83$^{+4}_{-7}$\%\ at $z<1$, and still significant,
  72$^{+5}_{-11}$\%, from $1<z<2$.
\item For the 331 radio-selected sources with spectroscopic
  confirmations, we measure the evolution in the FIR/radio correlation
  and find $q_{\rm IR}=(1+z)^{-0.30\pm0.02}$.  This evolution is
  stronger than some previous measurements, however the sample is
  biased by being radio-detected, thus the intrinsic evolution of
  $q_{\rm IR}$ with redshift is likely to be shallower.
\item We compute the luminosity functions for HSGs in four redshift
  bins across $0<z<1.6$ and find agreement with predictions from
  photometric redshifts.  The luminosity functions are well-fit by
  Schechter functions, with evolving parameters $\Phi^{\star}$
  (3.2$\times$10$^{3}\,h^3\,$Mpc$^{-3}$ at $z=0.2$ to
  3.1$\times$10$^{-4}\,h^{3}\,$Mpc$^{-3}$ at $z=1.4$) and $L^{\star}$
  (5$\times$10$^{10}$\lsun\ at $z=0.2$ to 2$\times$10$^{12}$\lsun\ at
  $z=1.4$).  The faint-end of the luminosity function is not well
  probed by our data and would only be accessible from {\it Herschel}
  data through stacking analyses.
\item We find the HSG contribution to the cosmic star formation rate
  density to be substantial at $z<2$, essentially comparable to the star
  formation rate density from optical/near-IR surveys
  \simgt1$\times$10$^{-2}$\sfr\,$h^{3}\,$Mpc$^{-3}$.  Without relying on
  extrapolations from the mid-infrared or model template libraries, we
  measure the contribution from LIRGs (10$^{11}<L<$10$^{12}$\lsun) to
  the SFRD peaks at $z\sim1$, and the ULIRG ($>$10$^{12}$\lsun)
  contribution increases with $z$, nearly surpassing the LIRG
  contribution at $z\sim1.4$.
\end{itemize}

This work has demonstrated that infrared-luminous galaxies form an
integral part of galaxy formation across a wide range of epochs,
particularly at $0<z<2$.  With spectroscopic redshifts in hand,
follow-up studies can be more efficiently carried out to determine the
{\it physical} and {\it evolutionary} origins of HSGs as a function of
infrared luminosity, dust temperature, dust mass, and AGN content.
This work will then begin to shed light on the dominant mechanisms
producing obscured star formation in the Universe.

\section*{Acknowledgements}

We thank the anonymous reviewer for a very useful, constructive report
which improved the manuscript.
CMC is generously supported by a Hubble Fellowship from Space
Telescope Science Institute, grant HST-HF-51268.01-A.
The data presented herein were obtained at the W.M. Keck Observatory,
which is operated as a scientific partnership among the California
Institute of Technology, the University of California and the National
Aeronautics and Space Administration. The Observatory was made
possible by the generous financial support of the W.M. Keck
Foundation.  The authors wish to recognize and acknowledge the very
significant cultural role and reverence that the summit of Mauna Kea
has always had within the indigenous Hawaiian community.  We are most
fortunate to have the opportunity to conduct observations from this
mountain. This work would not be possible without the hard work and
dedication of the Keck Observatory night and day staff; special thanks
to Marc Kassis, Luca Rizzi and Greg Wirth for help and advice while
observing.  The analysis pipeline used to reduce the DEIMOS data was
developed at UC Berkeley with support from NSF grant AST-0071048.
Thanks to Nick Scoville and Brian Siana for useful discussions which
improved the paper.

{\sc Spire} has been developed by a consortium of institutes led by
Cardiff Univ. (UK) and including: Univ. Lethbridge (Canada); NAOC
(China); CEA, LAM (France); IFSI, Univ. Padua (Italy); IAC (Spain);
Stockholm Observatory (Sweden); Imperial College London, RAL,
UCL-MSSL, UKATC, Univ. Sussex (UK); and Caltech, JPL, NHSC,
Univ. Colorado (USA).  This development has been supported by national
funding agencies: CSA (Canada); NAOC (China); CEA, CNES, CNRS
(France); ASI (Italy); MCINN (Spain); SNSB (Sweden); STFC, UKSA (UK);
and NASA (USA).

This research has made use of data from the HerMES project
(http://hermes.sussex.ac.uk/).  HerMES is a {\it Herschel} Key
Programme utilizing Guaranteed Time from the {\sc Spire} instrument
team, ESAC scientists and a mission scientist.  HerMES is described in
\citet{oliver12a}.  The {\sc Spire} data presented in this paper will
be released through the HerMES Database in Marseille, HeDaM
(http://hedam.oamp.fr/HerMES).

\clearpage
\begin{landscape}
\begin{deluxetable}{l@{ }c@{\,}c@{\,}c@{ }c@{ }c@{ }c@{ }c@{ }c@{ }c@{ }cc@{ }c}
\tablecolumns{13}
\tablecaption{Spectroscopically identified 0$<z<$2 {\sc Spire}-selected Galaxies}
 
\tablehead{
\colhead{NAME} & \colhead{$z_{\rm spec}$} & \colhead{{\sc Conf}} & \colhead{$z_{\rm phot}$} & \colhead{$S_{\rm 24}$} 
&  \colhead{$S_{\rm 100}$} & \colhead{$S_{\rm 160}$} & \colhead{$S_{\rm 250}$} & \colhead{$S_{\rm 350}$} & \colhead{$S_{\rm 500}$} 
&  \colhead{$S_{\rm 1.4GHz}$} & \colhead{L$_{\rm IR}$} & \colhead{T$_{\rm dust}$} \\
\colhead{} & & & & (\uJy) & (mJy) & (mJy) & (mJy) & (mJy) & (mJy) & (\uJy) &  (\lsun) & (K) \\
 }
\startdata

1HERMES\,X1.4\,J021856.18$-$043546.5$^{*}$... & 1.102$^{{\rm D}}$ & 1 & ...$^{{\rm U}}$ & 275.2$\pm$18.1 & $-$ & $-$ & 16.4$\pm$12.4 & 5.7$\pm$8.2 & ... & $-$ & (2.1$^{+2.4}_{-1.1}$)$\times$10$^{12}$ & 57.8$\pm$7.9 \\
1HERMES\,X1.4\,J021856.37$-$043538.0$^{*}$... & 0.470$^{{\rm D}}$ & 3 & 0.47$^{{\rm U}}$ & 336.3$\pm$18.6 & $-$ & $-$ & 10.2$\pm$12.4 & 35.9$\pm$8.2 & 40.6$\pm$5.5 & $-$ & (1.1$^{+1.2}_{-0.5}$)$\times$10$^{11}$ & 8.90$\pm$1.7 \\
1HERMES\,X1.4\,J021901.73$-$043449.0$^{*}$... & 1.010$^{{\rm D}}$ & 3 & ...$^{{\rm U}}$ & 676.3$\pm$17.3 & $-$ & $-$ & 31.0$\pm$12.4 & 25.9$\pm$7.0 & 8.5$\pm$5.3 & $-$ & (1.3$^{+5.9}_{-1.0}$)$\times$10$^{12}$ & 28.4$\pm$30.6 \\
1HERMES\,X1.4\,J021921.93$-$043335.2$^{*}$... & 0.922$^{{\rm D}}$ & 2 & ...$^{{\rm U}}$ & 337.2$\pm$16.1 & $-$ & $-$ & 32.6$\pm$12.4 & 27.1$\pm$7.1 & 12.1$\pm$5.5 & $-$ & (8.7$^{+25.}_{-6.5}$)$\times$10$^{11}$ & 25.2$\pm$19.7 \\
1HERMES\,X24\,J021926.00$-$043435.4 ... & 1.302$^{{\rm D}}$ & 4 & ...$^{{\rm U}}$ & 342.6$\pm$19.4 & $-$ & $-$ & 24.1$\pm$12.4 & 14.6$\pm$7.1 & 27.9$\pm$5.2 & $-$ & (1.1$^{+1.6}_{-0.6}$)$\times$10$^{12}$ & 16.3$\pm$5.0 \\
1HERMES\,X24\,J021926.85$-$043044.0 ... & 1.029$^{{\rm D}}$ & 3 & ...$^{{\rm U}}$ & 368.5$\pm$21.0 & $-$ & $-$ & 34.2$\pm$12.4 & 29.4$\pm$7.1 & 20.1$\pm$5.2 & $-$ & (9.1$^{+13.}_{-5.5}$)$\times$10$^{11}$ & 21.2$\pm$9.0 \\
1HERMES\,X24\,J021933.91$-$043335.3 ... & 0.582$^{{\rm D}}$ & 1 & ...$^{{\rm U}}$ & 1164.$\pm$21.1 & $-$ & $-$ & 48.2$\pm$12.4 & 50.8$\pm$7.0 & 29.9$\pm$5.4 & $-$ & (4.0$^{+3.4}_{-1.8}$)$\times$10$^{11}$ & 16.3$\pm$3.9 \\
1HERMES\,X1.4\,J021937.20$-$042931.2$^{*}$... & 0.295$^{{\rm D}}$ & 2 & ...$^{{\rm U}}$ & 191.0$\pm$17.0 & $-$ & $-$ & 38.8$\pm$12.4 & 56.8$\pm$7.0 & 58.5$\pm$5.3 & $-$ & (5.1$^{+2.5}_{-1.6}$)$\times$10$^{10}$ & 9.10$\pm$0.9 \\
\enddata
\label{tab_giant}
\tablecomments{
Source names indicate both detection band in the XID catalog (24\um\ or 1.4\,GHz) 
and source position based on near-infrared/optical counterparts.  Asterisks after 
the name indicate the source was observed in near-photometric conditions.  Superscripts
in the $z_{\rm spec}$ column indicate the instrument with which observations were 
obtained, LRIS (L) or DEIMOS (D).  The {\sc Conf} column indicates the confidence 
in the spectroscopic redshift identification, and ranges here from 1 to 5, 5 being 
the most confident (further identifications at {\sc Conf}$<$1 have been excluded
from this paper).  Photometric redshift superscript indicates which field the data 
are in, UDS (U), CDFS (S), COSMOS (C), LHN (L), GOODS-N (G), or Elais-N1 (E).  
{\sc Pacs} flux densities come from PEP \citep{lutz11a} in COSMOS and from HerMES 
\citep{oliver12a} in LHN.  Ellipsis (...) denote that the given source is undetected 
at the corresponding wavelength or has no photometric redshift despite having the 
necessary imaging or coverage whereas dashes ($-$) denote that no data exist.}
\end{deluxetable}
\clearpage
\begin{deluxetable}{l@{ }c@{\,}c@{\,}c@{ }c@{ }c@{ }c@{ }c@{ }c@{ }c@{ }cc@{ }c}
\tablecolumns{13}
\tablecaption{Spectroscopically identified 0$<z<$2 {\sc Spire}-selected Galaxies -- Continued.}
 
\tablehead{
\colhead{NAME} & \colhead{$z_{\rm spec}$} & \colhead{{\sc Conf}} & \colhead{$z_{\rm phot}$} & \colhead{$S_{\rm 24}$} 
&  \colhead{$S_{\rm 100}$} & \colhead{$S_{\rm 160}$} & \colhead{$S_{\rm 250}$} & \colhead{$S_{\rm 350}$} & \colhead{$S_{\rm 500}$} 
&  \colhead{$S_{\rm 1.4GHz}$} & \colhead{L$_{\rm IR}$} & \colhead{T$_{\rm dust}$} \\
\colhead{} & & & & (\uJy) & (mJy) & (mJy) & (mJy) & (mJy) & (mJy) & (\uJy) &  (\lsun) & (K) \\
 }
\startdata

1HERMES\,X24\,J033210.60$-$273405.5 ... & 0.248$^{{\rm L}}$ & 3 & 0.21$^{{\rm S}}$ & 846.5$\pm$18.1 & $-$ & $-$ & 13.5$\pm$3.8 & 6.0$\pm$3.8 & 5.6$\pm$4.6 & ... & (2.7$^{+35.}_{-2.5}$)$\times$10$^{10}$ & 20.0$\pm$25.1 \\
1HERMES\,X24\,J033211.75$-$275138.8 ... & 0.710$^{{\rm L}}$ & 1 & 1.32$^{{\rm S}}$ & 100.1$\pm$6.9 & $-$ & $-$ & 6.3$\pm$1.9 & ... & ... & ... & (3.4$^{+1.6}_{-1.1}$)$\times$10$^{11}$ & 45.2$\pm$3.2 \\
1HERMES\,X24\,J033218.26$-$275224.9 ... & 0.738$^{{\rm L}}$ & 5 & 0.72$^{{\rm S}}$ & 155.2$\pm$10.3 & $-$ & $-$ & 7.0$\pm$1.9 & 5.2$\pm$3.1 & 3.1$\pm$4.1 & ... & (1.0$^{+7.0}_{-0.9}$)$\times$10$^{11}$ & 21.6$\pm$18.6 \\
1HERMES\,X24\,J033219.27$-$275402.5 ... & 0.962$^{{\rm L}}$ & 1 & 1.27$^{{\rm S}}$ & 179.0$\pm$8.6 & $-$ & $-$ & 6.5$\pm$1.9 & 6.8$\pm$2.9 & ... & ... & (1.8$^{+15.}_{-1.6}$)$\times$10$^{11}$ & 19.6$\pm$15.1 \\
1HERMES\,X24\,J033222.75$-$275223.8 ... & 0.126$^{{\rm L}}$ & 5 & 0.13$^{{\rm S}}$ & 116.4$\pm$7.6 & $-$ & $-$ & 38.0$\pm$1.9 & 35.6$\pm$2.9 & 26.3$\pm$3.8 & ... & (6.4$^{+1.9}_{-1.5}$)$\times$10$^{9}$ & 11.2$\pm$0.8 \\
1HERMES\,X1.4\,J033224.52$-$275442.8$^{*}$... & 0.122$^{{\rm L}}$ & 5 & 0.13$^{{\rm S}}$ & 36.2$\pm$7.2 & $-$ & $-$ & 9.0$\pm$1.9 & ... & ... & 90.0$\pm$17.7 & (1.2$^{+-18}_{-1.2}$)$\times$10$^{9}$ & 8.90$\pm$17.0 \\
1HERMES\,X24\,J033230.91$-$275311.3 ... & 0.642$^{{\rm L}}$ & 5 & 0.64$^{{\rm S}}$ & 472.2$\pm$11.6 & $-$ & $-$ & 11.7$\pm$1.9 & 16.0$\pm$2.9 & 15.9$\pm$3.8 & ... & (1.5$^{+1.2}_{-0.6}$)$\times$10$^{11}$ & 12.3$\pm$1.7 \\
1HERMES\,X24\,J033234.17$-$281451.3 ... & 1.890$^{{\rm L}}$ & 4 & ...$^{{\rm S}}$ & 493.4$\pm$11.5 & $-$ & $-$ & 13.5$\pm$3.8 & 3.0$\pm$3.7 & ... & ... & (8.7$^{+3.3}_{-2.4}$)$\times$10$^{12}$ & 86.6$\pm$5.1 \\
\enddata
\tablecomments{Table~\ref{tab_giant} Continued.}
\end{deluxetable}
\clearpage
\begin{deluxetable}{l@{ }c@{\,}c@{\,}c@{ }c@{ }c@{ }c@{ }c@{ }c@{ }c@{ }cc@{ }c}
\tablecolumns{13}
\tablecaption{Spectroscopically identified 0$<z<$2 {\sc Spire}-selected Galaxies -- Continued.}
 
\tablehead{
\colhead{NAME} & \colhead{$z_{\rm spec}$} & \colhead{{\sc Conf}} & \colhead{$z_{\rm phot}$} & \colhead{$S_{\rm 24}$} 
&  \colhead{$S_{\rm 100}$} & \colhead{$S_{\rm 160}$} & \colhead{$S_{\rm 250}$} & \colhead{$S_{\rm 350}$} & \colhead{$S_{\rm 500}$} 
&  \colhead{$S_{\rm 1.4GHz}$} & \colhead{L$_{\rm IR}$} & \colhead{T$_{\rm dust}$} \\
\colhead{} & & & & (\uJy) & (mJy) & (mJy) & (mJy) & (mJy) & (mJy) & (\uJy) &  (\lsun) & (K) \\
 }
\startdata

1HERMES\,X24\,J095839.36+015611.9$^{*}$... & 0.945$^{{\rm L}}$ & 2 & 0.96$^{{\rm C}}$ & 310.0$\pm$17.0 & ... & ... & 8.5$\pm$2.7 & 5.1$\pm$3.2 & ... & ... & (5.0$^{+139}_{-4.9}$)$\times$10$^{11}$ & 37.6$\pm$99.7 \\
1HERMES\,X1.4\,J095839.84+015450.2$^{*}$... & 1.461$^{{\rm L}}$ & 1 & 1.58$^{{\rm C}}$ & 209.0$\pm$17.9 & 12.7$\pm$1.8 & 36.5$\pm$3.8 & 37.8$\pm$2.6 & 30.0$\pm$3.9 & 23.5$\pm$4.9 & 95.0$\pm$15.0 & (3.0$^{+0.5}_{-0.4}$)$\times$10$^{12}$ & 31.7$\pm$2.3 \\
1HERMES\,X24\,J095840.08+013047.4 ... & 0.653$^{{\rm D}}$ & 4 & ...$^{{\rm C}}$ & 444.0$\pm$23.0 & 15.9$\pm$2.3 & ... & 12.2$\pm$2.6 & 22.9$\pm$3.8 & 11.3$\pm$6.6 & ... & (3.8$^{+1.1}_{-0.8}$)$\times$10$^{11}$ & 26.0$\pm$2.5 \\
1HERMES\,X24\,J095840.79+013048.8 ... & 0.652$^{{\rm D}}$ & 2 & 0.56$^{{\rm C}}$ & 755.0$\pm$17.9 & 34.0$\pm$2.0 & 76.8$\pm$6.3 & 41.0$\pm$2.6 & 24.9$\pm$3.9 & 7.2$\pm$7.1 & ... & (7.9$^{+1.0}_{-0.9}$)$\times$10$^{11}$ & 28.6$\pm$1.5 \\
1HERMES\,X24\,J095842.23+013302.8 ... & 0.633$^{{\rm D}}$ & 4 & 0.65$^{{\rm C}}$ & 1640.$\pm$21.0 & 43.2$\pm$2.0 & 64.9$\pm$7.1 & 62.6$\pm$2.6 & 39.7$\pm$5.4 & 27.7$\pm$4.8 & ... & (9.9$^{+0.9}_{-0.8}$)$\times$10$^{11}$ & 26.7$\pm$0.9 \\
1HERMES\,X24\,J095843.91+013017.8 ... & 0.701$^{{\rm D}}$ & 2 & ...$^{{\rm C}}$ & 360.0$\pm$27.0 & ... & ... & 13.8$\pm$2.7 & 22.2$\pm$5.2 & 17.2$\pm$5.0 & ... & (2.0$^{+2.0}_{-1.0}$)$\times$10$^{11}$ & 12.9$\pm$2.2 \\
1HERMES\,X24\,J095844.39+013329.0 ... & 1.452$^{{\rm D}}$ & 3 & ...$^{{\rm C}}$ & $-$ & ... & ... & 11.4$\pm$2.6 & 5.0$\pm$3.3 & ... & ... & (9.9$^{+2.5}_{-2.0}$)$\times$10$^{12}$ & 100.3$\pm$0.0 \\
1HERMES\,X24\,J095844.39+013227.2 ... & 0.184$^{{\rm D}}$ & 5 & 0.17$^{{\rm C}}$ & 239.0$\pm$17.0 & ... & ... & 9.0$\pm$2.7 & 0.2$\pm$3.7 & 1.9$\pm$5.1 & ... & (2.8$^{+1.2}_{-0.8}$)$\times$10$^{10}$ & 33.7$\pm$2.1 \\
\enddata
\tablecomments{Table~\ref{tab_giant} Continued.}
\end{deluxetable}
\clearpage
\begin{deluxetable}{l@{ }c@{\,}c@{\,}c@{ }c@{ }c@{ }c@{ }c@{ }c@{ }c@{ }cc@{ }c}
\tablecolumns{13}
\tablecaption{Spectroscopically identified 0$<z<$2 {\sc Spire}-selected Galaxies -- Continued.}
 
\tablehead{
\colhead{NAME} & \colhead{$z_{\rm spec}$} & \colhead{{\sc Conf}} & \colhead{$z_{\rm phot}$} & \colhead{$S_{\rm 24}$} 
&  \colhead{$S_{\rm 100}$} & \colhead{$S_{\rm 160}$} & \colhead{$S_{\rm 250}$} & \colhead{$S_{\rm 350}$} & \colhead{$S_{\rm 500}$} 
&  \colhead{$S_{\rm 1.4GHz}$} & \colhead{L$_{\rm IR}$} & \colhead{T$_{\rm dust}$} \\
\colhead{} & & & & (\uJy) & (mJy) & (mJy) & (mJy) & (mJy) & (mJy) & (\uJy) &  (\lsun) & (K) \\
 }
\startdata

1HERMES\,X24\,J095909.84+013748.0 ... & 0.526$^{{\rm D}}$ & 3 & 0.52$^{{\rm C}}$ & 375.9$\pm$17.0 & ... & 24.5$\pm$6.2 & 25.6$\pm$2.6 & 26.6$\pm$7.4 & ... & ... & (1.7$^{+1.4}_{-0.8}$)$\times$10$^{11}$ & 20.2$\pm$4.8 \\
1HERMES\,X24\,J095910.79+013534.4 ... & 0.530$^{{\rm D}}$ & 4 & 0.47$^{{\rm C}}$ & 220.0$\pm$30.0 & 6.0$\pm$1.6 & ... & 11.9$\pm$2.7 & 11.6$\pm$3.3 & 4.6$\pm$5.0 & ... & (1.0$^{+0.5}_{-0.3}$)$\times$10$^{11}$ & 22.2$\pm$3.2 \\
1HERMES\,X1.4\,J095912.00+020824.1 ... & 0.352$^{{\rm L}}$ & 2 & 0.36$^{{\rm C}}$ & 1772.$\pm$55.0 & 57.8$\pm$5.3 & 80.5$\pm$4.2 & 62.1$\pm$2.6 & 27.1$\pm$3.5 & ... & 279.$\pm$25.0 & (2.8$^{+0.3}_{-0.3}$)$\times$10$^{11}$ & 24.9$\pm$1.4 \\
1HERMES\,X24\,J095912.00+013708.5 ... & 0.511$^{{\rm D}}$ & 2 & 1.44$^{{\rm C}}$ & 165.9$\pm$15.0 & ... & ... & 15.0$\pm$2.7 & 15.0$\pm$3.3 & 7.3$\pm$6.2 & ... & (7.4$^{+14.}_{-4.8}$)$\times$10$^{10}$ & 15.1$\pm$5.0 \\
1HERMES\,X24\,J095913.19+021147.3$^{*}$... & 1.155$^{{\rm L}}$ & 1 & 1.11$^{{\rm C}}$ & $-$ & ... & ... & 9.4$\pm$2.7 & ... & ... & ... & (2.4$^{+5.2}_{-1.6}$)$\times$10$^{11}$ & 25.0$\pm$15.0 \\
1HERMES\,X1.4\,J095915.12+013945.2$^{*}$... & 0.681$^{{\rm D}}$ & 3 & 0.68$^{{\rm C}}$ & 375.9$\pm$14.0 & ... & ... & 21.1$\pm$2.6 & 16.1$\pm$3.2 & 6.8$\pm$4.9 & 84.0$\pm$16.0 & (2.7$^{+4.9}_{-1.7}$)$\times$10$^{11}$ & 21.8$\pm$10.2 \\
1HERMES\,X1.4\,J095918.47+020859.5 ... & 0.281$^{{\rm L}}$ & 3 & 0.32$^{{\rm C}}$ & 187.9$\pm$57.0 & ... & ... & 11.7$\pm$2.6 & 13.7$\pm$4.1 & 4.6$\pm$8.3 & 61.0$\pm$10.0 & (1.8$^{+1.6}_{-0.8}$)$\times$10$^{10}$ & 15.5$\pm$3.0 \\
1HERMES\,X24\,J095918.72+013509.7 ... & 0.667$^{{\rm D}}$ & 3 & 0.54$^{{\rm C}}$ & 183.0$\pm$15.0 & ... & 11.0$\pm$2.5 & 8.9$\pm$2.6 & 6.6$\pm$3.3 & ... & ... & (1.5$^{+2.5}_{-0.9}$)$\times$10$^{11}$ & 27.2$\pm$12.6 \\
\enddata
\tablecomments{Table~\ref{tab_giant} Continued.}
\end{deluxetable}
\clearpage
\begin{deluxetable}{l@{ }c@{\,}c@{\,}c@{ }c@{ }c@{ }c@{ }c@{ }c@{ }c@{ }cc@{ }c}
\tablecolumns{13}
\tablecaption{Spectroscopically identified 0$<z<$2 {\sc Spire}-selected Galaxies -- Continued.}
 
\tablehead{
\colhead{NAME} & \colhead{$z_{\rm spec}$} & \colhead{{\sc Conf}} & \colhead{$z_{\rm phot}$} & \colhead{$S_{\rm 24}$} 
&  \colhead{$S_{\rm 100}$} & \colhead{$S_{\rm 160}$} & \colhead{$S_{\rm 250}$} & \colhead{$S_{\rm 350}$} & \colhead{$S_{\rm 500}$} 
&  \colhead{$S_{\rm 1.4GHz}$} & \colhead{L$_{\rm IR}$} & \colhead{T$_{\rm dust}$} \\
\colhead{} & & & & (\uJy) & (mJy) & (mJy) & (mJy) & (mJy) & (mJy) & (\uJy) &  (\lsun) & (K) \\
 }
\startdata

1HERMES\,X24\,J095932.64+014501.9 ... & 0.122$^{{\rm D}}$ & 1 & 0.11$^{{\rm C}}$ & 218.0$\pm$17.0 & 7.3$\pm$1.9 & ... & 9.0$\pm$2.6 & 3.8$\pm$4.1 & ... & ... & (3.5$^{+2.7}_{-1.5}$)$\times$10$^{9}$ & 20.3$\pm$3.8 \\
1HERMES\,X24\,J095932.87+014611.3 ... & 0.124$^{{\rm D}}$ & 3 & 0.11$^{{\rm C}}$ & 1003.$\pm$59.0 & 37.5$\pm$1.7 & 43.9$\pm$4.4 & 29.8$\pm$2.7 & 17.1$\pm$3.4 & 3.1$\pm$5.2 & ... & (1.6$^{+0.2}_{-0.2}$)$\times$10$^{10}$ & 22.7$\pm$1.2 \\
1HERMES\,X24\,J095933.36+021750.2$^{*}$... & 0.972$^{{\rm L}}$ & 5 & ...$^{{\rm C}}$ & 156.0$\pm$35.0 & ... & ... & 8.6$\pm$2.6 & 3.8$\pm$3.5 & ... & ... & (2.7$^{+3.2}_{-1.4}$)$\times$10$^{11}$ & 27.5$\pm$7.2 \\
1HERMES\,X1.4\,J095935.51+021950.4 ... & 0.940$^{{\rm L}}$ & 4 & 0.93$^{{\rm C}}$ & 999.0$\pm$30.0 & 26.2$\pm$2.6 & 32.2$\pm$6.4 & 39.3$\pm$2.7 & ... & 21.6$\pm$5.5 & 108.$\pm$13.0 & (1.5$^{+0.2}_{-0.2}$)$\times$10$^{12}$ & 30.5$\pm$2.2 \\
1HERMES\,X24\,J095935.99+024137.4$^{*}$... & 1.027$^{{\rm L}}$ & 2 & 1.00$^{{\rm C}}$ & 537.0$\pm$14.0 & ... & 13.8$\pm$3.6 & 15.8$\pm$2.7 & 14.8$\pm$3.8 & 4.0$\pm$5.4 & ... & (6.5$^{+5.7}_{-3.0}$)$\times$10$^{11}$ & 25.7$\pm$7.0 \\
1HERMES\,X24\,J095936.96+023240.0$^{*}$... & 0.721$^{{\rm L}}$ & 3 & ...$^{{\rm C}}$ & 349.0$\pm$14.0 & ... & 11.7$\pm$3.6 & 11.2$\pm$2.6 & 3.1$\pm$3.3 & 3.6$\pm$5.3 & ... & (2.3$^{+4.1}_{-1.4}$)$\times$10$^{11}$ & 28.6$\pm$15.4 \\
1HERMES\,X1.4\,J095936.96+014659.8$^{*}$... & 0.266$^{{\rm D}}$ & 4 & 0.27$^{{\rm C}}$ & 1382.$\pm$148. & 41.3$\pm$1.9 & 50.8$\pm$5.1 & 30.5$\pm$2.7 & 16.9$\pm$3.3 & 8.6$\pm$4.9 & 131.$\pm$12.0 & (1.0$^{+0.1}_{-0.1}$)$\times$10$^{11}$ & 26.0$\pm$1.5 \\
1HERMES\,X24\,J095937.44+021834.2$^{*}$... & 0.174$^{{\rm L}}$ & 4 & ...$^{{\rm C}}$ & 2328.$\pm$1117 & 72.9$\pm$2.1 & 68.5$\pm$2.9 & 32.4$\pm$2.7 & 12.5$\pm$3.4 & ... & ... & (7.7$^{+1.0}_{-0.8}$)$\times$10$^{10}$ & 29.0$\pm$1.6 \\
\enddata
\tablecomments{Table~\ref{tab_giant} Continued.}
\end{deluxetable}
\clearpage
\begin{deluxetable}{l@{ }c@{\,}c@{\,}c@{ }c@{ }c@{ }c@{ }c@{ }c@{ }c@{ }cc@{ }c}
\tablecolumns{13}
\tablecaption{Spectroscopically identified 0$<z<$2 {\sc Spire}-selected Galaxies -- Continued.}
 
\tablehead{
\colhead{NAME} & \colhead{$z_{\rm spec}$} & \colhead{{\sc Conf}} & \colhead{$z_{\rm phot}$} & \colhead{$S_{\rm 24}$} 
&  \colhead{$S_{\rm 100}$} & \colhead{$S_{\rm 160}$} & \colhead{$S_{\rm 250}$} & \colhead{$S_{\rm 350}$} & \colhead{$S_{\rm 500}$} 
&  \colhead{$S_{\rm 1.4GHz}$} & \colhead{L$_{\rm IR}$} & \colhead{T$_{\rm dust}$} \\
\colhead{} & & & & (\uJy) & (mJy) & (mJy) & (mJy) & (mJy) & (mJy) & (\uJy) &  (\lsun) & (K) \\
 }
\startdata

1HERMES\,X1.4\,J095943.91+023331.8$^{*}$... & 0.945$^{{\rm L}}$ & 2 & 0.91$^{{\rm C}}$ & 727.0$\pm$35.9 & 48.5$\pm$1.7 & 101.$\pm$5.8 & 74.4$\pm$2.6 & 47.5$\pm$9.3 & 38.1$\pm$4.8 & 336.$\pm$14.0 & (2.8$^{+0.2}_{-0.1}$)$\times$10$^{12}$ & 30.6$\pm$0.8 \\
1HERMES\,X24\,J095944.39+021755.3$^{*}$... & 0.929$^{{\rm L}}$ & 3 & 0.93$^{{\rm C}}$ & 206.0$\pm$14.0 & ... & ... & 23.1$\pm$2.7 & ... & ... & ... & (2.0$^{+0.3}_{-0.2}$)$\times$10$^{12}$ & 47.4$\pm$1.4 \\
1HERMES\,X1.4\,J095945.12+023439.1$^{*}$... & 0.123$^{{\rm L}}$ & 5 & ...$^{{\rm C}}$ & 794.0$\pm$270. & 19.4$\pm$1.6 & 31.3$\pm$4.3 & 16.0$\pm$2.6 & 5.6$\pm$3.3 & 4.2$\pm$4.9 & 346.$\pm$32.0 & (9.7$^{+3.1}_{-2.3}$)$\times$10$^{9}$ & 21.8$\pm$1.9 \\
1HERMES\,X1.4\,J095945.60+024117.5$^{*}$... & 0.985$^{{\rm L}}$ & 2 & 0.99$^{{\rm C}}$ & 1220.$\pm$88.0 & 33.1$\pm$1.7 & 53.7$\pm$4.7 & 57.2$\pm$2.6 & 40.7$\pm$3.3 & 16.6$\pm$5.3 & 170.$\pm$16.0 & (2.3$^{+0.2}_{-0.2}$)$\times$10$^{12}$ & 30.4$\pm$1.1 \\
1HERMES\,X24\,J095946.08+024212.5$^{*}$... & 0.988$^{{\rm L}}$ & 2 & ...$^{{\rm C}}$ & 193.0$\pm$14.0 & ... & ... & 13.9$\pm$2.6 & 10.8$\pm$3.4 & ... & ... & (4.8$^{+19.}_{-3.8}$)$\times$10$^{11}$ & 27.8$\pm$24.8 \\
1HERMES\,X24\,J095946.31+022331.4$^{*}$... & 1.127$^{{\rm L}}$ & 1 & 1.12$^{{\rm C}}$ & $-$ & ... & ... & 11.3$\pm$2.7 & 7.8$\pm$9.0 & ... & ... & (1.3$^{+44.}_{-1.2}$)$\times$10$^{12}$ & 45.1$\pm$135. \\
1HERMES\,X24\,J095948.24+022026.1$^{*}$... & 0.931$^{{\rm L}}$ & 5 & 0.93$^{{\rm C}}$ & 334.0$\pm$15.0 & ... & ... & 10.0$\pm$2.6 & 5.9$\pm$4.4 & ... & ... & (5.9$^{+235}_{-5.7}$)$\times$10$^{11}$ & 38.0$\pm$112. \\
1HERMES\,X24\,J095950.15+023653.6$^{*}$... & 0.887$^{{\rm L}}$ & 4 & 0.85$^{{\rm C}}$ & $-$ & ... & ... & 8.8$\pm$2.7 & 9.3$\pm$3.2 & 5.7$\pm$4.7 & ... & (1.6$^{+6.5}_{-1.2}$)$\times$10$^{11}$ & 18.1$\pm$9.8 \\
\enddata
\tablecomments{Table~\ref{tab_giant} Continued.}
\end{deluxetable}
\clearpage
\begin{deluxetable}{l@{ }c@{\,}c@{\,}c@{ }c@{ }c@{ }c@{ }c@{ }c@{ }c@{ }cc@{ }c}
\tablecolumns{13}
\tablecaption{Spectroscopically identified 0$<z<$2 {\sc Spire}-selected Galaxies -- Continued.}
 
\tablehead{
\colhead{NAME} & \colhead{$z_{\rm spec}$} & \colhead{{\sc Conf}} & \colhead{$z_{\rm phot}$} & \colhead{$S_{\rm 24}$} 
&  \colhead{$S_{\rm 100}$} & \colhead{$S_{\rm 160}$} & \colhead{$S_{\rm 250}$} & \colhead{$S_{\rm 350}$} & \colhead{$S_{\rm 500}$} 
&  \colhead{$S_{\rm 1.4GHz}$} & \colhead{L$_{\rm IR}$} & \colhead{T$_{\rm dust}$} \\
\colhead{} & & & & (\uJy) & (mJy) & (mJy) & (mJy) & (mJy) & (mJy) & (\uJy) &  (\lsun) & (K) \\
 }
\startdata

1HERMES\,X24\,J100013.68+021936.8$^{*}$... & 0.591$^{{\rm L}}$ & 5 & 0.59$^{{\rm C}}$ & $-$ & ... & ... & ... & 11.7$\pm$3.5 & ... & ... & (1.2$^{+2.8}_{-0.8}$)$\times$10$^{11}$ & 25.0$\pm$15.0 \\
1HERMES\,X1.4\,J100014.16+021311.8$^{*}$... & 1.138$^{{\rm L}}$ & 5 & ...$^{{\rm C}}$ & 324.0$\pm$17.0 & 7.0$\pm$1.5 & ... & 17.0$\pm$2.6 & 20.3$\pm$3.3 & 13.5$\pm$4.8 & 1685$\pm$69.0 & (9.5$^{+3.2}_{-2.4}$)$\times$10$^{11}$ & 26.7$\pm$3.3 \\
1HERMES\,X1.4\,J100014.39+020024.0$^{*}$... & 0.079$^{{\rm L}}$ & 5 & ...$^{{\rm C}}$ & 2669.$\pm$349. & 89.1$\pm$2.9 & 105.$\pm$6.9 & 53.3$\pm$2.6 & 21.8$\pm$3.3 & 6.9$\pm$4.9 & 367.$\pm$74.0 & (1.5$^{+0.1}_{-0.1}$)$\times$10$^{10}$ & 24.0$\pm$0.9 \\
1HERMES\,X24\,J100014.64+021413.5$^{*}$... & 1.173$^{{\rm L}}$ & 2 & 0.99$^{{\rm C}}$ & 158.0$\pm$16.0 & ... & ... & 11.6$\pm$2.7 & ... & ... & ... & (1.9$^{+0.6}_{-0.4}$)$\times$10$^{12}$ & 56.4$\pm$3.1 \\
1HERMES\,X24\,J100015.60+021355.8$^{*}$... & 1.315$^{{\rm L}}$ & 2 & 0.60$^{{\rm C}}$ & 156.0$\pm$26.0 & ... & ... & 12.4$\pm$2.6 & 19.3$\pm$3.6 & ... & ... & (7.1$^{+12.}_{-4.5}$)$\times$10$^{11}$ & 17.8$\pm$5.1 \\
1HERMES\,X1.4\,J100016.08+021237.3$^{*}$... & 0.186$^{{\rm L}}$ & 4 & ...$^{{\rm C}}$ & 2889.$\pm$659. & 87.2$\pm$2.0 & 106.$\pm$3.8 & 48.6$\pm$2.7 & 15.0$\pm$4.1 & 37.4$\pm$4.8 & 508.$\pm$35.0 & (1.0$^{+0.0}_{-0.0}$)$\times$10$^{11}$ & 24.6$\pm$0.7 \\
1HERMES\,X1.4\,J100016.31+021434.6$^{*}$... & 0.343$^{{\rm L}}$ & 2 & 0.33$^{{\rm C}}$ & 950.0$\pm$21.0 & 27.1$\pm$1.8 & 33.3$\pm$4.7 & 28.8$\pm$2.6 & 12.9$\pm$3.6 & 13.4$\pm$5.1 & 137.$\pm$29.0 & (1.2$^{+0.2}_{-0.2}$)$\times$10$^{11}$ & 24.7$\pm$1.5 \\
1HERMES\,X24\,J100016.56+022135.5$^{*}$... & 0.434$^{{\rm L}}$ & 5 & 0.43$^{{\rm C}}$ & 169.0$\pm$14.0 & ... & ... & 9.3$\pm$2.7 & ... & ... & ... & (1.7$^{+0.7}_{-0.5}$)$\times$10$^{11}$ & 38.7$\pm$2.4 \\
\enddata
\tablecomments{Table~\ref{tab_giant} Continued.}
\end{deluxetable}
\clearpage
\begin{deluxetable}{l@{ }c@{\,}c@{\,}c@{ }c@{ }c@{ }c@{ }c@{ }c@{ }c@{ }cc@{ }c}
\tablecolumns{13}
\tablecaption{Spectroscopically identified 0$<z<$2 {\sc Spire}-selected Galaxies -- Continued.}
 
\tablehead{
\colhead{NAME} & \colhead{$z_{\rm spec}$} & \colhead{{\sc Conf}} & \colhead{$z_{\rm phot}$} & \colhead{$S_{\rm 24}$} 
&  \colhead{$S_{\rm 100}$} & \colhead{$S_{\rm 160}$} & \colhead{$S_{\rm 250}$} & \colhead{$S_{\rm 350}$} & \colhead{$S_{\rm 500}$} 
&  \colhead{$S_{\rm 1.4GHz}$} & \colhead{L$_{\rm IR}$} & \colhead{T$_{\rm dust}$} \\
\colhead{} & & & & (\uJy) & (mJy) & (mJy) & (mJy) & (mJy) & (mJy) & (\uJy) &  (\lsun) & (K) \\
 }
\startdata

1HERMES\,X24\,J100027.60+015857.0$^{*}$... & 0.352$^{{\rm L}}$ & 4 & ...$^{{\rm C}}$ & $-$ & ... & ... & 8.7$\pm$2.6 & 5.8$\pm$3.3 & 8.9$\pm$5.0 & ... & (1.6$^{+5.0}_{-1.2}$)$\times$10$^{10}$ & 12.2$\pm$4.5 \\
1HERMES\,X24\,J100027.83+015936.9$^{*}$... & 0.355$^{{\rm L}}$ & 4 & ...$^{{\rm C}}$ & $-$ & ... & ... & 11.2$\pm$2.7 & 4.8$\pm$4.6 & ... & ... & (4.4$^{+175}_{-4.3}$)$\times$10$^{10}$ & 22.3$\pm$42.3 \\
1HERMES\,X1.4\,J100028.56+021323.1$^{*}$... & 0.832$^{{\rm L}}$ & 1 & ...$^{{\rm C}}$ & 536.0$\pm$46.0 & 9.6$\pm$1.9 & ... & 9.6$\pm$2.7 & 5.1$\pm$4.1 & 7.1$\pm$5.8 & 169.$\pm$37.0 & (4.1$^{+3.2}_{-1.8}$)$\times$10$^{11}$ & 33.1$\pm$7.8 \\
1HERMES\,X24\,J100029.75+022129.6$^{*}$... & 0.728$^{{\rm L}}$ & 5 & ...$^{{\rm C}}$ & 518.0$\pm$16.0 & 5.0$\pm$1.4 & ... & 10.5$\pm$2.7 & 7.7$\pm$3.8 & 2.8$\pm$6.0 & ... & (2.2$^{+1.7}_{-0.9}$)$\times$10$^{11}$ & 25.3$\pm$5.3 \\
1HERMES\,X24\,J100029.75+022129.6 ... & 0.296$^{{\rm D}}$ & 3 & ...$^{{\rm C}}$ & 518.0$\pm$16.0 & 5.0$\pm$1.4 & ... & 10.5$\pm$2.7 & 7.7$\pm$3.8 & 2.8$\pm$6.0 & ... & (2.7$^{+2.0}_{-1.1}$)$\times$10$^{10}$ & 19.4$\pm$3.5 \\
1HERMES\,X1.4\,J100030.00+024007.0 ... & 1.895$^{{\rm L}}$ & 1 & 1.37$^{{\rm C}}$ & 172.9$\pm$14.0 & 22.2$\pm$2.1 & 36.8$\pm$4.5 & 18.3$\pm$2.6 & 11.2$\pm$3.7 & ... & 58.0$\pm$17.0 & (6.4$^{+1.8}_{-1.4}$)$\times$10$^{12}$ & 56.8$\pm$7.7 \\
1HERMES\,X24\,J100030.47+024216.5$^{*}$... & 0.220$^{{\rm L}}$ & 5 & 0.22$^{{\rm C}}$ & $-$ & ... & ... & 12.5$\pm$2.7 & 7.6$\pm$3.9 & ... & ... & (1.3$^{+12.}_{-1.2}$)$\times$10$^{10}$ & 18.4$\pm$18.9 \\
1HERMES\,X24\,J100031.91+024123.2$^{*}$... & 1.192$^{{\rm L}}$ & 4 & 0.67$^{{\rm C}}$ & 520.0$\pm$16.0 & 10.4$\pm$1.5 & 27.0$\pm$3.9 & 20.6$\pm$2.6 & 18.5$\pm$3.8 & ... & ... & (1.4$^{+0.4}_{-0.3}$)$\times$10$^{12}$ & 31.5$\pm$3.3 \\
\enddata
\tablecomments{Table~\ref{tab_giant} Continued.}
\end{deluxetable}
\clearpage
\begin{deluxetable}{l@{ }c@{\,}c@{\,}c@{ }c@{ }c@{ }c@{ }c@{ }c@{ }c@{ }cc@{ }c}
\tablecolumns{13}
\tablecaption{Spectroscopically identified 0$<z<$2 {\sc Spire}-selected Galaxies -- Continued.}
 
\tablehead{
\colhead{NAME} & \colhead{$z_{\rm spec}$} & \colhead{{\sc Conf}} & \colhead{$z_{\rm phot}$} & \colhead{$S_{\rm 24}$} 
&  \colhead{$S_{\rm 100}$} & \colhead{$S_{\rm 160}$} & \colhead{$S_{\rm 250}$} & \colhead{$S_{\rm 350}$} & \colhead{$S_{\rm 500}$} 
&  \colhead{$S_{\rm 1.4GHz}$} & \colhead{L$_{\rm IR}$} & \colhead{T$_{\rm dust}$} \\
\colhead{} & & & & (\uJy) & (mJy) & (mJy) & (mJy) & (mJy) & (mJy) & (\uJy) &  (\lsun) & (K) \\
 }
\startdata

1HERMES\,X24\,J100115.60+023012.8$^{*}$... & 0.831$^{{\rm L}}$ & 3 & ...$^{{\rm C}}$ & 348.0$\pm$15.0 & 8.4$\pm$1.7 & 13.4$\pm$3.8 & 15.4$\pm$2.6 & 17.3$\pm$3.5 & 28.6$\pm$5.2 & ... & (4.7$^{+1.6}_{-1.2}$)$\times$10$^{11}$ & 19.8$\pm$2.0 \\
1HERMES\,X1.4\,J100115.60+022858.3 ... & 0.077$^{{\rm L}}$ & 2 & ...$^{{\rm C}}$ & -3194$\pm$247. & 106.$\pm$3.8 & 134.$\pm$5.9 & 74.5$\pm$2.7 & 32.9$\pm$3.4 & 14.8$\pm$4.8 & 422.$\pm$79.0 & (1.7$^{+0.1}_{-0.1}$)$\times$10$^{10}$ & 22.8$\pm$0.7 \\
1HERMES\,X24\,J100117.03+023131.5$^{*}$... & 0.651$^{{\rm L}}$ & 3 & 0.65$^{{\rm C}}$ & 263.0$\pm$16.0 & ... & 13.1$\pm$2.9 & 12.0$\pm$2.7 & 5.9$\pm$3.3 & 1.5$\pm$5.1 & ... & (1.8$^{+2.6}_{-1.1}$)$\times$10$^{11}$ & 27.2$\pm$11.9 \\
1HERMES\,X24\,J100117.03+023021.5$^{*}$... & 1.165$^{{\rm L}}$ & 4 & 1.13$^{{\rm C}}$ & $-$ & ... & ... & 3.7$\pm$2.7 & 21.2$\pm$3.5 & 1.6$\pm$10.1 & ... & (5.4$^{+290}_{-5.3}$)$\times$10$^{11}$ & 12.2$\pm$10.2 \\
1HERMES\,X24\,J100117.03+023218.5$^{*}$... & 0.803$^{{\rm L}}$ & 2 & 0.79$^{{\rm C}}$ & 164.0$\pm$14.0 & ... & ... & 10.8$\pm$2.6 & 17.2$\pm$3.4 & 17.1$\pm$5.2 & ... & (2.0$^{+2.3}_{-1.0}$)$\times$10$^{11}$ & 12.8$\pm$2.4 \\
1HERMES\,X1.4\,J100117.28+023016.1 ... & 0.220$^{{\rm L}}$ & 3 & ...$^{{\rm C}}$ & 710.0$\pm$22.0 & 22.3$\pm$1.9 & 27.6$\pm$4.5 & 12.2$\pm$2.7 & ... & 6.4$\pm$8.4 & 75.0$\pm$12.0 & (3.5$^{+1.5}_{-1.1}$)$\times$10$^{10}$ & 26.5$\pm$3.9 \\
1HERMES\,X24\,J100117.99+023256.7$^{*}$... & 0.612$^{{\rm L}}$ & 5 & 0.62$^{{\rm C}}$ & 200.0$\pm$16.0 & ... & ... & 8.1$\pm$2.6 & 1.5$\pm$3.5 & ... & ... & (3.5$^{+1.6}_{-1.1}$)$\times$10$^{11}$ & 44.8$\pm$3.1 \\
1HERMES\,X1.4\,J100118.47+022739.2 ... & 1.658$^{{\rm L}}$ & 2 & ...$^{{\rm C}}$ & 2243.$\pm$371. & 11.5$\pm$2.0 & 35.3$\pm$2.4 & 28.2$\pm$2.7 & 17.4$\pm$3.5 & 3.9$\pm$5.7 & 210.$\pm$39.0 & (5.4$^{+1.6}_{-1.2}$)$\times$10$^{12}$ & 39.5$\pm$4.9 \\
\enddata
\tablecomments{Table~\ref{tab_giant} Continued.}
\end{deluxetable}
\clearpage
\begin{deluxetable}{l@{ }c@{\,}c@{\,}c@{ }c@{ }c@{ }c@{ }c@{ }c@{ }c@{ }cc@{ }c}
\tablecolumns{13}
\tablecaption{Spectroscopically identified 0$<z<$2 {\sc Spire}-selected Galaxies -- Continued.}
 
\tablehead{
\colhead{NAME} & \colhead{$z_{\rm spec}$} & \colhead{{\sc Conf}} & \colhead{$z_{\rm phot}$} & \colhead{$S_{\rm 24}$} 
&  \colhead{$S_{\rm 100}$} & \colhead{$S_{\rm 160}$} & \colhead{$S_{\rm 250}$} & \colhead{$S_{\rm 350}$} & \colhead{$S_{\rm 500}$} 
&  \colhead{$S_{\rm 1.4GHz}$} & \colhead{L$_{\rm IR}$} & \colhead{T$_{\rm dust}$} \\
\colhead{} & & & & (\uJy) & (mJy) & (mJy) & (mJy) & (mJy) & (mJy) & (\uJy) &  (\lsun) & (K) \\
 }
\startdata

1HERMES\,X1.4\,J100208.40+014526.7$^{*}$... & 0.575$^{{\rm D}}$ & 2 & 0.17$^{{\rm C}}$ & 770.0$\pm$16.0 & 12.4$\pm$1.4 & ... & 9.5$\pm$2.6 & ... & ... & 95.0$\pm$16.0 & (2.1$^{+1.7}_{-0.9}$)$\times$10$^{11}$ & 32.9$\pm$8.0 \\
1HERMES\,X24\,J100212.48+014746.4 ... & 0.227$^{{\rm D}}$ & 4 & ...$^{{\rm C}}$ & 157.0$\pm$20.0 & ... & ... & 13.4$\pm$2.6 & 13.3$\pm$3.9 & ... & ... & (1.0$^{+3.2}_{-0.7}$)$\times$10$^{10}$ & 13.6$\pm$6.0 \\
1HERMES\,X24\,J100213.68+015048.9 ... & 0.374$^{{\rm D}}$ & 5 & ...$^{{\rm C}}$ & 357.0$\pm$19.0 & 6.9$\pm$1.6 & ... & 11.2$\pm$2.6 & 5.0$\pm$4.5 & 1.5$\pm$5.2 & ... & (5.0$^{+3.5}_{-2.0}$)$\times$10$^{10}$ & 22.6$\pm$4.0 \\
1HERMES\,X24\,J100213.68+015023.6 ... & 0.122$^{{\rm D}}$ & 5 & ...$^{{\rm C}}$ & 279.0$\pm$16.0 & 5.9$\pm$1.4 & ... & 9.3$\pm$2.7 & ... & ... & ... & (3.3$^{+2.7}_{-1.5}$)$\times$10$^{9}$ & 19.1$\pm$3.3 \\
1HERMES\,X24\,J100215.35+015036.6 ... & 0.470$^{{\rm D}}$ & 2 & ...$^{{\rm C}}$ & 412.0$\pm$53.0 & 16.3$\pm$1.4 & 21.8$\pm$4.3 & 16.1$\pm$2.7 & 3.6$\pm$3.3 & ... & ... & (1.5$^{+0.5}_{-0.4}$)$\times$10$^{11}$ & 27.8$\pm$3.3 \\
1HERMES\,X24\,J100248.47+013551.7 ... & 0.219$^{{\rm D}}$ & 4 & 0.22$^{{\rm C}}$ & 320.0$\pm$59.0 & 14.2$\pm$2.2 & 22.1$\pm$3.1 & 9.3$\pm$2.7 & ... & ... & ... & (2.1$^{+1.0}_{-0.7}$)$\times$10$^{10}$ & 21.9$\pm$2.8 \\
1HERMES\,X1.4\,J100252.80+013642.5$^{*}$... & 0.365$^{{\rm D}}$ & 5 & 0.36$^{{\rm C}}$ & 1862.$\pm$360. & 43.5$\pm$2.0 & 61.0$\pm$3.1 & 33.1$\pm$2.6 & 16.8$\pm$3.3 & 2.3$\pm$5.3 & 146.$\pm$24.0 & (2.4$^{+0.3}_{-0.2}$)$\times$10$^{11}$ & 27.3$\pm$1.4 \\
1HERMES\,X1.4\,J100253.28+013137.0$^{*}$... & 0.628$^{{\rm D}}$ & 3 & ...$^{{\rm C}}$ & 1886.$\pm$30.0 & 108.$\pm$2.7 & 118.$\pm$4.4 & 81.2$\pm$1.5 & 58.2$\pm$3.3 & 28.6$\pm$6.8 & 1843$\pm$118. & (1.9$^{+0.0}_{-0.0}$)$\times$10$^{12}$ & 32.9$\pm$0.8 \\
\enddata
\tablecomments{Table~\ref{tab_giant} Continued.}
\end{deluxetable}
\clearpage
\begin{deluxetable}{l@{ }c@{\,}c@{\,}c@{ }c@{ }c@{ }c@{ }c@{ }c@{ }c@{ }cc@{ }c}
\tablecolumns{13}
\tablecaption{Spectroscopically identified 0$<z<$2 {\sc Spire}-selected Galaxies -- Continued.}
 
\tablehead{
\colhead{NAME} & \colhead{$z_{\rm spec}$} & \colhead{{\sc Conf}} & \colhead{$z_{\rm phot}$} & \colhead{$S_{\rm 24}$} 
&  \colhead{$S_{\rm 100}$} & \colhead{$S_{\rm 160}$} & \colhead{$S_{\rm 250}$} & \colhead{$S_{\rm 350}$} & \colhead{$S_{\rm 500}$} 
&  \colhead{$S_{\rm 1.4GHz}$} & \colhead{L$_{\rm IR}$} & \colhead{T$_{\rm dust}$} \\
\colhead{} & & & & (\uJy) & (mJy) & (mJy) & (mJy) & (mJy) & (mJy) & (\uJy) &  (\lsun) & (K) \\
 }
\startdata

1HERMES\,X1.4\,J104406.32+585227.4$^{*}$... & 0.747$^{{\rm D}}$ & 5 & 0.58$^{{\rm L}}$ & 432.2$\pm$7.0 & ... & ... & 22.2$\pm$3.4 & 13.8$\pm$4.1 & 1.1$\pm$4.9 & 85.1$\pm$19.8 & (6.5$^{+31.}_{-5.4}$)$\times$10$^{11}$ & 32.5$\pm$42.2 \\
1HERMES\,X24\,J104407.12+585319.0 ... & 0.170$^{{\rm D}}$ & 1 & 0.20$^{{\rm L}}$ & 351.7$\pm$7.5 & ... & ... & 21.2$\pm$3.4 & 12.5$\pm$4.1 & ... & ... & (2.0$^{+9.8}_{-1.6}$)$\times$10$^{10}$ & 22.6$\pm$25.0 \\
1HERMES\,X1.4\,J104407.82+585511.1$^{*}$... & 0.119$^{{\rm D}}$ & 5 & ...$^{{\rm L}}$ & 1951.$\pm$8.3 & ... & ... & 64.3$\pm$3.5 & 26.6$\pm$5.9 & 35.0$\pm$6.1 & 316.$\pm$29.9 & (1.6$^{+0.8}_{-0.5}$)$\times$10$^{10}$ & 13.4$\pm$1.8 \\
1HERMES\,X1.4\,J104408.45+585258.5$^{*}$... & 0.606$^{{\rm D}}$ & 4 & 0.56$^{{\rm L}}$ & 550.9$\pm$7.6 & ... & ... & 18.0$\pm$3.5 & 7.2$\pm$4.1 & 3.2$\pm$5.6 & 129.$\pm$19.3 & (8.3$^{+2.1}_{-1.6}$)$\times$10$^{11}$ & 45.6$\pm$1.8 \\
1HERMES\,X1.4\,J104500.72+590802.3$^{*}$... & 0.246$^{{\rm L}}$ & 1 & 0.26$^{{\rm L}}$ & 608.4$\pm$7.3 & 16.1$\pm$2.6 & 15.4$\pm$4.6 & 19.9$\pm$3.5 & 11.2$\pm$4.3 & 5.6$\pm$5.0 & 69.2$\pm$10.6 & (3.6$^{+1.5}_{-1.0}$)$\times$10$^{10}$ & 22.2$\pm$2.7 \\
1HERMES\,X1.4\,J104504.55+590847.3$^{*}$... & 1.070$^{{\rm L}}$ & 1 & 0.58$^{{\rm L}}$ & 537.1$\pm$7.2 & 9.2$\pm$2.7 & ... & 12.4$\pm$3.5 & ... & ... & 45.8$\pm$8.4 & (7.7$^{+7.1}_{-3.7}$)$\times$10$^{11}$ & 36.2$\pm$11.0 \\
1HERMES\,X1.4\,J104509.36+591040.4$^{*}$... & 0.961$^{{\rm L}}$ & 1 & ...$^{{\rm L}}$ & 786.4$\pm$7.6 & 5.3$\pm$2.5 & ... & 26.1$\pm$3.4 & 14.9$\pm$4.4 & 3.8$\pm$4.9 & 63.8$\pm$7.4 & (8.0$^{+7.6}_{-3.9}$)$\times$10$^{11}$ & 25.2$\pm$7.5 \\
1HERMES\,X1.4\,J104552.00+590928.8$^{*}$... & 1.009$^{{\rm D}}$ & 3 & 1.40$^{{\rm L}}$ & 249.4$\pm$7.2 & 4.3$\pm$2.5 & ... & 21.2$\pm$3.5 & 17.8$\pm$3.9 & 8.0$\pm$5.0 & 32.6$\pm$4.4 & (6.1$^{+6.2}_{-3.0}$)$\times$10$^{11}$ & 23.6$\pm$7.2 \\
\enddata
\tablecomments{Table~\ref{tab_giant} Continued.}
\end{deluxetable}
\clearpage
\begin{deluxetable}{l@{ }c@{\,}c@{\,}c@{ }c@{ }c@{ }c@{ }c@{ }c@{ }c@{ }cc@{ }c}
\tablecolumns{13}
\tablecaption{Spectroscopically identified 0$<z<$2 {\sc Spire}-selected Galaxies -- Continued.}
 
\tablehead{
\colhead{NAME} & \colhead{$z_{\rm spec}$} & \colhead{{\sc Conf}} & \colhead{$z_{\rm phot}$} & \colhead{$S_{\rm 24}$} 
&  \colhead{$S_{\rm 100}$} & \colhead{$S_{\rm 160}$} & \colhead{$S_{\rm 250}$} & \colhead{$S_{\rm 350}$} & \colhead{$S_{\rm 500}$} 
&  \colhead{$S_{\rm 1.4GHz}$} & \colhead{L$_{\rm IR}$} & \colhead{T$_{\rm dust}$} \\
\colhead{} & & & & (\uJy) & (mJy) & (mJy) & (mJy) & (mJy) & (mJy) & (\uJy) &  (\lsun) & (K) \\
 }
\startdata

1HERMES\,X1.4\,J104626.86+590544.8$^{*}$... & 0.842$^{{\rm D}}$ & 2 & ...$^{{\rm L}}$ & 0.00$\pm$0.0 & ... & 47.4$\pm$4.7 & 18.3$\pm$3.4 & 19.7$\pm$5.1 & 11.1$\pm$4.7 & 35.5$\pm$6.3 & (1.9$^{+1.2}_{-0.7}$)$\times$10$^{12}$ & 42.1$\pm$14.8 \\
1HERMES\,X1.4\,J104630.24+590829.2$^{*}$... & 0.829$^{{\rm D}}$ & 3 & 0.86$^{{\rm L}}$ & 518.8$\pm$7.4 & 10.6$\pm$2.5 & 23.8$\pm$4.9 & 15.8$\pm$3.5 & 23.7$\pm$3.9 & 3.9$\pm$4.8 & 93.5$\pm$9.3 & (5.8$^{+2.1}_{-1.5}$)$\times$10$^{11}$ & 25.3$\pm$3.1 \\
1HERMES\,X1.4\,J104630.97+590820.7$^{*}$... & 0.533$^{{\rm D}}$ & 3 & 0.96$^{{\rm L}}$ & 445.9$\pm$6.9 & 5.0$\pm$2.5 & ... & 11.3$\pm$3.4 & ... & ... & 36.9$\pm$9.7 & (1.0$^{+1.4}_{-0.6}$)$\times$10$^{11}$ & 23.4$\pm$8.0 \\
1HERMES\,X1.4\,J104632.12+585729.0$^{*}$... & 0.801$^{{\rm D}}$ & 2 & 0.74$^{{\rm L}}$ & 306.0$\pm$7.4 & ... & ... & 16.8$\pm$3.5 & 10.2$\pm$3.9 & 2.7$\pm$4.8 & 19.5$\pm$3.4 & (7.0$^{+55.}_{-6.2}$)$\times$10$^{11}$ & 36.6$\pm$66.8 \\
1HERMES\,X1.4\,J104633.17+585815.4$^{*}$... & 0.858$^{{\rm D}}$ & 3 & 0.84$^{{\rm L}}$ & 1205.$\pm$8.6 & 8.6$\pm$2.7 & 35.8$\pm$5.1 & 27.8$\pm$3.4 & 14.2$\pm$4.0 & 4.2$\pm$4.7 & 266.$\pm$15.6 & (9.0$^{+4.1}_{-2.8}$)$\times$10$^{11}$ & 28.2$\pm$4.8 \\
1HERMES\,X1.4\,J104633.84+585924.3$^{*}$... & 0.233$^{{\rm L}}$ & 5 & 2.68$^{{\rm L}}$ & 220.9$\pm$7.2 & 3.5$\pm$2.6 & ... & 38.7$\pm$3.4 & 29.3$\pm$3.9 & 13.2$\pm$4.7 & 91.5$\pm$6.8 & (3.0$^{+2.0}_{-1.2}$)$\times$10$^{10}$ & 15.1$\pm$2.8 \\
1HERMES\,X1.4\,J104633.88+584911.3$^{*}$... & 1.398$^{{\rm D}}$ & 5 & 1.66$^{{\rm L}}$ & 175.1$\pm$6.9 & 5.6$\pm$2.7 & ... & 12.5$\pm$3.5 & 9.5$\pm$4.1 & 1.5$\pm$5.2 & 41.2$\pm$6.4 & (1.0$^{+0.9}_{-0.5}$)$\times$10$^{12}$ & 34.0$\pm$11.2 \\
1HERMES\,X1.4\,J104634.48+585942.6$^{*}$... & 0.608$^{{\rm D}}$ & 4 & 0.36$^{{\rm L}}$ & 215.1$\pm$6.2 & 2.8$\pm$2.6 & 14.4$\pm$5.0 & 12.2$\pm$3.4 & 3.3$\pm$4.1 & 0.7$\pm$5.6 & 24.6$\pm$5.7 & (1.6$^{+3.2}_{-1.0}$)$\times$10$^{11}$ & 27.8$\pm$16.3 \\
\enddata
\tablecomments{Table~\ref{tab_giant} Continued.}
\end{deluxetable}
\clearpage
\begin{deluxetable}{l@{ }c@{\,}c@{\,}c@{ }c@{ }c@{ }c@{ }c@{ }c@{ }c@{ }cc@{ }c}
\tablecolumns{13}
\tablecaption{Spectroscopically identified 0$<z<$2 {\sc Spire}-selected Galaxies -- Continued.}
 
\tablehead{
\colhead{NAME} & \colhead{$z_{\rm spec}$} & \colhead{{\sc Conf}} & \colhead{$z_{\rm phot}$} & \colhead{$S_{\rm 24}$} 
&  \colhead{$S_{\rm 100}$} & \colhead{$S_{\rm 160}$} & \colhead{$S_{\rm 250}$} & \colhead{$S_{\rm 350}$} & \colhead{$S_{\rm 500}$} 
&  \colhead{$S_{\rm 1.4GHz}$} & \colhead{L$_{\rm IR}$} & \colhead{T$_{\rm dust}$} \\
\colhead{} & & & & (\uJy) & (mJy) & (mJy) & (mJy) & (mJy) & (mJy) & (\uJy) &  (\lsun) & (K) \\
 }
\startdata

1HERMES\,X1.4\,J104653.63+585545.2$^{*}$... & 0.587$^{{\rm D}}$ & 4 & ...$^{{\rm L}}$ & 185.4$\pm$7.4 & ... & ... & 6.3$\pm$3.4 & 13.7$\pm$5.1 & 16.7$\pm$4.7 & 23.8$\pm$4.3 & (8.2$^{+19.}_{-5.8}$)$\times$10$^{10}$ & 9.40$\pm$2.6 \\
1HERMES\,X24\,J104654.04+584749.8 ... & 0.651$^{{\rm D}}$ & 2 & 0.58$^{{\rm L}}$ & 187.1$\pm$7.9 & 5.2$\pm$2.7 & ... & 16.3$\pm$3.4 & 24.5$\pm$6.4 & ... & ... & (1.8$^{+2.0}_{-0.9}$)$\times$10$^{11}$ & 18.8$\pm$5.1 \\
1HERMES\,X24\,J104654.47+585629.5$^{*}$... & 1.063$^{{\rm D}}$ & 3 & 0.96$^{{\rm L}}$ & 257.9$\pm$7.4 & 5.2$\pm$2.6 & ... & 16.5$\pm$3.5 & 15.1$\pm$4.0 & 6.8$\pm$4.8 & ... & (6.3$^{+5.5}_{-2.9}$)$\times$10$^{11}$ & 25.7$\pm$7.3 \\
1HERMES\,X1.4\,J104654.92+585541.0$^{*}$... & 0.071$^{{\rm D}}$ & 5 & ...$^{{\rm L}}$ & 538.5$\pm$7.5 & ... & ... & 12.6$\pm$3.5 & 6.3$\pm$4.5 & ... & 73.8$\pm$14.0 & (2.9$^{+117}_{-2.9}$)$\times$10$^{9}$ & 24.4$\pm$64.4 \\
1HERMES\,X1.4\,J104654.96+585541.1$^{*}$... & 0.088$^{{\rm L}}$ & 5 & 0.14$^{{\rm L}}$ & 478.1$\pm$5.8 & 6.6$\pm$2.6 & 9.5$\pm$5.1 & 36.0$\pm$3.4 & 20.6$\pm$3.9 & 7.6$\pm$4.7 & 77.7$\pm$5.3 & (3.6$^{+2.2}_{-1.3}$)$\times$10$^{9}$ & 13.6$\pm$2.1 \\
1HERMES\,X1.4\,J104655.08+584911.9$^{*}$... & 0.308$^{{\rm D}}$ & 5 & 0.30$^{{\rm L}}$ & 539.6$\pm$8.4 & 10.4$\pm$2.6 & ... & 15.5$\pm$3.5 & 7.4$\pm$4.8 & ... & 71.3$\pm$15.6 & (4.6$^{+3.0}_{-1.8}$)$\times$10$^{10}$ & 22.1$\pm$3.9 \\
1HERMES\,X1.4\,J104655.14+590212.3$^{*}$... & 1.271$^{{\rm D}}$ & 2 & 1.26$^{{\rm L}}$ & 178.2$\pm$6.9 & ... & 9.9$\pm$5.0 & 16.1$\pm$3.4 & 10.8$\pm$3.9 & 1.9$\pm$4.9 & 39.2$\pm$7.5 & (8.7$^{+17.}_{-5.7}$)$\times$10$^{11}$ & 29.6$\pm$16.8 \\
1HERMES\,X1.4\,J104655.44+585438.1$^{*}$... & 0.288$^{{\rm L}}$ & 5 & 1.24$^{{\rm L}}$ & 617.6$\pm$7.4 & ... & ... & 19.7$\pm$3.5 & 10.4$\pm$4.2 & 0.8$\pm$4.9 & 52.3$\pm$10.1 & (1.2$^{+21.}_{-1.2}$)$\times$10$^{11}$ & 32.7$\pm$85.3 \\
\enddata
\tablecomments{Table~\ref{tab_giant} Continued.}
\end{deluxetable}
\clearpage
\begin{deluxetable}{l@{ }c@{\,}c@{\,}c@{ }c@{ }c@{ }c@{ }c@{ }c@{ }c@{ }cc@{ }c}
\tablecolumns{13}
\tablecaption{Spectroscopically identified 0$<z<$2 {\sc Spire}-selected Galaxies -- Continued.}
 
\tablehead{
\colhead{NAME} & \colhead{$z_{\rm spec}$} & \colhead{{\sc Conf}} & \colhead{$z_{\rm phot}$} & \colhead{$S_{\rm 24}$} 
&  \colhead{$S_{\rm 100}$} & \colhead{$S_{\rm 160}$} & \colhead{$S_{\rm 250}$} & \colhead{$S_{\rm 350}$} & \colhead{$S_{\rm 500}$} 
&  \colhead{$S_{\rm 1.4GHz}$} & \colhead{L$_{\rm IR}$} & \colhead{T$_{\rm dust}$} \\
\colhead{} & & & & (\uJy) & (mJy) & (mJy) & (mJy) & (mJy) & (mJy) & (\uJy) &  (\lsun) & (K) \\
 }
\startdata

1HERMES\,X1.4\,J104715.58+590122.9$^{*}$... & 0.274$^{{\rm D}}$ & 3 & 0.56$^{{\rm L}}$ & 2260.$\pm$9.2 & ... & ... & 34.6$\pm$3.5 & 30.4$\pm$4.0 & 10.7$\pm$5.4 & 85.6$\pm$6.2 & (7.2$^{+6.3}_{-3.3}$)$\times$10$^{10}$ & 14.6$\pm$3.0 \\
1HERMES\,X1.4\,J104716.80+590650.7$^{*}$... & 0.386$^{{\rm L}}$ & 3 & 1.72$^{{\rm L}}$ & 641.1$\pm$6.3 & 4.7$\pm$2.6 & ... & 22.7$\pm$3.5 & 13.3$\pm$4.0 & ... & 70.1$\pm$6.3 & (7.9$^{+9.3}_{-4.3}$)$\times$10$^{10}$ & 18.5$\pm$5.6 \\
1HERMES\,X1.4\,J104717.57+585934.5$^{*}$... & 1.036$^{{\rm D}}$ & 3 & ...$^{{\rm L}}$ & 317.6$\pm$7.3 & 4.7$\pm$2.7 & 28.1$\pm$4.8 & 12.3$\pm$3.4 & 9.6$\pm$4.2 & 2.8$\pm$5.0 & 45.7$\pm$9.5 & (1.1$^{+1.1}_{-0.5}$)$\times$10$^{12}$ & 38.1$\pm$15.8 \\
1HERMES\,X1.4\,J104718.73+590116.3$^{*}$... & 0.288$^{{\rm D}}$ & 4 & 0.40$^{{\rm L}}$ & 1842.$\pm$9.2 & 32.7$\pm$2.6 & 83.3$\pm$5.3 & 43.3$\pm$3.4 & 33.3$\pm$3.9 & 15.9$\pm$5.1 & 274.$\pm$16.7 & (1.4$^{+0.2}_{-0.1}$)$\times$10$^{11}$ & 21.6$\pm$1.0 \\
1HERMES\,X24\,J104718.77+585425.3$^{*}$... & 0.922$^{{\rm D}}$ & 3 & ...$^{{\rm L}}$ & 167.3$\pm$6.7 & ... & ... & 10.1$\pm$3.5 & 20.5$\pm$4.2 & ... & ... & (2.9$^{+9.8}_{-2.2}$)$\times$10$^{11}$ & 13.8$\pm$5.1 \\
1HERMES\,X1.4\,J104719.02+585928.7$^{*}$... & 1.037$^{{\rm D}}$ & 3 & ...$^{{\rm L}}$ & 361.4$\pm$6.2 & ... & ... & 11.5$\pm$3.4 & 8.6$\pm$4.3 & ... & 61.4$\pm$16.5 & (7.1$^{+94.}_{-6.6}$)$\times$10$^{11}$ & 35.5$\pm$68.6 \\
1HERMES\,X24\,J104719.59+590505.5$^{*}$... & 0.119$^{{\rm D}}$ & 5 & 0.16$^{{\rm L}}$ & 444.9$\pm$6.8 & 11.1$\pm$2.7 & 30.1$\pm$4.5 & 12.4$\pm$3.5 & 9.4$\pm$3.9 & ... & ... & (6.8$^{+4.0}_{-2.5}$)$\times$10$^{9}$ & 19.6$\pm$3.1 \\
1HERMES\,X1.4\,J104720.72+585153.0$^{*}$... & 1.300$^{{\rm D}}$ & 1 & ...$^{{\rm L}}$ & $-$ & 7.2$\pm$2.6 & ... & 15.0$\pm$3.4 & ... & ... & 47.7$\pm$7.1 & (1.1$^{+0.9}_{-0.5}$)$\times$10$^{12}$ & 34.2$\pm$9.1 \\
\enddata
\tablecomments{Table~\ref{tab_giant} Continued.}
\end{deluxetable}
\clearpage
\begin{deluxetable}{l@{ }c@{\,}c@{\,}c@{ }c@{ }c@{ }c@{ }c@{ }c@{ }c@{ }cc@{ }c}
\tablecolumns{13}
\tablecaption{Spectroscopically identified 0$<z<$2 {\sc Spire}-selected Galaxies -- Continued.}
 
\tablehead{
\colhead{NAME} & \colhead{$z_{\rm spec}$} & \colhead{{\sc Conf}} & \colhead{$z_{\rm phot}$} & \colhead{$S_{\rm 24}$} 
&  \colhead{$S_{\rm 100}$} & \colhead{$S_{\rm 160}$} & \colhead{$S_{\rm 250}$} & \colhead{$S_{\rm 350}$} & \colhead{$S_{\rm 500}$} 
&  \colhead{$S_{\rm 1.4GHz}$} & \colhead{L$_{\rm IR}$} & \colhead{T$_{\rm dust}$} \\
\colhead{} & & & & (\uJy) & (mJy) & (mJy) & (mJy) & (mJy) & (mJy) & (\uJy) &  (\lsun) & (K) \\
 }
\startdata

1HERMES\,X1.4\,J123521.11+622713.5$^{*}$... & 0.387$^{{\rm D}}$ & 5 & $-$$^{{\rm G}}$ & $-$ & $-$ & $-$ & 33.9$\pm$4.4 & 15.1$\pm$4.7 & 7.7$\pm$4.0 & 161.$\pm$14.0 & (4.0$^{+15.}_{-3.2}$)$\times$10$^{11}$ & 32.7$\pm$42.3 \\
1HERMES\,X1.4\,J123525.88+622858.6$^{*}$... & 1.009$^{{\rm D}}$ & 5 & $-$$^{{\rm G}}$ & $-$ & $-$ & $-$ & 23.7$\pm$4.4 & 19.3$\pm$4.7 & 17.0$\pm$4.0 & 91.8$\pm$14.6 & (5.7$^{+5.0}_{-2.6}$)$\times$10$^{11}$ & 19.7$\pm$4.7 \\
1HERMES\,X1.4\,J123530.97+622053.5$^{*}$... & 0.298$^{{\rm D}}$ & 4 & $-$$^{{\rm G}}$ & 230.8$\pm$22.6 & $-$ & $-$ & 17.3$\pm$4.4 & 8.1$\pm$4.7 & 5.0$\pm$4.0 & 46.1$\pm$8.6 & (5.2$^{+62.}_{-4.8}$)$\times$10$^{10}$ & 24.0$\pm$38.9 \\
1HERMES\,X1.4\,J123531.17+622015.5$^{*}$... & 1.043$^{{\rm D}}$ & 1 & $-$$^{{\rm G}}$ & 165.8$\pm$31.1 & $-$ & $-$ & 29.4$\pm$4.4 & 21.7$\pm$4.7 & 9.6$\pm$4.0 & 83.7$\pm$12.0 & (6.3$^{+4.4}_{-2.6}$)$\times$10$^{11}$ & 19.7$\pm$3.2 \\
1HERMES\,X1.4\,J123531.23+622419.2$^{*}$... & 0.489$^{{\rm D}}$ & 5 & $-$$^{{\rm G}}$ & 396.9$\pm$35.0 & $-$ & $-$ & 24.6$\pm$4.4 & 27.8$\pm$4.7 & 24.9$\pm$4.0 & 90.5$\pm$10.9 & (1.1$^{+0.6}_{-0.4}$)$\times$10$^{11}$ & 12.4$\pm$1.6 \\
1HERMES\,X1.4\,J123531.38+622809.6$^{*}$... & 1.261$^{{\rm D}}$ & 4 & $-$$^{{\rm G}}$ & $-$ & $-$ & $-$ & 68.9$\pm$4.4 & 56.7$\pm$4.7 & 32.1$\pm$4.0 & 186.$\pm$35.1 & (3.6$^{+1.2}_{-0.9}$)$\times$10$^{12}$ & 27.1$\pm$4.1 \\
1HERMES\,X1.4\,J123531.55+621118.0$^{*}$... & 1.213$^{{\rm D}}$ & 1 & $-$$^{{\rm G}}$ & 166.9$\pm$14.4 & $-$ & $-$ & 15.8$\pm$4.4 & 7.8$\pm$4.8 & 3.8$\pm$4.3 & 61.0$\pm$6.3 & (2.5$^{+40.}_{-2.3}$)$\times$10$^{12}$ & 53.5$\pm$152. \\
1HERMES\,X1.4\,J123531.77+622453.0$^{*}$... & 1.106$^{{\rm D}}$ & 3 & $-$$^{{\rm G}}$ & 291.4$\pm$38.1 & $-$ & $-$ & 51.0$\pm$4.4 & 44.3$\pm$4.7 & 27.7$\pm$4.0 & 257.$\pm$9.6 & (1.6$^{+0.7}_{-0.4}$)$\times$10$^{12}$ & 23.1$\pm$3.8 \\
\enddata
\tablecomments{Table~\ref{tab_giant} Continued.}
\end{deluxetable}
\clearpage
\begin{deluxetable}{l@{ }c@{\,}c@{\,}c@{ }c@{ }c@{ }c@{ }c@{ }c@{ }c@{ }cc@{ }c}
\tablecolumns{13}
\tablecaption{Spectroscopically identified 0$<z<$2 {\sc Spire}-selected Galaxies -- Continued.}
 
\tablehead{
\colhead{NAME} & \colhead{$z_{\rm spec}$} & \colhead{{\sc Conf}} & \colhead{$z_{\rm phot}$} & \colhead{$S_{\rm 24}$} 
&  \colhead{$S_{\rm 100}$} & \colhead{$S_{\rm 160}$} & \colhead{$S_{\rm 250}$} & \colhead{$S_{\rm 350}$} & \colhead{$S_{\rm 500}$} 
&  \colhead{$S_{\rm 1.4GHz}$} & \colhead{L$_{\rm IR}$} & \colhead{T$_{\rm dust}$} \\
\colhead{} & & & & (\uJy) & (mJy) & (mJy) & (mJy) & (mJy) & (mJy) & (\uJy) &  (\lsun) & (K) \\
 }
\startdata

1HERMES\,X1.4\,J123651.12+621031.1$^{*}$... & 0.492$^{{\rm D}}$ & 4 & $-$$^{{\rm G}}$ & 482.6$\pm$18.9 & $-$ & $-$ & 21.0$\pm$4.4 & 1.7$\pm$4.9 & ... & 86.8$\pm$7.8 & (5.4$^{+1.5}_{-1.2}$)$\times$10$^{11}$ & 41.3$\pm$1.9 \\
1HERMES\,X1.4\,J123651.50+620644.1$^{*}$... & 0.840$^{{\rm D}}$ & 3 & $-$$^{{\rm G}}$ & 320.2$\pm$16.5 & $-$ & $-$ & 23.6$\pm$4.4 & 8.8$\pm$4.8 & ... & 57.3$\pm$8.2 & (1.8$^{+0.4}_{-0.3}$)$\times$10$^{12}$ & 47.8$\pm$1.9 \\
1HERMES\,X1.4\,J123652.25+620205.2$^{*}$... & 0.610$^{{\rm D}}$ & 3 & $-$$^{{\rm G}}$ & 177.7$\pm$14.3 & $-$ & $-$ & 20.5$\pm$4.4 & 18.3$\pm$4.9 & 8.6$\pm$4.7 & 62.5$\pm$7.2 & (1.6$^{+3.6}_{-1.1}$)$\times$10$^{11}$ & 18.6$\pm$8.7 \\
1HERMES\,X1.4\,J123653.08+620432.6$^{*}$... & 0.909$^{{\rm D}}$ & 3 & $-$$^{{\rm G}}$ & 239.3$\pm$84.2 & $-$ & $-$ & 32.3$\pm$4.4 & 17.3$\pm$4.8 & ... & 67.8$\pm$5.7 & (6.0$^{+6.6}_{-3.1}$)$\times$10$^{11}$ & 21.5$\pm$5.8 \\
1HERMES\,X24\,J123653.21+621117.0$^{*}$... & 0.938$^{{\rm D}}$ & 3 & $-$$^{{\rm G}}$ & 211.9$\pm$22.6 & $-$ & $-$ & 13.5$\pm$4.4 & 16.1$\pm$4.8 & 1.7$\pm$7.7 & ... & (3.2$^{+14.}_{-2.6}$)$\times$10$^{11}$ & 18.8$\pm$11.0 \\
1HERMES\,X1.4\,J123653.37+621139.9$^{*}$... & 1.267$^{{\rm D}}$ & 3 & $-$$^{{\rm G}}$ & 158.7$\pm$23.8 & $-$ & $-$ & 18.9$\pm$4.4 & 11.0$\pm$4.7 & 7.9$\pm$4.4 & 84.2$\pm$8.5 & (1.1$^{+4.0}_{-0.9}$)$\times$10$^{12}$ & 32.1$\pm$29.9 \\
1HERMES\,X1.4\,J123654.29+620548.6$^{*}$... & 0.558$^{{\rm D}}$ & 4 & $-$$^{{\rm G}}$ & 218.0$\pm$19.0 & $-$ & $-$ & 23.8$\pm$4.4 & 5.0$\pm$4.8 & ... & 43.0$\pm$5.4 & (6.4$^{+1.6}_{-1.3}$)$\times$10$^{11}$ & 38.8$\pm$1.7 \\
1HERMES\,X1.4\,J123654.81+620535.0$^{*}$... & 0.622$^{{\rm D}}$ & 2 & $-$$^{{\rm G}}$ & 217.4$\pm$24.2 & $-$ & $-$ & 29.3$\pm$4.4 & 20.5$\pm$4.8 & 7.9$\pm$4.0 & 122.$\pm$10.6 & (3.1$^{+5.5}_{-2.0}$)$\times$10$^{11}$ & 23.8$\pm$13.0 \\
\enddata
\tablecomments{Table~\ref{tab_giant} Continued.}
\end{deluxetable}
\clearpage
\begin{deluxetable}{l@{ }c@{\,}c@{\,}c@{ }c@{ }c@{ }c@{ }c@{ }c@{ }c@{ }cc@{ }c}
\tablecolumns{13}
\tablecaption{Spectroscopically identified 0$<z<$2 {\sc Spire}-selected Galaxies -- Continued.}
 
\tablehead{
\colhead{NAME} & \colhead{$z_{\rm spec}$} & \colhead{{\sc Conf}} & \colhead{$z_{\rm phot}$} & \colhead{$S_{\rm 24}$} 
&  \colhead{$S_{\rm 100}$} & \colhead{$S_{\rm 160}$} & \colhead{$S_{\rm 250}$} & \colhead{$S_{\rm 350}$} & \colhead{$S_{\rm 500}$} 
&  \colhead{$S_{\rm 1.4GHz}$} & \colhead{L$_{\rm IR}$} & \colhead{T$_{\rm dust}$} \\
\colhead{} & & & & (\uJy) & (mJy) & (mJy) & (mJy) & (mJy) & (mJy) & (\uJy) &  (\lsun) & (K) \\
 }
\startdata

1HERMES\,X1.4\,J123741.63+620147.9$^{*}$... & 0.958$^{{\rm D}}$ & 3 & $-$$^{{\rm G}}$ & 474.6$\pm$19.3 & $-$ & $-$ & 32.8$\pm$4.4 & 24.9$\pm$4.7 & 16.0$\pm$4.2 & 81.9$\pm$8.6 & (8.5$^{+7.9}_{-4.1}$)$\times$10$^{11}$ & 23.2$\pm$7.1 \\
1HERMES\,X1.4\,J123744.63+621219.1$^{*}$... & 0.230$^{{\rm D}}$ & 4 & $-$$^{{\rm G}}$ & 248.2$\pm$59.2 & $-$ & $-$ & 26.3$\pm$4.4 & 24.1$\pm$5.0 & 15.0$\pm$5.1 & 51.3$\pm$5.4 & (1.9$^{+1.2}_{-0.7}$)$\times$10$^{10}$ & 12.9$\pm$1.7 \\
1HERMES\,X1.4\,J123745.28+615830.2$^{*}$... & 0.773$^{{\rm D}}$ & 3 & $-$$^{{\rm G}}$ & 0.00$\pm$66.9 & $-$ & $-$ & 23.6$\pm$4.4 & 17.0$\pm$4.7 & 13.4$\pm$4.0 & 65.0$\pm$10.3 & (3.2$^{+3.8}_{-1.7}$)$\times$10$^{11}$ & 19.6$\pm$6.1 \\
1HERMES\,X1.4\,J123745.69+620429.0$^{*}$... & 0.824$^{{\rm D}}$ & 4 & $-$$^{{\rm G}}$ & 329.1$\pm$17.2 & $-$ & $-$ & 24.4$\pm$4.4 & 16.9$\pm$4.7 & 6.4$\pm$4.0 & 56.0$\pm$6.8 & (5.9$^{+15.}_{-4.2}$)$\times$10$^{11}$ & 27.5$\pm$21.0 \\
1HERMES\,X24\,J123746.19+621440.8$^{*}$... & 0.667$^{{\rm D}}$ & 3 & $-$$^{{\rm G}}$ & 236.7$\pm$21.3 & $-$ & $-$ & 13.6$\pm$4.4 & 9.2$\pm$4.9 & ... & ... & (2.9$^{+67.}_{-2.8}$)$\times$10$^{11}$ & 30.7$\pm$70.6 \\
1HERMES\,X1.4\,J123749.88+622217.3$^{*}$... & 1.205$^{{\rm D}}$ & 2 & $-$$^{{\rm G}}$ & 167.0$\pm$96.4 & $-$ & $-$ & 38.8$\pm$4.4 & 29.6$\pm$4.7 & 18.4$\pm$4.0 & 53.6$\pm$7.9 & (1.5$^{+0.9}_{-0.5}$)$\times$10$^{12}$ & 24.4$\pm$5.2 \\
1HERMES\,X1.4\,J123750.48+621143.0$^{*}$... & 0.188$^{{\rm D}}$ & 5 & $-$$^{{\rm G}}$ & 506.5$\pm$37.7 & $-$ & $-$ & 40.9$\pm$4.4 & 17.3$\pm$4.8 & 7.6$\pm$4.3 & 160.$\pm$11.5 & (1.0$^{+0.1}_{-0.1}$)$\times$10$^{11}$ & 31.3$\pm$0.8 \\
1HERMES\,X1.4\,J123750.48+621143.0$^{*}$... & 0.188$^{{\rm D}}$ & 4 & $-$$^{{\rm G}}$ & 506.5$\pm$37.7 & $-$ & $-$ & 40.9$\pm$4.4 & 17.3$\pm$4.8 & 7.6$\pm$4.3 & 160.$\pm$11.5 & (1.0$^{+0.1}_{-0.1}$)$\times$10$^{11}$ & 31.3$\pm$0.8 \\
\enddata
\tablecomments{Table~\ref{tab_giant} Continued.}
\end{deluxetable}
\clearpage
\begin{deluxetable}{l@{ }c@{\,}c@{\,}c@{ }c@{ }c@{ }c@{ }c@{ }c@{ }c@{ }cc@{ }c}
\tablecolumns{13}
\tablecaption{Spectroscopically identified 0$<z<$2 {\sc Spire}-selected Galaxies -- Continued.}
 
\tablehead{
\colhead{NAME} & \colhead{$z_{\rm spec}$} & \colhead{{\sc Conf}} & \colhead{$z_{\rm phot}$} & \colhead{$S_{\rm 24}$} 
&  \colhead{$S_{\rm 100}$} & \colhead{$S_{\rm 160}$} & \colhead{$S_{\rm 250}$} & \colhead{$S_{\rm 350}$} & \colhead{$S_{\rm 500}$} 
&  \colhead{$S_{\rm 1.4GHz}$} & \colhead{L$_{\rm IR}$} & \colhead{T$_{\rm dust}$} \\
\colhead{} & & & & (\uJy) & (mJy) & (mJy) & (mJy) & (mJy) & (mJy) & (\uJy) &  (\lsun) & (K) \\
 }
\startdata

1HERMES\,X1.4\,J123829.37+621649.1$^{*}$... & 0.277$^{{\rm D}}$ & 5 & $-$$^{{\rm G}}$ & 842.4$\pm$17.9 & $-$ & $-$ & 26.9$\pm$4.4 & 8.4$\pm$4.7 & 4.1$\pm$4.0 & 228.$\pm$15.9 & (2.1$^{+0.4}_{-0.3}$)$\times$10$^{11}$ & 36.7$\pm$1.2 \\
1HERMES\,X1.4\,J123830.17+621401.0$^{*}$... & 0.954$^{{\rm D}}$ & 3 & $-$$^{{\rm G}}$ & 449.5$\pm$14.4 & $-$ & $-$ & 74.1$\pm$4.4 & 47.9$\pm$4.7 & 20.0$\pm$4.0 & 322.$\pm$12.5 & (2.7$^{+1.5}_{-0.9}$)$\times$10$^{12}$ & 31.5$\pm$9.4 \\
1HERMES\,X1.4\,J123843.47+621226.9$^{*}$... & 0.708$^{{\rm D}}$ & 3 & $-$$^{{\rm G}}$ & 310.3$\pm$16.9 & $-$ & $-$ & 13.8$\pm$4.4 & 6.0$\pm$4.7 & 1.6$\pm$4.0 & 95.9$\pm$15.5 & (8.2$^{+3.7}_{-2.5}$)$\times$10$^{11}$ & 47.0$\pm$3.1 \\
1HERMES\,X1.4\,J123854.93+621137.0$^{*}$... & 0.831$^{{\rm D}}$ & 3 & $-$$^{{\rm G}}$ & 581.3$\pm$17.7 & $-$ & $-$ & 35.1$\pm$4.4 & 26.1$\pm$4.7 & 12.3$\pm$4.0 & 222.$\pm$9.2 & (7.7$^{+8.3}_{-4.0}$)$\times$10$^{11}$ & 24.9$\pm$9.5 \\
1HERMES\,X1.4\,J123855.34+621231.2$^{*}$... & 1.328$^{{\rm D}}$ & 3 & $-$$^{{\rm G}}$ & 241.6$\pm$30.7 & $-$ & $-$ & 78.5$\pm$4.4 & 57.7$\pm$4.7 & 32.3$\pm$4.0 & 304.$\pm$21.2 & (4.5$^{+1.5}_{-1.1}$)$\times$10$^{12}$ & 30.2$\pm$5.1 \\
1HERMES\,X24\,J160509.74+534119.7$^{*}$... & 1.313$^{{\rm D}}$ & 2 & ...$^{{\rm E}}$ & 224.5$\pm$17.1 & $-$ & $-$ & 17.9$\pm$4.0 & 35.5$\pm$4.3 & 3.9$\pm$5.2 & $-$ & (1.1$^{+1.7}_{-0.7}$)$\times$10$^{12}$ & 16.5$\pm$4.0 \\
1HERMES\,X24\,J160523.19+534321.5$^{*}$... & 0.866$^{{\rm D}}$ & 3 & 1.17$^{{\rm E}}$ & 436.5$\pm$15.7 & $-$ & $-$ & 14.7$\pm$4.4 & 6.6$\pm$4.3 & 4.0$\pm$4.4 & $-$ & (6.9$^{+66.}_{-6.2}$)$\times$10$^{11}$ & 36.1$\pm$68.3 \\
1HERMES\,X24\,J160530.14+535656.0$^{*}$... & 0.961$^{{\rm D}}$ & 3 & 1.25$^{{\rm E}}$ & 1120.$\pm$16.1 & $-$ & $-$ & 38.0$\pm$4.4 & 19.6$\pm$4.8 & 5.9$\pm$6.7 & $-$ & (5.1$^{+0.6}_{-0.6}$)$\times$10$^{12}$ & 55.6$\pm$1.2 \\
\enddata
\tablecomments{Table~\ref{tab_giant} Continued.}
\end{deluxetable}
\clearpage
\begin{deluxetable}{l@{ }c@{\,}c@{\,}c@{ }c@{ }c@{ }c@{ }c@{ }c@{ }c@{ }cc@{ }c}
\tablecolumns{13}
\tablecaption{Spectroscopically identified 0$<z<$2 {\sc Spire}-selected Galaxies -- Continued.}
 
\tablehead{
\colhead{NAME} & \colhead{$z_{\rm spec}$} & \colhead{{\sc Conf}} & \colhead{$z_{\rm phot}$} & \colhead{$S_{\rm 24}$} 
&  \colhead{$S_{\rm 100}$} & \colhead{$S_{\rm 160}$} & \colhead{$S_{\rm 250}$} & \colhead{$S_{\rm 350}$} & \colhead{$S_{\rm 500}$} 
&  \colhead{$S_{\rm 1.4GHz}$} & \colhead{L$_{\rm IR}$} & \colhead{T$_{\rm dust}$} \\
\colhead{} & & & & (\uJy) & (mJy) & (mJy) & (mJy) & (mJy) & (mJy) & (\uJy) &  (\lsun) & (K) \\
 }
\startdata

1HERMES\,X24\,J160607.26+541449.7$^{*}$... & 1.096$^{{\rm D}}$ & 3 & 1.19$^{{\rm E}}$ & 276.1$\pm$15.7 & $-$ & $-$ & 14.7$\pm$4.3 & 12.2$\pm$4.3 & ... & $-$ & (7.2$^{+47.}_{-6.2}$)$\times$10$^{11}$ & 29.9$\pm$36.1 \\
1HERMES\,X24\,J160608.48+541351.0$^{*}$... & 0.951$^{{\rm D}}$ & 1 & ...$^{{\rm E}}$ & 376.7$\pm$16.7 & $-$ & $-$ & 19.2$\pm$4.3 & 24.0$\pm$4.4 & 29.2$\pm$5.0 & $-$ & (5.2$^{+3.6}_{-2.1}$)$\times$10$^{11}$ & 13.8$\pm$1.9 \\
1HERMES\,X24\,J160611.46+541347.3$^{*}$... & 1.224$^{{\rm D}}$ & 2 & 3.28$^{{\rm E}}$ & 417.9$\pm$15.1 & $-$ & $-$ & 16.2$\pm$4.3 & 17.2$\pm$4.4 & 0.4$\pm$4.9 & $-$ & (8.4$^{+30.}_{-6.6}$)$\times$10$^{11}$ & 23.2$\pm$14.6 \\
1HERMES\,X24\,J160613.13+541159.8$^{*}$... & 0.855$^{{\rm D}}$ & 1 & ...$^{{\rm E}}$ & 348.0$\pm$16.4 & $-$ & $-$ & 23.8$\pm$4.2 & 33.3$\pm$4.6 & 33.1$\pm$6.0 & $-$ & (4.7$^{+2.9}_{-1.8}$)$\times$10$^{11}$ & 13.8$\pm$1.8 \\
1HERMES\,X24\,J160616.58+533929.2$^{*}$... & 1.010$^{{\rm D}}$ & 1 & ...$^{{\rm E}}$ & 991.5$\pm$18.1 & $-$ & $-$ & 49.9$\pm$4.3 & 43.4$\pm$4.3 & 29.9$\pm$4.7 & $-$ & (1.4$^{+0.6}_{-0.4}$)$\times$10$^{12}$ & 21.4$\pm$3.2 \\
1HERMES\,X24\,J160622.32+541343.3$^{*}$... & 1.299$^{{\rm D}}$ & 3 & ...$^{{\rm E}}$ & $-$ & $-$ & $-$ & 10.5$\pm$4.3 & 6.0$\pm$4.3 & 4.7$\pm$4.5 & $-$ & (6.4$^{+229}_{-6.3}$)$\times$10$^{11}$ & 30.2$\pm$58.8 \\
1HERMES\,X24\,J160632.31+533656.6$^{*}$... & 0.998$^{{\rm D}}$ & 1 & ...$^{{\rm E}}$ & 155.0$\pm$14.6 & $-$ & $-$ & 14.2$\pm$4.2 & 16.1$\pm$7.4 & ... & $-$ & (3.6$^{+34.}_{-3.3}$)$\times$10$^{11}$ & 20.1$\pm$16.6 \\
1HERMES\,X24\,J160636.87+541256.8$^{*}$... & 1.006$^{{\rm D}}$ & 4 & 1.28$^{{\rm E}}$ & 534.4$\pm$16.4 & $-$ & $-$ & 22.4$\pm$4.3 & 21.0$\pm$4.7 & 13.3$\pm$4.7 & $-$ & (6.8$^{+9.2}_{-3.9}$)$\times$10$^{11}$ & 20.9$\pm$7.0 \\
\enddata
\tablecomments{Table~\ref{tab_giant} Continued.}
\end{deluxetable}%
\clearpage
\end{landscape}

\label{lastpage}
\end{document}